\documentclass[onecolumn,draft]{IEEEtran}
\usepackage{cite}
\ifCLASSINFOpdf
\else
\fi
\usepackage[cmex10]{amsmath}
\usepackage{amssymb}
\usepackage{latexsym}
\usepackage{amsthm}
\usepackage{mathrsfs}
\usepackage{bbm}

\usepackage{url}
\theoremstyle{definition}
\newtheorem{defn}{Definition}[section] % definition
\newtheorem{exam}[defn]{Example} % example
\newtheorem{remk}[defn]{Remark} % remark
\theoremstyle{plain}
\newtheorem{thrm}[defn]{Theorem} % theorem
\newtheorem{lemm}[defn]{Lemma} % lemma
\newtheorem{prop}[defn]{Proposition} % proposition
\newtheorem{coro}[defn]{Corollary} % corollary

\newcommand{\refexam}[1]{Example \ref{#1}}
\newcommand{\refremk}[1]{Remark \ref{#1}}
\newcommand{\refthrm}[1]{Theorem \ref{#1}}
\newcommand{\reflemm}[1]{Lemma \ref{#1}}
\newcommand{\refprop}[1]{Proposition \ref{#1}}

\newcommand{\N}{\mathbb{N}}
\newcommand{\Z}{\mathbb{Z}}

\newcommand{\R}{\mathbb{R}}

\newcommand{\set}[1]{\left\{ #1 \right\}}
\newcommand{\setbc}[2]{\left\{ #1:#2 \right\}}

\newcommand{\abs}[1]{\left| #1 \right|}
\newcommand{\norm}[1]{\left\| #1 \right\|}
\newcommand{\ra}{\,\rightarrow\,}
\newcommand{\Epsilon}{\mathcal{E}}

\newcommand{\defas}{:=}
\newcommand{\seq}[1]{\left( #1 \right)}

\newcommand{\Ind}{\mathbbm{1}}

\begin{document}
\title{A Generalized Typicality for Abstract Alphabets}

\author{\IEEEauthorblockN{Junekey Jeon}\\
\IEEEauthorblockA{Department of EE\\KAIST, Dajeon, Korea\\
Email:jk\_jeon@kaist.ac.kr}}


%Telephone: (800) 555--1212, Fax: (888) 555--1212}


\maketitle

\begin{abstract}
%\mathbfmath
A new notion of typicality for arbitrary probability measures on standard Borel 
spaces is proposed, which encompasses the classical notions of weak and strong typicality 
as special cases. Useful lemmas about strong typical sets, including conditional typicality 
lemma, joint typicality lemma, and packing and covering lemmas, which are 
fundamental tools for deriving many inner bounds of various multi-terminal 
coding problems, are obtained in terms of the proposed notion. 
This enables us to directly generalize lots of results on finite alphabet problems to 
general problems involving abstract alphabets, without any complicated additional arguments. 
For instance, quantization procedure is no longer necessary to achieve such generalizations. 
Another fundamental lemma, Markov lemma, is also obtained but 
its scope of application is quite limited compared to others. Yet, an alternative theory 
of typical sets for Gaussian measures, free from this limitation, is also developed. 
Some remarks on a possibility to generalize the proposed notion for sources with memory 
are also given. 
\end{abstract}
% then you can use LaTeX's standard command \mathbfmath at the very start



% \fi
\IEEEpeerreviewmaketitle

\let\ORIGthefootnote\thefootnote
\renewcommand{\thefootnote}{}
\footnote{This paper was in part presented at the IEEE International 
Symposium on Information Theory (ISIT), Honolulu, HI, USA, July 2014.}
\let\thefootnote\ORIGthefootnote\addtocounter{footnote}{-1}

\section{Introduction}
\IEEEPARstart{T}{he} notion of typicality is one of the central concepts in information 
theory, especially for deriving inner bounds of various coding problems. 
There are many notions of typicality used these days. Among them, perhaps the most 
convenient one for network information theory is the notion of so called the strong 
typicality~\cite[p.326]{CoverElements:2006}, or its variants such as the robust 
typicality~\cite{OrlitskyRoche:2001}. The notion is based on the idea that a long samples 
from an i.i.d. source has a property that its empirical distribution is sufficiently close 
to the true distribution with high probability. However, it is not so clear how to represent 
this ``closeness'' when the source takes infinitely many values, so usually 
strong typical sets (and their variants) were only defined on finite alphabets. 
On the other hand, there is another notion called the 
weak typicality~\cite[p.59]{CoverElements:2006} which can be defined for a wide class of 
sources, including i.i.d. sources on the Euclidean space of certain dimension with a 
well-defined density functions. However, it turns out that this notion of typicality is 
not so useful for various multi-terminal coding problems, since it lacks many 
properties of strong typicality that are widely used; for example, the conditional typicality 
lemma does not hold for weak typicality~\cite[p.32]{ElGamalNetwork:2011}. 
Accordingly, many previous researches on deriving inner bounds primarily concerned only 
finite alphabet problems. Hence, generalizing the strong typicality to a 
more general class of sources is a key to solve various multi-terminal 
coding problems for general alphabets. 

There were several attempts in this direction. For example, in \cite{Mitran:2010}, 
a generalization of strong typicality was introduced, which can be applied when 
the alphabet is a Polish space (a separable completely metrizable topological space). 
The main idea was based on a famous duality between the space of continuous 
bounded functions and the space of countably-additive compact-regular Borel measures. 
Lots of useful properties of usual strong typicality still hold here, but 
there are some limitations; the natural class of functions under consideration 
in this typicality, is the set of continuous bounded functions, which is too 
restrictive. Even the average power constraint for Gaussian channels cannot be directly 
handled, so a kind of truncation arguments were needed; see \cite[Section VI]{Mitran:2010}. 
Also, in \cite{Raginsky:2013}, a more general notion of strong typicality which can 
be applied when the alphabet is a standard Borel space~\cite{CohnMeasure:1980} 
(which is essentially just a Polish space, but topology need not be explicitly given) 
was introduced. However, this notion of typicality lacks some crucial properties of the 
usual strong typicality, including the conditional typicality lemma. Another notion of 
typicality which is applicable when the alphabet is countable was introduced 
in \cite{Siu-Wai:2010}. 

In this paper, a new notion of typicality for an arbitrary probability measure 
on a standard Borel space is proposed. The class of those measurable 
spaces is fairly general; in particular, every separable Banach space 
endowed with the Borel $\sigma$-algebra belongs to the class. It turns out that both 
the classical notions of strong and weak typicality are special cases of the proposed 
notion of typicality. Lots of useful results about strong typicality for finite alphabets 
continue to hold in this generalization. Those results were the fundamental tools for 
proving achievability. For instances, asymptotic equipartition property, conditional 
typicality lemma, joint typicality lemma, packing and covering lemmas (as well as 
there ``mutual versions'') can be derived in this generalization. Hence, one do not 
need to do anything further (such as quantization arguments) to generalize a 
result for finite alphabet case into the general case, whenever the result is a 
consequence of those lemmas. Another fundamental tool called Markov lemma, is also 
obtained but its scope of application is quite limited compared to others. 
However, it is shown that we can develop an alternative theory of typical sets so that 
those restrictions disappear, when every involved probability measure is Gaussian. 

The rest of this paper is organized as follows. We first introduce the new definition 
of typical sets in Section II, and introduce some basic properties 
in Section III. Joint typicality lemma, packing lemma, and covering lemma are obtained in 
Section IV. Section V is devoted to applications to coding problems. Section VI deals with 
Markov lemma. In Section VII, we show that limitations on Markov lemma can be very 
much relaxed when considering problems involving only Gaussian measures. 
Finally, we discuss how to extend the proposed notion to sources with memory in Section VIII. 

%%%%%%%%%%%%%%%%%%%%%%%%%%%%%%%%%%%%%%%%%%%%%%%%%%%%%%%%%%%%%%%%%%%%%%%%%%%%%%%%%%%%%%%%%%%%%%%%%%
\section{Definition of Typical Sets}
Most of the useful results about the strong typicality in the case of finite alphabet 
are based on a simple lemma called \emph{typical average lemma}~\cite[p.26]{ElGamalNetwork:2011}. 
The lemma says that the sample average of any nonnegative function on the alphabet should be 
close to the true average, whenever the samples are typical. 
The main idea of the new definition is to make a list of ``test functions'' for which 
the typical average lemma should hold. This idea is similar to the notions of typicality 
in \cite{Mitran:2010} or \cite{Raginsky:2013}. However, those notions only utilize bounded 
measurable functions. The notion of typicality defined in this paper utilizes any integrable 
functions, and such a treatment is required because many useful functions in information theory 
are actually unbounded. Use of unbounded functions makes some proofs (for instance, the 
proof of \refthrm{thrm:divergence lemma}) much easier and intuitive. In \cite{Raginsky:2013}, 
it is claimed that boundedness condition can be removed by considering suitable finite moment 
conditions and some straightforward truncation processes. But such processes are often 
time-consuming and tedious. Basically, the approach of this paper does not rely on boundedness, and 
such truncation processes are required for only some basic results. One can perhaps completely 
forget about boundedness and truncation issues when applying the results to actual 
coding problems. 

There are three parameters to determine conventional typical sets: a probability distribution $\mu$, 
the number of samples $n$, and a positive real number $\epsilon>0$. 
The $\epsilon>0$ determines how the empirical distribution should be close to the 
true distribution $\mu$; hence, one may call this $\epsilon$ as a \emph{typicality criterion}. 
However, the new definition requires some extra information rather than 
just a positive real number $\epsilon$ to determine this ``closeness''. 
The first one of those extra information is the list of test functions which are integrable, and 
not necessarily nonnegative. Those functions are the candidates for the typical average lemma. 
The second is a set of points in the alphabet ``to be excluded''; this is added due to some 
technical reasons, because it is crucial when proving some theorems. One can think of this 
``set of excluded points'' as something similar to the set of points at which the probability 
mass function vanishes for the case of finite alphabet (see \refexam{exam:strong typicality}). 
In \cite{Mitran:2010}, the ``closeness'' is given with respect to a metrizable topology, 
so no extra information other than $\epsilon$ was necessary. On the other hand, 
\cite{Raginsky:2013} uses a similar typicality criteria to that used here. 

Before giving the precise definition of the new notion of typicality, first we 
define some notations which will be used throughout this paper. 
The set of positive integers (excluding $0$) is denoted as $\Z^{+}$, and 
any function is assumed to be extended real-valued, if not specified. The base of a logarithm 
is always taken to be $2$. The terminal object in the category of sets (that is, a singleton 
set whose actual value of the element is not important) will be denoted as 
$\set{*}$. This set will be served as the trivial alphabet admitting the only one probability 
measure. For any measure-theoretic terminologies and notations that is not defined in this paper, 
refer to \cite{RoydenReal:2010}, \cite{LangReal:1993}, or \cite{RobertProb:2000}. 
Every measure in this paper is assumed to be positive and countably-additive. 
We often omit to write the $\sigma$-algebra of a measurable space. 
For a measurable space $(Z,\mathscr{C})$, the set of every probability measure on $Z$ is denoted 
as $\Delta(Z)$, and the set of every measure on $Z$ is denoted as $\mathcal{P}(Z)$. 
The point-mass measure at a point $x$ is denoted as $\mathfrak{d}_{x}$. 
For a measure $\mu$, the set of every $\mu$-integrable function 
is denoted as $\mathscr{L}^{1}(\mu)$. This $\mathscr{L}^{1}(\mu)$ is a set of functions; 
it is not a set of equivalence classes of $\mu$-almost equivalent functions. This distinction 
was made because the empirical distribution is sensitive to pointwise behaviors. The indicator 
function of a set $A$ is denoted as $\Ind_{A}$. Given a measure $\mu\in\mathcal{P}(X)$ and a 
measurable mapping $f:X\ra Y$, the \emph{pushforward} of $\mu$ by $f$ means the 
measure $f_{*}\mu:B\mapsto\mu(f^{-1}[B])$ on $Y$. The set of integers 
$m$ such that $u\leq m\leq\lceil U\rceil$ ($u\leq m\leq\lfloor U\rfloor$, respectively) 
for some integer $u$ and a real number $U\geq u$ will be denoted as $[u:U]$ 
($[u:U)$, respectively). We denote by $a:=b$ to say $a$ is defined as $b$. 

Throughout this section, let $(X,\mathscr{A})$ be a 
measurable space. This space $X$ will be served as the alphabet. 

\begin{defn}[Typicality criteria]\ \\
	Let $\mu\in\Delta(X)$. A \mbox{\emph{$\mu$-typicality criterion}} $\mathcal{U}$ is an 
	ordered triple $(\mathscr{F};\epsilon;N)$, starting from a finite collection 
	$\mathscr{F}=\set{f_{1},\ \cdots\ ,f_{M}}\subseteq\mathscr{L}^{1}(\mu)$ 
	of \mbox{$\mu$-integrable functions} together with a positive real number $\epsilon>0$ 
	and a $\mu$-null set $N$. We also write $(f_{1},\ \cdots\ ,f_{M};\epsilon;N)$ to denote 
	$(\mathscr{F};\epsilon;N)$. 
\end{defn}

The set of every $\mu$-typicality criterion naturally becomes a 
lattice (a poset having the supremum and the infimum for any pair of elements); 
for $\mu$-typicality criteria 
$\mathcal{U}_{1}=(\mathscr{F}_{1};\epsilon_{1};N_{1})$ 
and $\mathcal{U}_{2}=(\mathscr{F}_{2};\epsilon_{2};N_{2})$, we denote 
$\mathcal{U}_{1}\leq\mathcal{U}_{2}$ if $\mathscr{F}_{1}\supseteq\mathscr{F}_{2}$, 
$\epsilon_{1}\leq\epsilon_{2}$, and $N_{1}\supseteq N_{2}$, so that 
$\mathcal{U}_{1}\lor\mathcal{U}_{2}\defas(\mathscr{F}_{1}\cap\mathscr{F}_{2};
\max\set{\epsilon_{1},\epsilon_{2}};N_{1}\cap N_{2})$ is the least upper bound and 
$\mathcal{U}_{1}\land\mathcal{U}_{2}\defas(\mathscr{F}_{1}\cup\mathscr{F}_{2};
\min\set{\epsilon_{1},\epsilon_{2}};N_{1}\cup N_{2})$ is the greatest lower bound.

\begin{defn}[Typical sets]\ \\
	Let $\mu\in\Delta(X)$ and $n\in\Z^{+}$. Let $\mathcal{U}=(\mathscr{F};\epsilon;N)$ 
	be a $\mu$-typicality criterion. The \emph{$\mu$-typical set} of length $n$ 
	with respect to $\mathcal{U}$ is defined as
	\begin{eqnarray*}
		\mathcal{T}_{\mathcal{U}}^{(n)}(\mu)&\defas&
		\Bigg\{x^{n}\in(X\setminus N)^{n}:\abs{\frac{1}{n}\sum_{i=1}^{n}f(x_{i})
		-\int f\,d\mu}\leq\epsilon\quad
		\textrm{for all $f\in\mathscr{F}$}\Bigg\}.
	\end{eqnarray*}
\end{defn} 

By definition, a function in $\mathscr{F}$ automatically satisfies 
the typical average lemma. 

\begin{remk}\ 
	\begin{enumerate}
		\item Note that $\mathcal{T}_{\mathcal{U}_{1}}^{(n)}(\mu)\subseteq
		\mathcal{T}_{\mathcal{U}_{2}}^{(n)}(\mu)$ whenever 
		$\mathcal{U}_{1}\leq\mathcal{U}_{2}$. 
		\item The collection $\mathscr{F}$ can be empty; in that case, 
		$\mathcal{T}_{\mathcal{U}}^{(n)}(\mu)$ becomes $(X\setminus N)^{n}$. 
	\end{enumerate}
\end{remk}

In \cite{Mitran:2010}, a sequence is declared to be typical, if its empirical 
distribution belongs to a weak-* neighborhood of the true distribution. 
A basic open neighborhood in the weak-* topology is characterized by 
integrations of a finite collection of bounded continuous functions, 
so the notion of typicality in \cite{Mitran:2010} is essentially a special case 
of the notion of typicality just introduced. 

Typical sets should be measurable sets; otherwise, a notion such as ``the probability 
that a sequence is typical'' does not make sense. 

\begin{prop}\ \\
	Let $\mu\in\Delta(X)$, $n\in\Z^{+}$, and $\mathcal{U}$ be a $\mu$-typicality 
	criterion. Then, $\mathcal{T}_{\mathcal{U}}^{(n)}(\mu)$ is a measurable subset of 
	$(X^{n},\mathscr{A}^{\otimes n})$. 
\end{prop}

\begin{IEEEproof}
	Let $\mathcal{U}=(\mathscr{F};\epsilon;N)$. Consider the function given by 
	\begin{displaymath}
		e_{f}:x^{n}\mapsto\frac{1}{n}\sum_{i=1}^{n}f(x_{i})-\int f\,d\mu
	\end{displaymath}
	for each $f\in\mathscr{F}$. Since $f\in\mathscr{F}$ is measurable, it follows that 
	$e_{f}$ is measurable. Therefore, 
	$\mathcal{T}_{\mathcal{U}}^{(n)}(\mu)=(X\setminus N)^{n}\cap\bigcap_{f\in\mathscr{F}}
	e_{f}^{-1}\left[-\epsilon,\epsilon\right]$ is measurable. 
\end{IEEEproof}

Now, we give two familiar examples of typical sets. 

\begin{exam}[Strong typicality]\label{exam:strong typicality}\ \\
	Assume that $X$ is a nonempty finite set and $\mathscr{A}$ is the power set of $X$. 
	Then a probability measure \mbox{$\mu\in\Delta(X)$} can be completely 
	characterized by a probability mass function $p_{X}$ on $X$. Define 
	$N\defas\setbc{x\in X}{p_{X}(x)=0}$. For given $\epsilon>0$, define 
	$\mathcal{U}=\left(\set{\Ind_{\set{x}}}_{x\in X};\frac{\epsilon}{|X|};N\right)$, 
	then $\mathcal{T}_{\mathcal{U}}^{(n)}(\mu)$ is exactly the strong typical set 
	appearing in \cite[p.326]{CoverElements:2006}. On the other hand, 
	we get the robust typical set used in \cite{ElGamalNetwork:2011} by letting 
	$\mathcal{U}=\left(\set{\Ind_{\set{x}}/p_{X}(x)}_{x\in X\setminus N};
	\epsilon;N\right)$. 
\end{exam}

\begin{exam}[Weak typicality]\ \\
	Assume that $(X,\mathscr{A})$ is the real line with the Borel $\sigma$-algebra. 
	Let $\mu$ be a Borel probability measure having a density function $f_{X}$, 
	and assume that the differential entropy $h(\mu)\defas-\int\log f_{X}\,d\mu$ 
	exists and finite. Hence, $\log f_{X}\in\mathscr{L}^{1}(\mu)$, so 
	$\mathcal{U}\defas(\log f_{X};\epsilon;\emptyset)$ is a $\mu$-typicality criterion. 
	Then, $\mathcal{T}_{\mathcal{U}}^{(n)}(\mu)$ is exactly the weak typical set 
	appearing in \cite[p.59]{CoverElements:2006}. 
\end{exam}

Let $(Y,\mathscr{B})$ be another measurable space. We show that typical sets 
in $X$ can be related to typical sets in $Y$ by a measurable mapping 
from $X$ to $Y$. 

\begin{defn}[Pullback of typicality criteria]\ \\
	Let $\nu\in\Delta(Y)$ and $\mathcal{V}=(\mathscr{G};\epsilon;K)$ be a 
	$\nu$-typicality criterion. Let $\phi:X\ra Y$ be a measurable mapping. 
	Then the \emph{pullback} of $\mathcal{V}$ under $\phi$ is defined as 
	\begin{displaymath}
		\phi^{*}\mathcal{V}\defas\left(\setbc{g\circ\phi}{g\in\mathscr{G}};
		\epsilon;\phi^{-1}[K]\right).
	\end{displaymath}
\end{defn}

\begin{prop}\label{prop:typ pullback}\ \\
	Let $\mu\in\Delta(X)$ and $\phi:X\ra Y$ be a measurable mapping. 
	Let $\mathcal{V}$ be a $\phi_{*}\mu$-typicality criterion. Then, 
	$\phi^{*}\mathcal{V}$ is a $\mu$-typicality criterion, and 
	$\mathcal{T}_{\phi^{*}\mathcal{V}}^{(n)}(\mu)=(\phi^{n})^{-1}\left[
	\mathcal{T}_{\mathcal{V}}^{(n)}(\phi_{*}\mu)\right]$ for any $n\in\Z^{+}$, 
	where $\phi^{n}:X^{n}\ra Y^{n}$ is defined as 
	$\phi^{n}:x^{n}\mapsto\seq{\phi(x_{i})}_{i=1}^{n}$. 
\end{prop}

\begin{IEEEproof}
	Let $\mathcal{V}=(\mathscr{G};\epsilon;K)$. Note that for each $g\in\mathscr{G}$, 
	$\int|g\circ\phi|\,d\mu=\int|g|\,d\phi_{*}\mu<\infty$, thus 
	$\setbc{g\circ\phi}{g\in\mathscr{G}}\subseteq\mathscr{L}^{1}(\mu)$. 
	Also, $\mu(\phi^{-1}[K])=\phi_{*}\mu(K)=0$. Hence, $\phi^{*}\mathcal{V}$ is a 
	$\mu$-typicality criterion. Next, note that $x^{n}\in(\phi^{n})^{-1}\left[
	\mathcal{T}_{\mathcal{V}}^{(n)}(\phi_{*}\mu)\right]$ if and only if 
	$\seq{\phi(x_{i})}_{i=1}^{n}\in\mathcal{T}_{\mathcal{V}}^{(n)}(\phi_{*}\mu)$ 
	if and only if $\phi(x_{i})\notin K$ for all $i=1,\ \cdots\ ,n$ and 
	\begin{displaymath}
		\abs{\frac{1}{n}\sum_{i=1}^{n}(g\circ\phi)(x_{i})-\int(g\circ\phi)d\mu}
		=\abs{\frac{1}{n}\sum_{i=1}^{n}g(\phi(x_{i}))-\int g\,d\phi_{*}\mu}
		\leq\epsilon\quad\textrm{for all $g\in\mathscr{G}$},
	\end{displaymath}
	if and only if $x^{n}\in\mathcal{T}_{\phi^{*}\mathcal{V}}^{(n)}(\mu)$. 
\end{IEEEproof}

Hence, one can say that \emph{if a sequence in $X$ is typical with respect to 
$\mu$, then its image under $\phi$ in $Y$ is also typical with respect to 
$\phi_{*}\mu$}. In particular, this fact is important when $\phi$ is a projection. 
Consider $\mu\in\Delta(X\times Y)$ and the canonical projection 
$\phi=\pi_{X}:X\times Y\ra X$. By applying the proposition to this case, one can say 
that \emph{if a sequence in $X\times Y$ is typical with respect to $\mu$, then its 
$X$-components are also typical with respect to the marginal 
distribution of $\mu$}. 

%%%%%%%%%%%%%%%%%%%%%%%%%%%%%%%%%%%%%%%%%%%%%%%%%%%%%%%%%%%%%%%%%%%%%%%%%%%%%%%%%%%%%%%%%%%%%%%%%%
\section{Basic Properties}
In this section, we will explore some important properties of the proposed notion of 
typicality. Let $(X,\mathscr{A})$ and $(Y,\mathscr{B})$ be measurable spaces. 

The following is a simple consequence of the weak law of large numbers: 

\begin{thrm}[Asymptotic equipartition property]\ \\
	Let \mbox{$\mu\in\Delta(X)$} and $\mathcal{U}$ be a $\mu$-typicality criterion. Then, 
	\begin{displaymath}
		\lim_{n\ra\infty}\mu^{n}\left(\mathcal{T}_{\mathcal{U}}^{(n)}(\mu)\right)=1.
	\end{displaymath}
\end{thrm}

Here $\mu^{n}$ denotes the $n$-fold product measure of $\mu$. 
Two main reasons why the above theorem holds is: first, any function in $\mathscr{F}$ is 
$\mu$-integrable, and second, $\mathscr{F}$ is a finite set. 

\begin{IEEEproof}
	Let $\mathcal{U}=(\mathscr{F};\epsilon;N)$, then for each $f\in\mathscr{F}$, by 
	the weak law of large numbers, 
	\begin{displaymath}
		\lim_{n\ra\infty}\mu^{n}\left(\setbc{x^{n}\in X^{n}}
		{\left|\frac{1}{n}\sum_{i=1}^{n}f(x_{i})
		-\int f\,d\mu\right|\leq\epsilon}\right)=1.
	\end{displaymath}
	Note that the weak law of large numbers still holds 
	without the assumption of finite variance. Let $\delta>0$ be given, then for any 
	$f\in\mathscr{F}$, it follows that 
	\begin{displaymath}
		\mu^{n}\left(\setbc{x^{n}\in X^{n}}{\abs{\frac{1}{n}\sum_{i=1}^{n}
		f(x_{i})-\int f\,d\mu}>\epsilon}\right)\leq
		\frac{\delta}{\abs{\mathscr{F}}+1}
	\end{displaymath}
	for sufficiently large $n$; say, $n\geq n_{f}\in\Z^{+}$. 
	Since $\mu^{n}\left(X^{n}\setminus(X\setminus N)^{n}\right)=0$, 
	\begin{displaymath}
		\mu^{n}\left(X^{n}\setminus\mathcal{T}_{\mathcal{U}}^{(n)}(\mu)\right)
		\leq\frac{\delta\abs{\mathscr{F}}}
		{\abs{\mathscr{F}}+1}\leq\delta
	\end{displaymath}
	for $n\geq\max_{f\in\mathscr{F}}n_{f}$. Hence, 
	it follows that $\lim_{n\ra\infty}\mu^{n}
	\left(\mathcal{T}_{\mathcal{U}}^{(n)}(\mu)\right)=1$. 
\end{IEEEproof}

Using the result above, we will prove an important statement about the 
size of a typical set. In the next theorem, we use the notation $D(\mu\|\nu)$ for a 
probability measure $\mu$ and a $\sigma$-finite measure $\nu$ 
to denote the following quantity: 
\begin{displaymath}
	D(\mu\|\nu)\defas\begin{cases}\int\log\frac{d\mu}{d\nu}\,d\mu
	&\textrm{if $\mu\ll\nu$}\\\infty&\textrm{otherwise}\end{cases}
\end{displaymath}
provided that the integral exists, for the case when $\mu\ll\nu$. Here, 
$\mu\ll\nu$ means that $\mu$ is absolutely continuous with respect to $\nu$; 
that is, whenever $\nu(A)=0$ for some $A\in\mathscr{A}$, then $\mu(A)=0$. For $\mu\ll\nu$, 
$\frac{d\mu}{d\nu}$ is the Radon-Nikodym derivative~\cite{LangReal:1993} of $\mu$ with 
respect to $\nu$. If $\nu$ is a probability measure, then $D(\mu\|\nu)$ becomes the usual 
Kullback-Leibler divergence~\cite{GrayEntropy:2011}, but here we allow $\nu$ to be an 
arbitrary \mbox{$\sigma$-finite} measure. Hence, $D(\mu\|\nu)$ can be negative. 
In particular, when $\nu$ is the counting measure, $D(\mu\|\nu)$ is $-H(\mu)$, where $H(\mu)$ 
is the entropy of $\mu$, and when $\nu$ is the Lebesgue measure on $\R^{d}$, $D(\mu\|\nu)$ is 
$-h(\mu)$, where $h(\mu)$ is the differential entropy of $\mu$. 

\begin{thrm}[Divergence lemma]\label{thrm:divergence lemma}\ \\
	Let $\mu\in\Delta(X)$ and $\nu\in\mathcal{P}(X)$ be 
	$\sigma$-finite. Assume that $D(\mu\|\nu)$ exists; it can be either 
	finite, $+\infty$, or $-\infty$. 
	\begin{enumerate}
		\item If $D(\mu\|\nu)$ is finite, then for any $\epsilon>0$, there is a 
		$\mu$-typicality criterion $\mathcal{U}_{0}$ such that for any 
		$\mu$-typicality criterion $\mathcal{U}\leq\mathcal{U}_{0}$, we have 
		\begin{displaymath}
			\nu^{n}\left(\mathcal{T}_{\mathcal{U}}^{(n)}(\mu)\right)
			\leq2^{-n(D(\mu\|\nu)-\epsilon)}
		\end{displaymath}
		for all $n$ and 
		\begin{displaymath}
			\nu^{n}\left(\mathcal{T}_{\mathcal{U}}^{(n)}(\mu)\right)
			\geq2^{-n(D(\mu\|\nu)+\epsilon)}
		\end{displaymath}
		for all sufficiently large $n$. 
		\item If $\mu\not\ll\nu$, then there exists a $\mu$-typicality criterion 
		$\mathcal{U}_{0}$ such that for any $\mu$-typicality criterion 
		$\mathcal{U}\leq\mathcal{U}_{0}$, we have 
		\begin{displaymath}
			\nu^{n}\left(\mathcal{T}_{\mathcal{U}}^{(n)}(\mu)\right)=0
		\end{displaymath}
		for all $n$. 
		\item If $\mu\ll\nu$ and $D(\mu\|\nu)=+\infty$, then for any $M\geq0$, there is a 
		$\mu$-typicality criterion $\mathcal{U}_{0}$ such that for any $\mu$-typicality 
		criterion $\mathcal{U}\leq\mathcal{U}_{0}$, we have 
		\begin{displaymath}
			\nu^{n}\left(\mathcal{T}_{\mathcal{U}}^{(n)}(\mu)\right)\leq2^{-nM}
		\end{displaymath}
		for all $n$. 
		\item If $\mu\ll\nu$ and $D(\mu\|\nu)=-\infty$, then for any $M\geq0$, there is a 
		$\mu$-typicality criterion $\mathcal{U}_{0}$ such that for any $\mu$-typicality 
		criterion $\mathcal{U}\leq\mathcal{U}_{0}$, we have 
		\begin{displaymath}
			\nu^{n}\left(\mathcal{T}_{\mathcal{U}}^{(n)}(\mu)\right)\geq2^{nM}
		\end{displaymath}
		for all sufficiently large $n$. 
	\end{enumerate}
\end{thrm}

For the special case when $\nu$ is the counting measure (the Lebesgue measure, respectively), 
one can conclude that \emph{the cardinality (the volume, respectively) of a typical set is 
approximately the exponential of the entropy (the differential entropy, respectively)}. 

\begin{IEEEproof}
	\begin{enumerate}
		\item Choose $g=\frac{d\mu}{d\nu}$ and define
		$\mathcal{U}_{0}\defas(\log g;\epsilon';\emptyset)$, where $0<\epsilon'<\epsilon$. 
		Fix $n\in\Z^{+}$ and a $\mu$-typicality criterion $\mathcal{U}\leq\mathcal{U}_{0}$, 
		then by the definition of typical sets, 
		\begin{displaymath}
			D(\mu\|\nu)-\epsilon'\leq
			\frac{1}{n}\sum_{i=1}^{n}\log g(x_{i})\leq
			D(\mu\|\nu)+\epsilon'
		\end{displaymath}
		for $x^{n}\in\mathcal{T}_{\mathcal{U}}^{(n)}(\mu)$, so for that case we have 
		\begin{displaymath}
			2^{n(D(\mu\|\nu)-\epsilon')}\leq\prod_{i=1}^{n}g(x_{i})
			\leq2^{n(D(\mu\|\nu)+\epsilon')}.
		\end{displaymath}
		Consider the following identity: 
		\begin{displaymath}
			\mu^{n}\left(\mathcal{T}_{\mathcal{U}}^{(n)}(\mu)\right)
			=\int_{\mathcal{T}_{\mathcal{U}}^{(n)}(\mu)}\,d\mu^{n}
			=\int_{\mathcal{T}_{\mathcal{U}}^{(n)}(\mu)}\frac{d\mu^{n}}
			{d\nu^{n}}\,d\nu^{n}=\int_{\mathcal{T}_{\mathcal{\mathcal{U}}}^{(n)}(\mu)}
			\left(\prod_{i=1}^{n}\frac{d\mu}{d\nu}(x_{i})\right)d\nu^{n}(x^{n}).
		\end{displaymath}
		Hence, it follows that 
		\begin{displaymath}
			2^{n(D(\mu\|\nu)-\epsilon')}\nu^{n}\left(
			\mathcal{T}_{\mathcal{U}}^{(n)}(\mu)\right)\leq\mu^{n}\left(
			\mathcal{T}_{\mathcal{U}}^{(n)}(\mu)\right)\leq
			2^{n(D(\mu\|\nu)+\epsilon')}\nu^{n}\left(
			\mathcal{T}_{\mathcal{U}}^{(n)}(\mu)\right).
		\end{displaymath}
		Since we have $\mu^{n}\left(\mathcal{T}_{\mathcal{U}}^{(n)}(\mu)\right)\leq1$ 
		for all $n$, it follows that 
		\begin{displaymath}
			\nu^{n}\left(\mathcal{T}_{\mathcal{U}}^{(n)}(\mu)\right)
			\leq2^{-n(D(\mu\|\nu)-\epsilon')}\leq2^{-n(D(\mu\|\nu)-\epsilon)}
		\end{displaymath}
		for all $n$. Also, for sufficiently large $n$, we have 
		$\mu^{n}\left(\mathcal{T}_{\mathcal{U}}^{(n)}(\mu)\right)\geq1-\delta$ 
		for any given small $\delta\in(0,1)$, by the asymptotic equipartition property. 
		Therefore, for such $n$, 
		\begin{displaymath}
			\nu^{n}\left(\mathcal{T}_{\mathcal{U}}^{(n)}(\mu)\right)
			\geq(1-\delta)2^{-n(D(\mu\|\nu)+\epsilon')}
			=2^{-n(D(\mu\|\nu)+\epsilon'-\frac{1}{n}\log(1-\delta))}.
		\end{displaymath}
		By taking $n$ sufficiently large, we can assume that 
		$\epsilon'+\frac{1}{n}\log\frac{1}{1-\delta}<\epsilon$, thus 
		\begin{displaymath}
			\nu^{n}\left(\mathcal{T}_{\mathcal{U}}^{(n)}(\mu)\right)
			\geq2^{-n(D(\mu\|\nu)+\epsilon)}
		\end{displaymath}
		for sufficiently large $n$. 
			
		\item Pick $A\in\mathscr{A}$ such that $\nu(A)=0$ while $\mu(A)>0$. 
		Pick $\epsilon>0$ with $\epsilon<\mu(A)$ and 
		define $\mathcal{U}_{0}\defas(\Ind_{A};\epsilon;\emptyset)$. 
		Fix $n\in\Z^{+}$ and a $\mu$-typicality criterion 
		$\mathcal{U}\leq\mathcal{U}_{0}$, then for any 
		$x^{n}\in\mathcal{T}_{\mathcal{U}}^{(n)}(\mu)$, 
		\begin{displaymath}
			0<\mu(A)-\epsilon\leq\frac{1}{n}\sum_{i=1}^{n}\Ind_{A}(x_{i}), 
		\end{displaymath}
		so at least one $x_{i}$ should belong to $A$, concluding that 
		$\mathcal{T}_{\mathcal{U}}^{(n)}(\mu)\cap(X\setminus A)^{n}=\emptyset$. 
		Since $\nu^{n}\left((X\setminus A)^{n}\right)=1$, it follows that 
		$\nu^{n}\left(\mathcal{T}_{\mathcal{U}}^{(n)}(\mu)\right)=0$. 
			
		\item Choose $g=\frac{d\mu}{d\nu}$. For each $k\in\Z^{+}$, define a measurable 
		function $f_{k}$ on $X$ as 
		\begin{displaymath}
			f_{k}(x)\defas\begin{cases}
			\log g(x)&\textrm{if $g(x)\leq k$}\\0&\textrm{otherwise}\end{cases}
		\end{displaymath}
		for each $x\in X$. Note that $f_{k}^{-}=(\log g)^{-}$ for each $k\in\Z^{+}$ and 
		$(\log g)^{-}$ is $\mu$-integrable, since $D(\mu\|\nu)>0$. Since $f_{k}^{+}$ is 
		bounded, $f_{k}$ is $\mu$-integrable. Also, $\seq{f_{k}^{+}}_{k\in\Z^{+}}$ is an 
		increasing sequence of nonnegative measurable functions converging pointwise to 
		$(\log g)^{+}$ $\mu$-almost everywhere, since we know that 
		\begin{displaymath}
			\mu(\setbc{x\in X}{g(x)=\infty})=\nu(\setbc{x\in X}{g(x)=\infty})=0
		\end{displaymath}
		to have $\mu(X)<\infty$. So by monotone convergence theorem~\cite{RoydenReal:2010}, 
		$\int f_{k}^{+}\,d\mu\ra\int(\log g)^{+}\,d\mu=+\infty$ as $k\ra\infty$. 
		Since $f_{k}^{-}=(\log g)^{-}$ is integrable for all $k$, it follows that 
		$\int f_{k}\,d\mu\ra+\infty$ as $k\ra\infty$. Take $k\in\Z^{+}$ so that 
		$\int f_{k}\,d\mu\geq M+1$. Define $\mathcal{U}_{0}\defas(f_{k};1;\emptyset)$. 
		Fix $n\in\Z^{+}$ and a $\mu$-typicality criterion 
		$\mathcal{U}\leq\mathcal{U}_{0}$, then 
		\begin{displaymath}
			M\leq\int f_{k}\,d\mu-1\leq\frac{1}{n}\sum_{i=1}^{n}f_{k}(x_{i})
			\leq\frac{1}{n}\sum_{i=1}^{n}\log g(x_{i})
		\end{displaymath}
		for $x^{n}\in\mathcal{T}_{\mathcal{U}}^{(n)}(\mu)$, so for that case we have 
		\begin{displaymath}
			2^{nM}\leq\prod_{i=1}^{n}g(x_{i}).
		\end{displaymath}
		Proceeding as the same as the first part of the case 1, we get the result. 
			
		\item Choose $g=\frac{d\mu}{d\nu}$. For each $k\in\Z^{+}$, define a measurable 
		function $f_{k}$ on $X$ as 
		\begin{displaymath}
			f_{k}(x)\defas\begin{cases}\log g(x)&
			\textrm{if $\frac{1}{k}\leq g(x)$}\\0&\textrm{otherwise}\end{cases}
		\end{displaymath}
		for each $x\in X$. Note that $f_{k}^{+}=(\log g)^{+}$ for each $k\in\Z^{+}$ and 
		$(\log g)^{+}$ is $\mu$-integrable, since $D(\mu\|\nu)<0$. Since $f_{k}^{-}$ is 
		bounded, $f_{k}$ is $\mu$-integrable. Also, $\seq{f_{k}^{-}}_{k\in\Z^{+}}$ is an 
		increasing sequence of nonnegative measurable functions converging pointwise to 
		$(\log g)^{-}$ $\mu$-almost everywhere, since we know that 
		\begin{displaymath}
			\mu(\setbc{x\in X}{g(x)=0})=\int_{\setbc{x\in X}{g(x)=0}}g\,d\nu=0.
		\end{displaymath}
		So by monotone convergence theorem, we get $\int f_{k}\,d\mu\ra\int\log g\,d\mu=-\infty$ 
		as $k\ra\infty$ by considering positive parts and negative parts separately. 
		Take $k\in\Z^{+}$ so that $\int f_{k}\,d\mu\leq-M-2$. Define 
		$\mathcal{U}_{0}\defas(f_{k};1;\emptyset)$. Fix $n\in\Z^{+}$ and a $\mu$-typicality 
		criterion $\mathcal{U}\leq\mathcal{U}_{0}$, then 
		\begin{displaymath}
			-M-1\geq\int f_{k}\,d\mu+1
			\geq\frac{1}{n}\sum_{i=1}^{n}f_{k}(x_{i})
			\geq\frac{1}{n}\sum_{i=1}^{n}\log g(x_{i})
		\end{displaymath}
		for $x^{n}\in\mathcal{T}_{\mathcal{U}}^{(n)}(\mu)$, so for that case we have 
		\begin{displaymath}
			2^{-n(M+1)}\geq\prod_{i=1}^{n}g(x_{i}).
		\end{displaymath}
		Proceeding as the same as the second part of the case 1, we get the result. 
	\end{enumerate}
\end{IEEEproof}

Although the proofs are much delicate, conditional versions of above 
theorems are also true. Before stating them, let us look at some definitions. 
The following definition is from \cite[Chapter 4]{PollardUserGuide:1981}, but 
notations used here are different from it: 

\begin{defn}[Measure kernels]\ \\
	A \emph{measure kernel} from $X$ to $Y$ is a mapping 
	$\kappa:X\ra\mathcal{P}(Y)$ such that 
	$x\mapsto\kappa(x)(B)$ is a measurable function for all $B\in\mathscr{B}$. We write 
	$\kappa(B|x)$ to denote $\kappa(x)(B)$ for each $x\in X$ and $B\in\mathscr{B}$. 
	The integration of a function $g:Y\ra\R$ with respect to the measure $\kappa(x)$ is denoted as 
	$\int g(y)\,d\kappa(y|x)$ where $y$ is a dummy variable. 
	If $\kappa(x)\in\Delta(Y)$ for all $x\in X$, we call $\kappa$ a \emph{probability kernel}. 
	The set of every probability kernel from $X$ to $Y$ is denoted as $\mathcal{K}(X;Y)$. If 
	there exists a countable partition $\seq{A_{k}\times B_{k}}_{k\in\N^{+}}$ of $X\times Y$ 
	by measurable rectangles such that $\kappa(B_{k}|x)<\infty$ for all $x\in A_{k}$ for each 
	$k\in\Z^{+}$, then $\kappa$ is said to be \emph{$\sigma$-finite}. 
\end{defn}

The conditional distribution of a random variable with respect to another random variable 
is an example of probability kernels. One can also view a probability kernel $\kappa$ as a 
channel with the input alphabet $X$ and the output alphabet $Y$. 

For $\mu\in\mathcal{P}(X)$ and a $\sigma$-finite measure kernel 
$\kappa:X\ra\mathcal{P}(Y)$, one can construct a measure 
$\mu\rtimes\kappa\in\mathcal{P}(X\times Y)$ 
with the following property: for any $f\in\mathscr{L}^{1}(\mu\rtimes\kappa)$, 
the function $x\mapsto\int f(x,y)d\kappa(y|x)$ is measurable and 
\begin{displaymath}
	\int_{X\times Y}f\,d(\mu\rtimes\kappa)=\int_{X}\left[\int_{Y}
	f(x,y)\,d\kappa(y|x)\right]d\mu(x).
\end{displaymath}
If $\mu$ is $\sigma$-finite, then $\mu\rtimes\kappa$ is also $\sigma$-finite, and 
if $\mu\in\Delta(X)$ and $\kappa\in\mathcal{K}(X;Y)$, then 
$\mu\rtimes\kappa\in\Delta(X\times Y)$. If there is no potential confusion, we will denote 
$\mu\rtimes\kappa$ simply as $\mu\kappa$. Let $(Z,\mathscr{C})$ be another measurable space and 
$\kappa\in\mathcal{K}(X;Y)$, $\lambda\in\mathcal{K}(X\times Y;Z)$. Then we can define another 
probability kernel $\kappa\rtimes\lambda$ (or simply $\kappa\lambda$) from $X$ to 
$Y\times Z$ as $\kappa\rtimes\lambda:=x\mapsto\kappa(x)\rtimes\lambda(x,\cdot)$, and 
we have an identity $(\mu\kappa)\lambda=\mu(\kappa\lambda)$. 
For example, let $\sigma\in\mathcal{K}(X;Z)$ and treat it as an element in 
$\mathcal{K}(X\times Y;Z)$, then $(\mu\kappa)\sigma=\mu(\kappa\times\sigma)$; on the 
right-hand side, $\kappa\times\sigma:=x\ra\kappa(x)\times\sigma(x)$ is a kernel 
from $X$ to $Y\times Z$. Note that, if $\sigma$ is considered as a kernel from 
$X\times Y$ to $Z$, then $\kappa\times\sigma=\kappa\rtimes\sigma$. 
For details about kernels, refer to \cite[Chapter 4]{PollardUserGuide:1981}. 

\begin{remk}\label{remk:marginals}\ \\
	Let $\pi_{X}:X\times Y\ra X$ and $\pi_{Y}:X\times Y\ra Y$ be the canonical 
	projections. Note that $\pi_{X*}(\mu\kappa)=\mu$. We will denote 
	$\pi_{Y*}(\mu\kappa)$ as $\kappa_{*}\mu$. 
\end{remk}

The following notion is useful for discussions from now on. 

\begin{defn}[Conditional typical sets]\ \\
	Let $\nu\in\Delta(X\times Y)$ and $\mathcal{V}$ be a 
	$\nu$-typicality criterion. For $n\in\Z^{+}$ and \mbox{$x^{n}\in X^{n}$}, 
	we define the \emph{conditional $\mu$-typical set} of length $n$ with 
	respect to $\mathcal{V}$ given $x^{n}$ as 
	\begin{displaymath}
		\mathcal{T}_{\mathcal{V}}^{(n)}(\nu|x^{n})\defas\setbc{y^{n}\in Y^{n}}
		{(x^{n},y^{n})\in\mathcal{T}_{\mathcal{V}}^{(n)}(\nu)}.
	\end{displaymath}
\end{defn}

Note that a conditional typical set $\mathcal{T}_{\mathcal{V}}^{(n)}(\nu|x^{n})$ 
is always measurable, since it is a section of the joint typical set 
$\mathcal{T}_{\mathcal{V}}^{(n)}(\nu)$, which is 
$(\mathscr{A}^{\otimes n}\otimes\mathscr{B}^{\otimes n})$-measurable. 

Now, we will prove conditional typicality lemma of \cite[p.27]{ElGamalNetwork:2011} in 
our setting. 

\begin{defn}[Bounded typicality criteria]\ \\
	Let $\mu\in\Delta(X)$ and $\mathcal{U}:=(\mathscr{F};\epsilon;N)$ be a 
	$\mu$-typicality criterion. If $\mathscr{F}$ consists of 
	$\mu$-essentially bounded functions~\cite{CohnMeasure:1980}, we call $\mathcal{U}$ a 
	\emph{$\mu$-bounded typicality criterion}. 
\end{defn}

In the below, $\kappa^{n}:X^{n}\ra\Delta(Y^{n})$ denotes the probability 
kernel defined as $\kappa^{n}(x^{n})=\prod_{i=1}^{n}\kappa(x_{i})$ for each 
$x^{n}\in X^{n}$. 

\begin{thrm}[Bounded conditional typicality lemma]\label{thrm:bdd ctl}\ \\
	Let $\mu\in\Delta(X)$ and $\kappa\in\mathcal{K}(X;Y)$. Then, for any 
	$\mu\kappa$-bounded typicality criterion $\mathcal{V}$, there exists a 
	\mbox{$\mu$-bounded} typicality criterion $\mathcal{U}$ and a positive 
	number $c>0$ such that 
	\begin{displaymath}
		\sup_{x^{n}\in\mathcal{T}_{\mathcal{U}}^{(n)}(\mu)}
		\kappa^{n}\left(Y^{n}\setminus\mathcal{T}_{\mathcal{V}}^{(n)}
		(\mu\kappa|x^{n})\Big|x^{n}\right)\leq2^{-cn}
	\end{displaymath}
	for all sufficiently large $n\in\Z^{+}$. 
\end{thrm}

The above theorem says that whenever the criterion of being jointly typical
consists of essentially bounded functions, the probability that a random sequence 
$\bold{y}^{n}$, which is generated conditionally i.i.d. given a typical sequence $x^{n}$, 
is jointly typical with $x^{n}$, converges to $1$ exponentially fast. 

\begin{IEEEproof}
	Let $\mathcal{V}=(\mathscr{G};\epsilon;K)$. We may assume that $\mathscr{G}$ consists of a 
	single measurable function $g:X\times Y\ra\R$; one can easily modify the proof a little 
	bit to deal with the general case. We can also assume that $g$ is bounded on 
	$(X\times Y)\setminus K$ by enlarging $K$ if necessary. For each $x\in X$, let 
	$K_{x}:=\setbc{y\in Y}{(x,y)\in K}$, then 
	\begin{displaymath}
		0=\mu\kappa(K)=\int\kappa(K_{x}|x)\,d\mu(x),
	\end{displaymath}
	so there exists a $\mu$-null set $N$ so that $\kappa(K_{x}|x)=0$ for all $x\in X\setminus N$. 
	Define a function $f:X\ra\R$ as 
	\begin{displaymath}
		f:x\mapsto\int g(x,y)\,d\kappa(y|x),
	\end{displaymath}
	and let $\mathcal{U}:=(f;\epsilon';N)$ for some $\epsilon'\in(0,\epsilon)$, then 
	$\mathcal{U}$ is a $\mu$-bounded typicality criterion. 
	Fix $n\in\Z^{+}$ and $x^{n}\in\mathcal{T}_{\mathcal{U}}^{(n)}(\mu)$. Consider a set 
	\begin{displaymath}
		Z\defas\setbc{y^{n}\in Y^{n}}{\abs{\frac{1}{n}\sum_{i=1}^{n}g(x_{i},y_{i})
		-\frac{1}{n}\sum_{i=1}^{n}f(x_{i})}\geq\epsilon-\epsilon'}.
	\end{displaymath}
	From Hoeffding's inequality~\cite[Theorem 2]{Hoeffding:1963}, it follows that 
	\begin{displaymath}
		\kappa^{n}(Z|x^{n})\leq2\exp\left(-\frac{2n(\epsilon-\epsilon')^{2}}{M}\right)
	\end{displaymath}
	where $M>0$ is chosen so that $\abs{g}\leq M$. Since 
	$x^{n}\in\mathcal{T}_{\mathcal{U}}^{(n)}(\mu)$, one can easily verify that 
	\begin{displaymath}
		(Y^{n}\setminus Z)\cap\prod_{i=1}^{n}(Y\setminus K_{x_{i}})\subseteq
		\mathcal{T}_{\mathcal{V}}^{(n)}(\mu\kappa|x^{n}).
	\end{displaymath}
	Therefore, 
	\begin{eqnarray*}
		\kappa^{n}\left(Y^{n}\setminus\mathcal{T}_{\mathcal{V}}^{(n)}(\mu\kappa|x^{n})
		\Big|x^{n}\right)&\leq&\kappa^{n}(Z|x^{n})+\kappa^{n}\left(Y^{n}\setminus
		\prod_{i=1}^{n}(Y\setminus K_{x_{i}})\Bigg|x^{n}\right)\\&=&\kappa^{n}(Z|x^{n})\leq
		2\exp\left(-\frac{2n(\epsilon-\epsilon')^{2}}{M}\right).
	\end{eqnarray*}
	The above inequality holds for any $x^{n}\in\mathcal{T}_{\mathcal{U}}^{(n)}(\mu)$, 
	proving the theorem. 
\end{IEEEproof}

A similar result for general typicality criteria is also true: 

\begin{thrm}[Conditional typicality lemma]\label{thrm:cond typ lemma}\ \\
	Let $\mu\in\Delta(X)$ and $\kappa\in\mathcal{K}(X;Y)$. Then, for any 
	$\mu\kappa$-typicality criterion $\mathcal{V}$ and $\delta\in(0,1)$, 
	there exists a $\mu$-typicality criterion $\mathcal{U}$ such that 
	\begin{displaymath}
		\liminf_{n\ra\infty}\inf_{x^{n}\in\mathcal{T}_{\mathcal{U}}^{(n)}(\mu)}
		\kappa^{n}\left(\mathcal{T}_{\mathcal{V}}^{(n)}
		(\mu\kappa|x^{n})\Big|x^{n}\right)\geq1-\delta.
	\end{displaymath}
\end{thrm}

\begin{IEEEproof}
	As in the proof of the previous theorem, we may assume that $\mathcal{V}=(g;\epsilon;K)$ for some 
	$g\in\mathscr{L}^{1}(\mu\kappa)$, $\epsilon>0$, and a $\mu\kappa$-null set $K$. 
	We use a truncation argument; for each $k\in\Z^{+}$, define 
	\begin{displaymath}
		g_{k}(x,y):=\begin{cases}g(x,y)&\textrm{if $\abs{g(x,y)}\leq k$}
		\\0&\textrm{otherwise}\end{cases}
	\end{displaymath}
	for each $(x,y)\in X\times Y$, then $\seq{g_{k}}_{k\in\Z^{+}}$ is a sequence of bounded 
	measurable functions converging pointwise $\mu\kappa$-almost everywhere to $g$. 
	
	Define 
	\begin{displaymath}
		h_{k}:x\mapsto\int\abs{g(x,y)-g_{k}(x,y)}\,d\kappa(y|x),
	\end{displaymath}
	then from Lebesgue dominated convergence theorem~\cite{RoydenReal:2010}, 
	we know that $g_{k}\ra g$ in $\mathscr{L}^{1}(\mu\kappa)$ and 
	$h_{k}\ra0$ in $\mathscr{L}^{1}(\mu)$. Choose $k\in\Z^{+}$ such that 
	\begin{displaymath}
		\abs{\int g_{k}\,d\mu\kappa-\int g\,d\mu\kappa}\leq\frac{\epsilon}{3}
		\quad\textrm{and}\quad\int h_{k}\,d\mu\leq\frac{\epsilon\delta}{12}.
	\end{displaymath}
	Let $\mathcal{V}_{k}\defas\left(g_{k};\frac{\epsilon}{3};K\right)$, then 
	from \refthrm{thrm:bdd ctl}, we know that 
	\begin{displaymath}
		\lim_{n\ra\infty}\sup_{x^{n}\in\mathcal{T}_{\mathcal{U}_{k}}^{(n)}(\mu)}
		\kappa^{n}\left(Y^{n}\setminus\mathcal{T}_{\mathcal{V}_{k}}^{(n)}
		(\mu\kappa|x^{n})\Big|x^{n}\right)=0
	\end{displaymath}
	for some $\mu$-typicality criterion $\mathcal{U}_{k}$, 
	since $\mathcal{V}_{k}$ is a $\mu\kappa$-bounded typicality criterion. 
	Define $\mathcal{U}:=\mathcal{U}_{k}\wedge
	\left(h_{k};\frac{\epsilon\delta}{12};\emptyset\right)$. 
	
	Choose a sufficiently large $n\in\Z^{+}$ so that 
	\begin{displaymath}
		\kappa^{n}\left(\Epsilon_{1}(n)|x^{n}\right)\leq\frac{\delta}{2}\quad
		\textrm{where}\quad\Epsilon_{1}(n):=Y^{n}\setminus\mathcal{T}_{\mathcal{V}_{k}}^{(n)}
		(\mu\kappa|x^{n})
	\end{displaymath}
	for any given $x^{n}\in\mathcal{T}_{\mathcal{U}}^{(n)}(\mu)$. On the other hand, define 
	\begin{displaymath}
		\Epsilon_{2}(n)\defas\setbc{y^{n}\in Y^{n}}{\abs{\frac{1}{n}\sum_{i=1}^{n}g(x_{i},y_{i})
		-\frac{1}{n}\sum_{i=1}^{n}g_{k}(x_{i},y_{i})}>\frac{\epsilon}{3}}, 
	\end{displaymath}
	then since $x^{n}\in\mathcal{T}_{\mathcal{U}_{k}}^{(n)}(\mu)$, we have 
	\begin{displaymath}
		\frac{1}{n}\sum_{i=1}^{n}\int_{Y^{n}}
		\abs{g(x_{i},y_{i})-g_{k}(x_{i},y_{i})}d\kappa^{n}(y^{n}|x^{n})=
		\frac{1}{n}\sum_{i=1}^{n}h_{k}(x_{i})
		\leq\int h_{k}\,d\mu+\frac{\epsilon\delta}{12}\leq\frac{\epsilon\delta}{6}, 
	\end{displaymath}
	thus we can deduce 
	\begin{eqnarray*}
		\kappa^{n}\left(\Epsilon_{2}(n)|x^{n}\right)&\leq&
		\frac{3}{\epsilon}\cdot\frac{1}{n}\sum_{i=1}^{n}\int_{Y^{n}}
		\abs{g(x_{i},y_{i})-g_{k}(x_{i},y_{i})}d\kappa^{n}(y^{n}|x^{n})
		\leq\frac{\delta}{2}
	\end{eqnarray*}
	by Chevychev's inequality~\cite{RoydenReal:2010}. Note that if 
	$y^{n}\notin\Epsilon_{1}(n)\cup\Epsilon_{2}(n)$, then 
	$(x_{i},y_{i})\notin K$ for all $i=1,\ \cdots\ ,n$ and  
	\begin{eqnarray*}
		&&\abs{\frac{1}{n}\sum_{i=1}^{n}g(x_{i},y_{i})-\int g\,d\mu\kappa}\leq
		\abs{\frac{1}{n}\sum_{i=1}^{n}g(x_{i},y_{i})-
		\frac{1}{n}\sum_{i=1}^{n}g_{k}(x_{i},y_{i})}\\
		&&\quad\quad\quad\quad
		+\abs{\frac{1}{n}\sum_{i=1}^{n}g_{k}(x_{i},y_{i})-
		\int g_{k}\,d\mu\kappa}+\abs{\int g_{k}\,d\mu\kappa-\int g\,d\mu\kappa}\\
		&&\quad\quad\quad\leq
		\frac{\epsilon}{3}+\frac{\epsilon}{3}+\frac{\epsilon}{3}=\epsilon
	\end{eqnarray*}
	so $(x^{n},y^{n})\in\mathcal{T}_{\mathcal{V}}^{(n)}(\mu\kappa)$, 
	concluding that $\kappa^{n}\left(\mathcal{T}_{\mathcal{V}}^{(n)}
	(\mu\kappa|x^{n})\Big|x^{n}\right)\geq1-\delta$.
\end{IEEEproof}

\begin{remk}\ 
	\begin{enumerate}
		\item In the proof, the choice of $\mathcal{U}$ depends on $\delta$. However, 
		even if when test functions in $\mathcal{V}$ are not bounded, one can prove that 
		there exists $\mathcal{U}$ so that 
		\begin{displaymath}
			\lim_{n\ra\infty}\inf_{x^{n}\in\mathcal{T}_{\mathcal{U}}^{(n)}(\mu)}
			\kappa^{n}\left(\mathcal{T}_{\mathcal{V}}^{(n)}(\mu\kappa|x^{n})
			\Big|x^{n}\right)=1
		\end{displaymath}
		if test functions satisfy some finite moment conditions. 
		In Chapter 6, an argument of this kind is stated in detail. 
		\item We have seen in Chapter 2 that, if $(x^{n},y^{n})$ are jointly typical 
		for some $y^{n}$, then $x^{n}$ should be typical. Conditional 
		typicality lemma can be seen as a kind of converse to this. 
	\end{enumerate}
\end{remk}

Now, a conditional version of the divergence lemma also can be proved by using 
this conditional typicality lemma instead of the asymptotic equipartition property. 
Up to here, we did not impose any assumptions on measurable spaces; therefore, 
all theorems we have stated are true for arbitrary alphabets (that is, arbitrary measurable 
spaces). However, the proof given here of the following theorem relies on a lemma 
(see \reflemm{lemm:joint RND}) which uses the assumption that $(Y,\mathscr{B})$ is 
\emph{countably-generated}; that is, there exists a countable subset $\mathscr{G}$ of 
$\mathscr{B}$ so that $\mathscr{B}$ is the smallest $\sigma$-algebra containing 
$\mathscr{G}$. Hence, from now on we assume that $(Y,\mathscr{B})$ is countably-generated. 
Except the lemma, the whole procedure of the proof 
is similar to that of \mbox{\refthrm{thrm:divergence lemma}}. 

\begin{thrm}[Conditional divergence lemma]\label{thrm:cond div lemma}\ \\
	Let $\mu\in\Delta(X)$ and $\kappa\in\mathcal{K}(X;Y)$. 
	Let $\lambda:X\ra\mathcal{P}(Y)$ be a 
	\mbox{$\sigma$-finite} measure kernel such that 
	\mbox{$D(\mu\kappa\|\mu\lambda)$} exists. 
	\begin{enumerate}
		\item If $D(\mu\kappa\|\mu\lambda)$ is finite, then 
		for any $\epsilon>0$, there is a $\mu\kappa$-typicality criterion 
		$\mathcal{V}_{0}$ such that for any $\mu\kappa$-typicality criterion 
		$\mathcal{V}\leq\mathcal{V}_{0}$, we have 
		\begin{eqnarray*}
			&&\sup_{x^{n}\in X^{n}}\lambda^{n}\left(
			\mathcal{T}_{\mathcal{V}}^{(n)}(\mu\kappa|x^{n})\Big|x^{n}\right)
			\leq2^{-n\left(D\left(\mu\kappa\|\mu\lambda\right)-\epsilon\right)}
		\end{eqnarray*}
		for all $n$, and there exists a $\mu$-typicality criterion 
		$\mathcal{U}$ (depending on $\mathcal{V}$) so that 
		\begin{eqnarray*}
			&&\inf_{x^{n}\in\mathcal{T}_{\mathcal{U}}^{(n)}(\mu)}
			\lambda^{n}\left(\mathcal{T}_{\mathcal{V}}^{(n)}
			(\mu\kappa|x^{n})\Big|x^{n}\right)\geq2^{-n\left(D\left(
			\mu\kappa\|\mu\lambda\right)+\epsilon\right)}
		\end{eqnarray*}
		for all sufficiently large $n$. 
		\item If $\mu\kappa\not\ll\mu\lambda$, then there is a 
		$\mu\kappa$-typicality criterion $\mathcal{V}_{0}$ such that for any 
		$\mu\kappa$-typicality criterion $\mathcal{V}\leq\mathcal{V}_{0}$, we have 
		\begin{eqnarray*}
			&&\sup_{x^{n}\in X^{n}}\lambda^{n}\left(
			\mathcal{T}_{\mathcal{V}}^{(n)}(\mu\kappa|x^{n})\Big|x^{n}\right)=0
		\end{eqnarray*}
		for all $n$.
		\item If $\mu\kappa\ll\mu\lambda$ and $D(\mu\kappa\|\mu\lambda)=+\infty$, then 
		for any $M\geq0$, there is a $\mu\kappa$-typicality criterion 
		$\mathcal{V}_{0}$ such that for any $\mu\kappa$-typicality criterion 
		$\mathcal{V}\leq\mathcal{V}_{0}$, we have 
		\begin{eqnarray*}
			&&\sup_{x^{n}\in X^{n}}\lambda^{n}\left(\mathcal{T}_{\mathcal{V}}^{(n)}
			(\mu\kappa|x^{n})\Big|x^{n}\right)\leq2^{-nM}
		\end{eqnarray*}
		for all $n$. 
		\item If $D(\mu\kappa\|\mu\lambda)=-\infty$, then 
		for any $M\geq0$, there is a $\mu\kappa$-typicality criterion 
		$\mathcal{V}_{0}$ such that for any $\mu\kappa$-typicality criterion 
		$\mathcal{V}\leq\mathcal{V}_{0}$, there exists a $\mu$-typicality criterion 
		$\mathcal{U}$ such that 
		\begin{displaymath}
			\inf_{x^{n}\in\mathcal{T}_{\mathcal{U}}^{(n)}(\mu)}
			\lambda^{n}\left(\mathcal{T}_{\mathcal{V}}^{(n)}
			(\mu\kappa|x^{n})\Big|x^{n}\right)\geq2^{nM}
		\end{displaymath}
		for all sufficiently large $n$. 
	\end{enumerate}
\end{thrm}

For the special case when $\lambda$ is identically the counting measure 
(the Lebesgue measure, respectively), one can conclude that \emph{a typical cardinality 
(volume, respectively) of a conditional typical set is approximately the exponential of 
the conditional entropy (conditional differential entropy, respectively)}. 
Many other statements about the size of typical sets are also simple corollaries of this lemma. 

\begin{IEEEproof}
	\begin{enumerate}
		\item Choose $g=\frac{d\mu\kappa}{d\mu\lambda}$, then 
		there exists a $\mu$-null set $N$ so that $\kappa(x)\ll\lambda(x)$ and 
		$g(x,\cdot)=\frac{d\kappa(x)}{d\lambda(x)}$ for all 
		$x\in X\setminus N$ by \reflemm{lemm:joint RND}. Define $\mathcal{V}_{0}\defas
		(\log g;\epsilon';N\times Y)$ for some $\epsilon'\in(0,\epsilon)$. Fix a 
		$\mu\kappa$-typicality criterion $\mathcal{V}\leq\mathcal{V}_{0}$, then 
		\begin{displaymath}
			D(\mu\kappa\|\mu\lambda)-\epsilon'\leq
			\frac{1}{n}\sum_{i=1}^{n}\log g(x_{i},y_{i})\leq
			D(\mu\kappa\|\mu\lambda)+\epsilon'
		\end{displaymath}
		for $(x^{n},y^{n})\in\mathcal{T}_{\mathcal{V}}^{(n)}(\mu\kappa)$, 
		so for that case we have 
		\begin{displaymath}
			2^{n(D(\mu\kappa\|\mu\lambda)-\epsilon')}
			\leq\prod_{i=1}^{n}g(x_{i},y_{i})\leq
			2^{n(D(\mu\kappa\|\mu\lambda)+\epsilon')}.
		\end{displaymath}
		Note that if $(x^{n},y^{n})\in\mathcal{T}_{\mathcal{V}}^{(n)}(\mu\kappa)$, then 
		\begin{displaymath}
			\prod_{i=1}^{n}g(x_{i},y_{i})=
			\frac{d\kappa^{n}(x^{n})}{d\lambda^{n}(x^{n})}(y^{n}). 
		\end{displaymath}
		Thus, we get 
		\begin{displaymath}
			1\geq\kappa^{n}\left(\mathcal{T}_{\mathcal{V}}^{(n)}
			(\mu\kappa|x^{n})\Big|x^{n}\right)\geq2^{n(
			D(\mu\kappa\|\mu\lambda)-\epsilon')}
			\lambda^{n}\left(\mathcal{T}_{\mathcal{V}}^{(n)}
			(\mu\kappa|x^{n})\Big|x^{n}\right)
		\end{displaymath}
		concluding that 
		\begin{displaymath}
			\lambda^{n}\left(\mathcal{T}_{\mathcal{V}}^{(n)}(\mu\kappa|x^{n})\Big|x^{n}
			\right)\leq2^{-n\left(D(\mu\kappa\|\mu\lambda\right)-\epsilon')}
			\leq2^{-n\left(D(\mu\kappa\|\mu\lambda\right)-\epsilon)}
		\end{displaymath}
		for all $x^{n}\in X^{n}$, for all $n$. Note that if 
		$x^{n}\notin(X\setminus N)^{n}$, then the inequality trivially holds, because 
		$\mathcal{T}_{\mathcal{V}}^{(n)}(\mu\kappa|x^{n})$ is the empty set. 
		
		For the second part of the theorem, note that from the conditional typicality 
		lemma, we get a $\mu$-typicality criterion $\mathcal{U}$ for some 
		$\delta\in(0,1)$ so that 
		\begin{displaymath}
			\inf_{x^{n}\in\mathcal{T}_{\mathcal{U}}^{(n)}(\mu)}
			\kappa^{n}\left(\mathcal{T}_{\mathcal{V}}^{(n)}
			(\mu\kappa)\Big|x^{n}\right)\geq1-\delta
		\end{displaymath}
		for any sufficiently large $n$. Since we know 
		\begin{displaymath}
			\kappa^{n}\left(\mathcal{T}_{\mathcal{V}}^{(n)}
			(\mu\kappa|x^{n})\Big|x^{n}\right)\leq2^{n(
			D(\mu\kappa\|\mu\lambda)+\epsilon')}\lambda^{n}\left(
			\mathcal{T}_{\mathcal{V}}^{(n)}(\mu\kappa|x^{n})\Big|x^{n}\right)
		\end{displaymath}
		for every $x^{n}\in X^{n}$, it follows that for sufficiently large $n$, 
		\begin{displaymath}
			\lambda^{n}\left(\mathcal{T}_{\mathcal{V}}^{(n)}(\mu\kappa|x^{n})
			\Big|x^{n}\right)\geq(1-\delta)2^{-n(D(\mu\kappa\|\mu\lambda)+\epsilon')}
		\end{displaymath}
		for all $x^{n}\in\mathcal{T}_{\mathcal{U}}^{(n)}(\mu)$. Take $n$ large enough 
		to satisfy $\epsilon'+\frac{1}{n}\log\frac{1}{1-\delta}<\epsilon$, then we get 
		\begin{displaymath}
			\lambda^{n}\left(\mathcal{T}_{\mathcal{V}}^{(n)}
			(\mu\kappa|x^{n})\Big|x^{n}\right)\geq2^{-n(D(\mu\kappa\|\mu\lambda)+\epsilon)}
		\end{displaymath}
		for all $x^{n}\in\mathcal{T}_{\mathcal{U}}^{(n)}(\mu)$, 
		for sufficiently large $n$. 
		
		\item Pick $C\in\mathscr{A}\otimes\mathscr{B}$ such that $\mu\lambda(C)=0$ while 
		$\mu\kappa(C)>0$. Pick $\epsilon>0$ with $\epsilon<\mu\kappa(C)$. For $x\in X$, 
		let $C_{x}\defas\setbc{y\in Y}{(x,y)\in C}$, then there exists a $\mu$-null set 
		$N$ such that $\lambda(C_{x}|x)=0$ for all $x\in X\setminus N$, since 
		$\mu\lambda(C)=\int\lambda(C_{x}|x)\,d\mu(x)=0$. 
		Define $\mathcal{V}_{0}\defas(\Ind_{C};\epsilon;N\times Y)$. Fix $n\in\Z^{+}$ 
		and a $\mu\kappa$-typicality criterion $\mathcal{V}\leq\mathcal{V}_{0}$, 
		then for any $(x^{n},y^{n})\in\mathcal{T}_{\mathcal{V}}^{(n)}(\mu\kappa)$, 
		\begin{displaymath}
			0<\mu\kappa(C)-\epsilon\leq\frac{1}{n}\sum_{i=1}^{n}\Ind_{C}(x_{i},y_{i})
		\end{displaymath}
		so at least one $(x_{i},y_{i})$ should belong to $C$, concluding that 
		\begin{displaymath}
			\mathcal{T}_{\mathcal{V}}^{(n)}(\mu\kappa|x^{n})\cap
			\prod_{i=1}^{n}(Y\setminus C_{x_{i}})=\emptyset.
		\end{displaymath}
		Since $\lambda^{n}\left(\prod_{i=1}^{n}(Y\setminus C_{x_{i}})\big|x^{n}\right)
		=\prod_{i=1}^{n}\lambda(Y\setminus C_{x_{i}}|x_{i})=1$ whenever 
		$x_{i}\notin N$ for all $i=1,\ \cdots\ ,n$, it 
		follows that $\lambda^{n}\left(\mathcal{T}_{\mathcal{V}}^{(n)}
		(\mu\kappa|x^{n})\Big|x^{n}\right)=0$ for all $x^{n}$. 
		
		\item Choose $g=\frac{d\mu\kappa}{d\mu\lambda}$, and 
		take $N$ as the case 1. For each $k\in\Z^{+}$, define a measurable 
		function $f_{k}$ on $X\times Y$ as 
		\begin{displaymath}
			f_{k}(x,y)\defas\begin{cases}\log g(x,y)&\textrm{if $g(x,y)\leq k$}\\
			0&\textrm{otherwise}\end{cases}
		\end{displaymath}
		for each $(x,y)\in X\times Y$. Note that $f_{k}^{-}=(\log g)^{-}$ for each 
		$k\in\Z^{+}$ and $(\log g)^{-}$ is $\mu\kappa$-integrable, since 
		$D(\mu\kappa\|\mu\lambda)>0$. Since $f_{k}^{+}$ is bounded, $f_{k}$ is 
		$\mu\kappa$-integrable. Also, $\seq{f_{k}^{+}}_{k\in\Z^{+}}$ 
		is an increasing sequence of nonnegative measurable functions converging 
		pointwise to $(\log g)^{+}$ $\mu\kappa$-almost everywhere. So by monotone 
		convergence theorem, we get $\int f_{k}\,d\mu\kappa\ra\int\log g\,d\mu\kappa=+\infty$ 
		as $k\ra\infty$ by considering positive parts and negative parts separately. 
		Take $k\in\Z^{+}$ so that $\int f_{k}\,d\mu\kappa\geq M+1$. 
		Define $\mathcal{V}_{0}\defas(f_{k};1;N\times Y)$. Fix $n\in\Z^{+}$ and a 
		$\mu\kappa$-typicality criterion $\mathcal{V}\leq\mathcal{V}_{0}$, then 
		\begin{displaymath}
			M\leq\int f_{k}\,d\mu\kappa-1\leq\frac{1}{n}\sum_{i=1}^{n}f_{k}(x_{i},y_{i})
			\leq\frac{1}{n}\sum_{i=1}^{n}\log g(x_{i},y_{i})
		\end{displaymath}
		for $(x^{n},y^{n})\in\mathcal{T}_{\mathcal{V}}^{(n)}(\mu\kappa)$. Since 
		\begin{displaymath}
			\prod_{i=1}^{n}g(x_{i},y_{i})
			=\frac{d\kappa^{n}(x^{n})}{d\lambda^{n}(x^{n})}(y^{n}). 
		\end{displaymath}
		for $(x^{n},y^{n})\in\mathcal{T}_{\mathcal{V}}^{(n)}(\mu\kappa)$, 
		so for that case we have 
		\begin{displaymath}
			2^{nM}\leq\frac{d\kappa^{n}(x^{n})}{d\lambda^{n}(x^{n})}(y^{n}). 
		\end{displaymath}
		Proceeding as the same as the first part of the case 1, we get the result. 
		
		\item Choose $g=\frac{d\mu\kappa}{d\mu\lambda}$, and 
		take $N$ as the case 1. For each $k\in\Z^{+}$, define a measurable 
		function $f_{k}$ on $X\times Y$ as 
		\begin{displaymath}
			f_{k}(x,y)\defas\begin{cases}\log g(x,y)&
			\textrm{if $\frac{1}{k}\leq g(x,y)$}\\0&\textrm{otherwise}\end{cases}
		\end{displaymath}
		for each $(x,y)\in X\times Y$. Note that $f_{k}^{+}=(\log g)^{+}$ for each 
		$k\in\Z^{+}$ and $(\log g)^{+}$ is $\mu\kappa$-integrable, since 
		$D(\mu\kappa\|\mu\lambda)<0$. Since $f_{k}^{-}$ is bounded, $f_{k}$ is 
		$\mu\kappa$-integrable. Also, $\seq{f_{k}^{-}}_{k\in\Z^{+}}$ 
		is an increasing sequence of nonnegative measurable functions converging 
		pointwise to $(\log g)^{-}$ $\mu\kappa$-almost everywhere. So by monotone 
		convergence theorem, we get $\int f_{k}\,d\mu\kappa\ra\int\log g\,d\mu\kappa=-\infty$ 
		as $k\ra\infty$ by considering positive parts and negative parts separately. 
		Take $k\in\Z^{+}$ so that $\int f_{k}\,d\mu\kappa\leq-M-2$. 
		Define $\mathcal{V}_{0}\defas(f_{k};1;N\times Y)$. Fix $n\in\Z^{+}$ and a 
		$\mu\kappa$-typicality criterion $\mathcal{V}\leq\mathcal{V}_{0}$, then 
		\begin{displaymath}
			-M-1\geq\int f_{k}\,d\mu+1\geq\frac{1}{n}\sum_{i=1}^{n}f_{k}(x_{i},y_{i})
			\geq\frac{1}{n}\sum_{i=1}^{n}\log g(x_{i},y_{i})
		\end{displaymath}
		for $(x^{n},y^{n})\in\mathcal{T}_{\mathcal{V}}^{(n)}(\mu\kappa)$. Since 
		\begin{displaymath}
			\prod_{i=1}^{n}g(x_{i},y_{i})
			=\frac{d\kappa^{n}(x^{n})}{d\lambda^{n}(x^{n})}(y^{n}). 
		\end{displaymath}
		for $(x^{n},y^{n})\in\mathcal{T}_{\mathcal{V}}^{(n)}(\mu\kappa)$, 
		so for that case we have 
		\begin{displaymath}
			2^{-n(M+1)}\geq\frac{d\kappa^{n}(x^{n})}{d\lambda^{n}(x^{n})}(y^{n}). 
		\end{displaymath}
		Proceeding as the same as the second part of the case 1, we get the result. 
	\end{enumerate}
\end{IEEEproof}

%%%%%%%%%%%%%%%%%%%%%%%%%%%%%%%%%%%%%%%%%%%%%%%%%%%%%%%%%%%%%%%%%%%%%%%%%%%%%%%%%%%%%%%%%%%%%%%%%%
\section{Packing and Covering Lemmas}
In this section, we will prove some fundamental tools to be used for various achievability 
proofs. Let $(X,\mathscr{A}_{X})$, $(Y,\mathscr{A}_{Y})$, and 
$(Z,\mathscr{A}_{Z})$ be standard Borel spaces, and let $\mu_{XYZ}\in\Delta(X\times Y\times Z)$. 
Define $\mu_{X}$ to denote the pushforward of $\mu_{XYZ}$ onto $X$ under the projection 
(which is, the marginal distribution on $X$), and similarly define $\mu_{XY}$ and 
$\mu_{XZ}$. Since we are dealing with standard Borel spaces, there exist a probability 
kernel $\kappa_{Y|X}\in\mathcal{K}(X;Y)$ such that $\mu_{XY}=\mu_{X}\kappa_{Y|X}$, 
and similarly $\kappa_{Z|X}$, $\kappa_{Y|XZ}$, and 
$\kappa_{Z|XY}$~\cite[Chapter 5]{ParthasarathyMeas:1967}. If $\mathbf{x},\mathbf{y},\mathbf{z}$ 
are random variables taking values in $X$, $Y$, and $Z$, 
respectively, with the joint distribution $\mu_{XYZ}$, then one can think of 
the Kullback-Leibler divergence 
\begin{displaymath}
	D(\mu_{XYZ}\|\mu_{XY}\kappa_{Z|X})
	=D(\mu_{XYZ}\|\mu_{X}(\kappa_{Y|X}\times\kappa_{Z|X}))
	=D(\mu_{XZY}\|\mu_{XZ}\kappa_{Y|X})
\end{displaymath}
as the conditional mutual information $I(\mathbf{y};\mathbf{z}|\mathbf{x})$ 
(in the expression $\mu_{XY}\kappa_{Z|X}$, $\kappa_{Z|X}$ is treated as a kernel 
from $X\times Y$ to $Z$, and similarly in $\mu_{XZ}\kappa_{Y|X}$, 
$\kappa_{Y|X}$ is treated as a kernel from $X\times Z$ to $Y$). 
For a general definition of conditional mutual information 
for arbitrary alphabets, see \cite{Wyner:1978}. Then the following theorem is just a 
specialized result of the conditional divergence lemma: 

\begin{thrm}[Joint typicality lemma]\ 
	\begin{enumerate}
		\item If $I(\mathbf{y};\mathbf{z}|\mathbf{x})<\infty$, then for any $\epsilon>0$, there 
		is a $\mu_{XYZ}$-typicality criterion $\mathcal{W}_{0}$ such that for any 
		\mbox{$\mu_{XYZ}$-typicality} criterion $\mathcal{W}\leq\mathcal{W}_{0}$, we have 
		\begin{eqnarray*}
			&&\sup_{(x^{n},y^{n})\in X^{n}\times Y^{n}}\kappa_{Z|X}^{n}\left(
			\mathcal{T}_{\mathcal{W}}^{(n)}(\mu_{XYZ}|x^{n},y^{n})\Big|x^{n}\right)
			\leq2^{-n(I(\mathbf{y};\mathbf{z}|\mathbf{x})-\epsilon)}
		\end{eqnarray*}
		for all $n$, and there exists a $\mu_{XY}$-typicality criterion 
		$\mathcal{V}$ (depending on $\mathcal{W}$) such that 
		\begin{eqnarray*}
			&&\inf_{(x^{n},y^{n})\in\mathcal{T}_{\mathcal{V}}^{(n)}(\mu_{XY})}
			\kappa_{Z|X}^{n}\left(\mathcal{T}_{\mathcal{W}}^{(n)}
			(\mu_{XYZ}|x^{n},y^{n})\Big|x^{n}\right)
			\geq2^{-n(I(\mathbf{y};\mathbf{z}|\mathbf{x})+\epsilon)}
		\end{eqnarray*}
		for all sufficiently large $n$. 
		\item If $I(\mathbf{y};\mathbf{z}|\mathbf{x})=\infty$, then for any $M\geq0$, there 
		is a $\mu_{XYZ}$-typicality criterion $\mathcal{W}_{0}$ such that 
		for any \mbox{$\mu_{XYZ}$-typicality} criterion 
		$\mathcal{W}\leq\mathcal{W}_{0}$, we have 
		\begin{eqnarray*}
			&&\sup_{(x^{n},y^{n})\in X^{n}\times Y^{n}}\kappa_{Z|X}^{n}\left(
			\mathcal{T}_{\mathcal{W}}^{(n)}(\mu_{XYZ}|x^{n},y^{n})\Big|x^{n}\right)\leq2^{-nM}
		\end{eqnarray*}
		for all $n$.
	\end{enumerate}
\end{thrm}

\begin{IEEEproof}
	Apply the conditional divergence lemma with $\mu\leftarrow\mu_{XY}$, 
	$\kappa\leftarrow\kappa_{Z|XY}$, and $\lambda\leftarrow\kappa_{Z|X}$. 
\end{IEEEproof}

\begin{remk}\ \\
	By symmetry, the same theorem holds when the role of $Y$ and $Z$ are interchanged. 
\end{remk}

This theorem is a generalization of the theorem with the same name 
found in \cite[p.29]{ElGamalNetwork:2011}. Since packing lemma, covering lemma, 
and ``mutual versions'' of these lemmas in \cite{ElGamalNetwork:2011} are all basically 
derived from joint typicality lemma, it follows that almost the same proof procedure is 
also valid in the generalized setting. Proofs of these generalizations that are directly 
following those in \cite{ElGamalNetwork:2011} are given from now on. We will often use 
abstract conditional expectations in the proofs; details about abstract conditional 
expectations can be found in \cite{RobertProb:2000}. We will denote a conditional expectation 
of a real-valued random variable $\mathbf{w}$ with respect to the $\sigma$-algebra generated by 
another random variable $\mathbf{x}$ as $\mathrm{E}[\mathbf{w}|\mathbf{x}]$; here, $\mathbf{x}$ 
inside the bracket does not mean the value of $\mathbf{x}$ but the mapping $\mathbf{x}$ 
itself. Let us define the following terminology: 

\begin{defn}[Conditional distribution]\ \\
	Let $\kappa\in\mathcal{K}(X;Y)$ and $\mathbf{y}$ be a random variable taking values 
	in $Y$. Then we say $\mathbf{y}$ follows a \emph{conditional distribution} $\kappa$ 
	given $\mathbf{x}$ for another random variable $\mathbf{x}$ taking values in $X$, if 
	$(\mathbf{x},\mathbf{y})_{*}\Pr=(\mathbf{x}_{*}\Pr)\rtimes\kappa$. In that case, we write 
	$\Pr(\mathbf{y}\in B|\mathbf{x}=x)=\kappa(B|x)$ for each $B\in\mathscr{A}_{Y}$ 
	and $x\in X$. 
\end{defn}

Now we state and prove the main theorems of this section: 

\begin{thrm}[Packing lemma]\ \\
	Let $R\geq0$ be a nonnegative real number such that $R<I(\mathbf{y};\mathbf{z}|\mathbf{x})$. 
	Then, there exists a $\mu_{XYZ}$-typicality criterion $\mathcal{W}_{0}$ and a positive 
	number $c>0$ such that, for any $\mu_{XYZ}$-typicality criterion 
	$\mathcal{W}\leq\mathcal{W}_{0}$, we have the following for all $n\in\Z^{+}$: 
		
	Let $I_{n}$ be a finite set with $|I_{n}|\leq2^{nR}$. 
	Let $(\mathbf{x}^{n},\mathbf{y}^{n})$ be a random variable taking values in 
	$X^{n}\times Y^{n}$, and for each $m\in I_{n}$, let $\mathbf{z}^{n}(m)$ be a random 
	variable taking values in $Z^{n}$. Assume that each $\mathbf{z}^{n}(m)$ follows a 
	conditional distribution $\kappa_{Z|X}^{n}$ given 
	$(\mathbf{x}^{n},\mathbf{y}^{n})$. Then, 
	\begin{displaymath}
		\Pr\left((\mathbf{x}^{n},\mathbf{y}^{n},\mathbf{z}^{n}(m))
		\in\mathcal{T}_{\mathcal{W}}^{(n)}(\mu_{XYZ})\quad
		\textrm{for some $m\in I_{n}$}\right)\leq2^{-cn}.
	\end{displaymath}
\end{thrm}

\begin{IEEEproof}
	Assume first that $I(\mathbf{y};\mathbf{z}|\mathbf{x})<\infty$. Choose $\epsilon>0$ with 
	$R<I(\mathbf{y};\mathbf{z}|\mathbf{x})-\epsilon$, and take $\mathcal{W}_{0}$ obtained from the 
	joint typicality lemma. Fix $n\in\Z^{+}$ and an $\mu_{XYZ}$-typicality criterion $\mathcal{W}
	\leq\mathcal{W}_{0}$. Take $\mathbf{x}^{n}$, $\mathbf{y}^{n}$, $I_{n}$, 
	$\mathbf{z}^{n}(m)$'s as above, then for given $m\in I_{n}$, by the 
	joint typicality lemma, 
	\begin{eqnarray*}
		&&\Pr\left((\mathbf{x}^{n},\mathbf{y}^{n},\mathbf{z}^{n}(m))
		\in\mathcal{T}_{\mathcal{W}}^{(n)}(\mu_{XYZ})\right)\\&&\quad\quad\quad
		=\int_{X^{n}\times Y^{n}}\kappa_{Z|X}^{n}\left(
		\mathcal{T}_{\mathcal{W}}^{(n)}(\mu_{XYZ}|x^{n},y^{n})\Big|x^{n}
		\right)\,d(\mathbf{x}^{n},\mathbf{y}^{n})_{*}\Pr(x^{n},y^{n})\\
		&&\quad\quad\quad\leq\int_{X^{n}\times Y^{n}}
		2^{-n(I(\mathbf{y};\mathbf{z}|\mathbf{x})-\epsilon)}
		\,d(\mathbf{x}^{n},\mathbf{y}^{n})_{*}\Pr
		=2^{-n(I(\mathbf{y};\mathbf{z}|\mathbf{x})-\epsilon)}, 
	\end{eqnarray*}
	thus 
	\begin{eqnarray*}
		&&\Pr\left((\mathbf{x}^{n},\mathbf{y}^{n},\mathbf{z}^{n}(m))
		\in\mathcal{T}_{\mathcal{W}}^{(n)}(\mu_{XYZ})\quad
		\textrm{for some $m\in I_{n}$}\right)\\&&\quad\quad\quad\quad
		\leq\sum_{m\in I_{n}}\Pr\left((\mathbf{x}^{n},\mathbf{y}^{n},\mathbf{z}^{n}(m))
		\in\mathcal{T}_{\mathcal{W}}^{(n)}(\mu_{XYZ})\right)\\
		&&\quad\quad\quad\quad\leq2^{nR}\times
		2^{-n(I(\mathbf{y};\mathbf{z}|\mathbf{x})-\epsilon)}
		=2^{-n(I(\mathbf{y};\mathbf{z}|\mathbf{x})-R-\epsilon)}.
	\end{eqnarray*}
	The case $I(\mathbf{y};\mathbf{z}|\mathbf{x})=\infty$ also can be proved similarly. 
\end{IEEEproof}

Clearly, we can get the same conclusion (with a minor modification of ``for all $n$'' to 
``for all sufficiently large $n$'') when $\abs{I_{n}}\leq f(n)2^{nR}$ for a fixed function 
$f$ such that $\lim_{n\ra\infty}f(n)2^{-\delta n}=0$ for all $\delta>0$. 

\begin{thrm}[Covering lemma]\ \\
	Let $R\geq0$ be a nonnegative real number such that 
	$R>I(\mathbf{y};\mathbf{z}|\mathbf{x})$. Then, there exists a $\mu_{XYZ}$-typicality criterion 
	$\mathcal{W}_{0}$ and a positive number $c>0$ such that, for any $\mu_{XYZ}$-typicality criterion 
	$\mathcal{W}\leq\mathcal{W}_{0}$, there exists a $\mu_{XY}$-typicality criterion 
	$\mathcal{V}$ so that we have the following for all sufficiently large $n\in\Z^{+}$: 
	
	Let $I_{n}$ be a finite set with $|I_{n}|\geq2^{nR}$. 
	Let $(\mathbf{x}^{n},\mathbf{y}^{n})$ be a random variable taking values in 
	$X^{n}\times Y^{n}$, and for each $m\in I_{n}$, let $\mathbf{z}^{n}(m)$ be a random 
	variable taking values in $Z^{n}$. Assume that for $m,m'\in I_{n}$ with 
	$m\neq m'$, $(\mathbf{z}^{n}(m),\mathbf{z}^{n}(m'))$ follows a conditional 
	distribution $\kappa_{Z|X}^{n}\times\kappa_{Z|X}^{n}$ given 
	$(\mathbf{x}^{n},\mathbf{y}^{n})$. Then we have 
	\begin{displaymath}
		\Pr\left((\mathbf{x}^{n},\mathbf{y}^{n})
		\in\mathcal{T}_{\mathcal{V}}^{(n)}(\mu_{XY})\ \textrm{and}\ 
		(\mathbf{x}^{n},\mathbf{y}^{n},\mathbf{z}^{n}(m))
		\notin\mathcal{T}_{\mathcal{W}}^{(n)}(\mu_{XYZ})\quad
		\textrm{for all $m\in I_{n}$}\right)\leq2^{-cn}.
	\end{displaymath}
\end{thrm}

\begin{IEEEproof}
	From the assumption, it should be the case that 
	$I(\mathbf{y};\mathbf{z}|\mathbf{x})<\infty$ and $R>0$. Choose $\epsilon>0$ with 
	$R>I(\mathbf{y};\mathbf{z}|\mathbf{x})+\epsilon$, and by using the joint typicality lemma, 
	take a $\mu_{XYZ}$-typicality criterion $\mathcal{W}_{0}$ such that for any 
	$\mu_{XYZ}$-typicality criterion $\mathcal{W}\leq\mathcal{W}_{0}$, there exists a 
	$\mu_{XY}$-typicality criterion $\mathcal{V}$ so that 
	\begin{displaymath}
		2^{-n(I(\mathbf{y};\mathbf{z}|\mathbf{x})+\epsilon)}\leq
		\kappa_{Z|X}^{n}\left(\mathcal{T}_{\mathcal{W}}^{(n)}
		(\mu_{XYZ}|x^{n},y^{n})\Big|x^{n}\right)\leq
		2^{-n(I(\mathbf{y};\mathbf{z}|\mathbf{x})-\epsilon)}
	\end{displaymath}
	for all $(x^{n},y^{n})\in\mathcal{T}_{\mathcal{V}}^{(n)}(\mu_{XY})$, 
	whenever $n\geq n_{0}$ for some $n_{0}\in\Z^{+}$. 
	
	Fix $\mathcal{W}\leq\mathcal{W}_{0}$ and find such a $\mathcal{V}$ and $n_{0}$. 
	Let $n\geq n_{0}$, and take $\mathbf{x}^{n}$, $\mathbf{y}^{n}$, $I_{n}$, 
	$\mathbf{z}^{n}(m)$'s as above. For each $m\in I_{n}$, define $\mathbf{e}(m)$ 
	to be the indicator random variable of the event 
	\begin{displaymath}
		\set{(\mathbf{x}^{n},\mathbf{y}^{n},\mathbf{z}^{n}(m))
		\in\mathcal{T}_{\mathcal{W}}^{(n)}(\mu_{XYZ})}
	\end{displaymath}
	and define $\mathbf{N}\defas\sum_{m}\mathbf{e}(m)$. Since $R>0$, we have 
	$|I_{n}|\geq 2$. For each $(x^{n},y^{n})\in X^{n}\times Y^{n}$, define 
	\begin{eqnarray*}
		p_{1}(x^{n},y^{n})&\defas&\Pr\left(\mathbf{z}^{n}(1)
		\in\mathcal{T}_{\mathcal{W}}^{(n)}(\mu_{XYZ}|x^{n},y^{n})\bigg|
		\mathbf{x}^{n}=x^{n},\mathbf{y}^{n}=y^{n}\right)=\kappa_{Z|X}^{n}
		\left(\mathcal{T}_{\mathcal{W}}^{(n)}(\mu_{XYZ}|x^{n},y^{n})\Big|x^{n}\right),\\
		p_{2}(x^{n},y^{n})&\defas&\Pr\left(\mathbf{z}^{n}(1),\mathbf{z}^{n}(2)
		\in\mathcal{T}_{\mathcal{W}}^{(n)}(\mu_{XYZ}|x^{n},y^{n})\bigg|
		\mathbf{x}^{n}=x^{n},\mathbf{y}^{n}=y^{n}\right)=p_{1}(x^{n},y^{n})^{2}.
	\end{eqnarray*}
	Note that $p_{1},p_{2}$ are measurable functions. Then, 
	\begin{eqnarray*}
		p_{1}(\mathbf{x}^{n},\mathbf{y}^{n})&=&\mathrm{E}[\mathbf{e}(m)|\mathbf{x}^{n},\mathbf{y}^{n}]
		=\mathrm{E}[\mathbf{e}(m)^{2}|\mathbf{x}^{n},\mathbf{y}^{n}], \\
		p_{2}(\mathbf{x}^{n},\mathbf{y}^{n})&=&\mathrm{E}[\mathbf{e}(m)\mathbf{e}(m')|
		\mathbf{x}^{n},\mathbf{y}^{n}]
	\end{eqnarray*}
	almost surely, for $m,m'\in I_{n}$ with $m\neq m'$. By Chevyshev's inequality, 
	\begin{eqnarray*}
		\Pr(\mathbf{N}=0|\mathbf{x}^{n},\mathbf{y}^{n})&\leq&
		\Pr\left((\mathbf{N}-\mathrm{E}[\mathbf{N}|\mathbf{x}^{n},\mathbf{y}^{n}])^{2}
		\geq(\mathrm{E}[\mathbf{N}|\mathbf{x}^{n},\mathbf{y}^{n}])^{2}
		\Big|\mathbf{x}^{n},\mathbf{y}^{n}\right)\\&\leq&
		\frac{\mathrm{E}\left[(\mathbf{N}-\mathrm{E}[\mathbf{N}
		|\mathbf{x}^{n},\mathbf{y}^{n}])^{2}\Big|\mathbf{x}^{n},\mathbf{y}^{n}\right]}
		{(\mathrm{E}[\mathbf{N}|\mathbf{x}^{n},\mathbf{y}^{n}])^{2}}=
		\frac{\mathrm{E}[\mathbf{N}^{2}|\mathbf{x}^{n},\mathbf{y}^{n}]-
		(\mathrm{E}[\mathbf{N}|\mathbf{x}^{n},\mathbf{y}^{n}])^{2}}
		{(\mathrm{E}[\mathbf{N}|\mathbf{x}^{n},\mathbf{y}^{n}])^{2}}
	\end{eqnarray*}
	almost surely. We compute $\mathrm{E}[\mathbf{N}|\mathbf{x}^{n},\mathbf{y}^{n}]$ and 
	$\mathrm{E}[\mathbf{N}^{2}|\mathbf{x}^{n},\mathbf{y}^{n}]$ as follows: 
	\begin{eqnarray*}
		\mathrm{E}[\mathbf{N}|\mathbf{x}^{n},\mathbf{y}^{n}]&=&\sum_{m}
		\mathrm{E}[\mathbf{e}(m)|\mathbf{x}^{n},\mathbf{y}^{n}]
		=|I_{n}|p_{1}(\mathbf{x}^{n},\mathbf{y}^{n}),\\
		\mathrm{E}[\mathbf{N}^{2}|\mathbf{x}^{n},\mathbf{y}^{n}]&=&\sum_{m}
		\mathrm{E}[\mathbf{e}(m)^{2}|\mathbf{x}^{n},\mathbf{y}^{n}]
		+\sum_{m}\sum_{m'\neq m}\mathrm{E}[\mathbf{e}(m)
		\mathbf{e}(m')|\mathbf{x}^{n},\mathbf{y}^{n}]\\&\leq&
		|I_{n}|p_{1}(\mathbf{x}^{n},\mathbf{y}^{n})+|I_{n}|^{2}p_{2}(\mathbf{x}^{n},\mathbf{y}^{n})
	\end{eqnarray*}
	almost surely, thus 
	\begin{displaymath}
		\Pr(\mathbf{N}=0|\mathbf{x}^{n},\mathbf{y}^{n})\leq
		\frac{1}{|I_{n}|p_{1}(\mathbf{x}^{n},\mathbf{y}^{n})}
		\leq\frac{2^{-nR}}{p_{1}(\mathbf{x}^{n},\mathbf{y}^{n})}
	\end{displaymath}
	almost surely. Note that 
	\begin{displaymath}
		p_{1}(x^{n},y^{n})=\kappa_{Z|X}^{n}\left(
		\mathcal{T}_{\mathcal{W}}^{(n)}(\mu_{XYZ}|x^{n},y^{n})\Big|x^{n}\right)\\
		\geq2^{-n(I(\mathbf{y};\mathbf{z}|\mathbf{x})+\epsilon)}
	\end{displaymath}
	whenever $(x^{n},y^{n})\in\mathcal{T}_{\mathcal{V}}^{(n)}(\mu_{XY})$, 
	thus it follows that 			
	\begin{eqnarray*}
		\Pr\left(\mathbf{N}=0,\ (\mathbf{x}^{n},\mathbf{y}^{n})\in
		\mathcal{T}_{\mathcal{V}}^{(n)}(\mu_{XY})\right)&=&
		\int_{(\mathbf{x}^{n},\mathbf{y}^{n})\in\mathcal{T}_{\mathcal{V}}^{(n)}(\mu_{XY})}
		\Pr\left(\mathbf{N}=0|\mathbf{x}^{n},\mathbf{y}^{n}\right)\,d\Pr\\&\leq&
		2^{-n(R-I(\mathbf{y};\mathbf{z}|\mathbf{x})-\epsilon)}.
	\end{eqnarray*}
\end{IEEEproof}

Again, we can get the same conclusion when $\abs{I_{n}}\geq f(n)2^{nR}$ for a fixed function 
$f$ such that $\lim_{n\ra\infty}f(n)2^{\delta n}=\infty$ for all $\delta>0$. 
Next we prove ``mutual versions'' of packing and covering lemmas. Of course, similar remarks 
about estimates on sizes of index sets $I_{n},J_{n}$ are also true. 

\begin{thrm}[Mutual packing lemma]\ \\
	Let $R_{1},R_{2}\geq0$ be nonnegative real numbers such that 
	$R_{1}+R_{2}<I(\mathbf{y};\mathbf{z}|\mathbf{x})$. Then, 
	there exists a $\mu_{XYZ}$-typicality criterion $\mathcal{W}_{0}$ and a positive number $c>0$ 
	such that, for any $\mu_{XYZ}$-typicality criterion $\mathcal{W}\leq\mathcal{W}_{0}$, 
	we have the following for all $n\in\Z^{+}$: 
		
	Let $I_{n},J_{n}$ be a finite sets with $|I_{n}|\leq2^{nR_{1}}$ 
	and $|J_{n}|\leq2^{nR_{2}}$. Let $\mathbf{x}^{n}$ be a random variable taking values in 
	$X^{n}$, and for each $m_{1}\in I_{n}$ and $m_{2}\in J_{n}$, let $\mathbf{y}^{n}(m_{1})$ 
	and $\mathbf{z}^{n}(m_{2})$ be random variables taking values in $Y^{n}$ and $Z^{n}$, 
	respectively. Assume that for each $m_{1}\in I_{n}$ and $m_{2}\in J_{n}$, 
	$(\mathbf{y}^{n}(m_{1}),\mathbf{z}^{n}(m_{2}))$ follows a conditional distribution 
	$\kappa_{Y|X}^{n}\times\kappa_{Z|X}^{n}$ given $\mathbf{x}^{n}$. Then, 
	\begin{displaymath}
		\Pr\left((\mathbf{x}^{n},\mathbf{y}^{n}(m_{1}),\mathbf{z}^{n}(m_{2}))
		\in\mathcal{T}_{\mathcal{W}}^{(n)}(\mu_{XYZ})\quad
		\textrm{for some $m_{1}\in I_{n}$, $m_{2}\in J_{n}$}\right)\leq2^{-cn}.
	\end{displaymath}
\end{thrm}

\begin{IEEEproof}
	Assume first that $I(\mathbf{y};\mathbf{z}|\mathbf{x})<\infty$. 
	Choose $\epsilon>0$ with $R_{1}+R_{2}<I(\mathbf{y};\mathbf{z}|\mathbf{x})-\epsilon$ and 
	take $\mathcal{W}_{0}$ obtained from the joint typicality lemma. Fix $n\in\Z^{+}$ 
	and a $\mu_{XYZ}$-typicality criterion $\mathcal{W}\leq\mathcal{W}_{0}$. 
	Take $\mathbf{x}^{n}$, $I_{n}$, $\mathbf{y}^{n}(m_{1})$'s, $J_{n}$, 
	$\mathbf{z}^{n}(m_{2})$'s as above, then for given $m_{1}\in I_{n}$ and 
	$m_{2}\in J_{n}$, by the joint typicality lemma, 
	\begin{eqnarray*}
		&&\Pr\left((\mathbf{x}^{n},\mathbf{y}^{n}(m_{1}),\mathbf{z}^{n}(m_{2})))
		\in\mathcal{T}_{\mathcal{W}}^{(n)}(\mu_{XYZ})\right)\\&&\quad\quad\quad
		=\int_{X^{n}}(\kappa_{Y|X}^{n}\times\kappa_{Z|X}^{n})\left(
		\mathcal{T}_{\mathcal{W}}^{(n)}(\mu_{XYZ}|x^{n})\Big|x^{n}\right)\,
		d(\mathbf{x}_{*}^{n}\Pr)(x^{n})\\&&\quad\quad\quad=\int_{X^{n}}
		\left[\int_{Y^{n}}\kappa_{Z|X}^{n}\left(\mathcal{T}_{\mathcal{W}}^{(n)}
		(\mu_{XYZ}|x^{n},y^{n})\Big|x^{n}\right)\,d\kappa_{Y|X}^{n}(y^{n}|x^{n})\right]
		d(\mathbf{x}_{*}^{n}\Pr)(x^{n})\\		
		&&\quad\quad\quad\leq\int_{X^{n}}\left[\int_{Y^{n}}
		2^{-n(I(\mathbf{y};\mathbf{z}|\mathbf{x})-\epsilon)}
		d\kappa_{Y|X}^{n}(y^{n}|x^{n})\right]d(\mathbf{x}_{*}^{n}\Pr)(x^{n})
		=2^{-n(I(\mathbf{y};\mathbf{z}|\mathbf{x})-\epsilon)}, 
	\end{eqnarray*}
	thus 
	\begin{eqnarray*}
		&&\Pr\left((\mathbf{x}^{n},\mathbf{y}^{n}(m_{1}),\mathbf{z}^{n}(m_{2}))
		\in\mathcal{T}_{\mathcal{W}}^{(n)}(\mu_{XYZ})\quad\textrm{for some 
		$m_{1}\in I_{n}$, $m_{2}\in J_{n}$}\right)\\&&\quad\quad\quad\quad
		\leq\sum_{m_{1}\in I_{n},m_{2}\in J_{n}}\Pr\left(
		(\mathbf{x}^{n},\mathbf{y}^{n}(m_{1}),\mathbf{z}^{n}(m_{2}))
		\in\mathcal{T}_{\mathcal{W}}^{(n)}(\mu_{XYZ})\right)\\
		&&\quad\quad\quad\quad\leq2^{n(R_{1}+R_{2})}\times
		2^{-n(I(\mathbf{y};\mathbf{z}|\mathbf{x})-\epsilon)}
		=2^{-n(I(\mathbf{y};\mathbf{z}|\mathbf{x})-R_{1}-R_{2}-\epsilon)}.
	\end{eqnarray*}
	The case $I(\mathbf{y};\mathbf{z}|\mathbf{x})=\infty$ also can be proved similarly. 
\end{IEEEproof}

\begin{thrm}[Mutual covering lemma]\ \\
	Let $R_{1},R_{2}\geq0$ be nonnegative real numbers such that 
	$R_{1}+R_{2}>I(\mathbf{y};\mathbf{z}|\mathbf{x})$. Then, 
	there exists a $\mu_{XYZ}$-typicality criterion $\mathcal{W}_{0}$ and a positive number 
	$c>0$ such that, for any $\mu_{XYZ}$-typicality criterion 
	$\mathcal{W}\leq\mathcal{W}_{0}$, there exists a $\mu_{X}$-typicality criterion 
	$\mathcal{U}$ so that we have the following for all sufficiently large $n\in\Z^{+}$: 
	
	Let $I_{n},J_{n}$ be finite sets with $|I_{n}|\geq2^{nR_{1}}$ and $|J_{n}|\geq2^{nR_{2}}$. 
	Let $\mathbf{x}^{n}$ be a random variable taking values in $X^{n}$, and for each 
	$m_{1}\in I_{n}$ and $m_{2}\in J_{n}$, let $\mathbf{y}^{n}(m_{1})$ and $\mathbf{z}^{n}(m_{2})$ 
	be random variables taking values in $Y^{n}$ and $Z^{n}$, respectively. Assume followings: 
	\begin{enumerate}
		\item For each $m_{1}\in I_{n}$ and $m_{2}\in J_{n}$, 
		$(\mathbf{y}^{n}(m_{1}),\mathbf{z}^{n}(m_{2}))$ follows a conditional distribution 
		$\kappa_{Y|X}^{n}\times\kappa_{Z|X}^{n}$ given $\mathbf{x}^{n}$. 
		\item For each $m_{1},m_{1}'\in I_{n}$ and $m_{2}\in J_{n}$ with 
		$m_{1}\neq m_{1}'$, $(\mathbf{y}^{n}(m_{1}),\mathbf{y}^{n}(m_{1}'),\mathbf{z}^{n}(m_{2}))$ 
		follows a conditional distribution $\kappa_{Y|X}^{n}\times\kappa_{Y|X}^{n}\times
		\kappa_{Z|X}^{n}$ given $\mathbf{x}^{n}$. 
		\item For each $m_{1}\in I_{n}$ and $m_{2},m_{2}'\in J_{n}$ with 
		$m_{2}\neq m_{2}'$, $(\mathbf{y}^{n}(m_{1}),\mathbf{z}^{n}(m_{2}),\mathbf{z}^{n}(m_{2}'))$ 
		follows a conditional distribution $\kappa_{Y|X}^{n}\times\kappa_{Z|X}^{n}\times
		\kappa_{Z|X}^{n}$ given $\mathbf{x}^{n}$. 
		\item For each $m_{1},m_{1}'\in I_{n}$ and $m_{2},m_{2}'\in J_{n}$ with 
		$m_{1}\neq m_{1}'$ and $m_{2}\neq m_{2}'$, 
		$(\mathbf{y}^{n}(m_{1}),\mathbf{y}^{n}(m_{1}'),\mathbf{z}^{n}(m_{2}),
		\mathbf{z}^{n}(m_{2}'))$ follows a conditional distribution 
		$\kappa_{Y|X}^{n}\times\kappa_{Y|X}^{n}\times
		\kappa_{Z|X}^{n}\times\kappa_{Z|X}^{n}$ given $\mathbf{x}^{n}$. 
	\end{enumerate}
	Then we have 
	\begin{displaymath}
		\Pr\left(\mathbf{x}^{n}\in\mathcal{T}_{\mathcal{U}}^{(n)}(\mu_{X})\ 
		\textrm{and}\ (\mathbf{x}^{n},\mathbf{y}^{n}(m_{1}),\mathbf{z}^{n}(m_{2}))
		\notin\mathcal{T}_{\mathcal{W}}^{(n)}(\mu_{XYZ})\quad
		\textrm{for all $m_{1}\in I_{n}$, $m_{2}\in J_{n}$}\right)\leq2^{-cn}.
	\end{displaymath}
\end{thrm}

\begin{IEEEproof}
	We may assume that $I(\mathbf{y};\mathbf{z}|\mathbf{x})<\infty$. 
	We also assume that $R_{1},R_{2}>0$. A proof for the case $R_{1}=0$ or $R_{2}=0$ can be 
	written similarly. Choose $\epsilon>0$ with 
	$R_{1}+R_{2}>I(\mathbf{y};\mathbf{z}|\mathbf{x})+\epsilon$ and 
	$R_{1},R_{2}>4\epsilon$. Using the joint typicality lemma, take a $\mu_{XYZ}$-typicality 
	criterion $\mathcal{W}_{0}$ so that 
	\begin{eqnarray*}
		\kappa_{Y|X}^{n}\left(\mathcal{T}_{\mathcal{W}}^{(n)}
		(\mu_{XYZ}|x^{n},z^{n})\Big|x^{n}\right)&\leq&
		2^{-n(I(\mathbf{y};\mathbf{z}|\mathbf{x})-\epsilon)},\\
		\kappa_{Z|X}^{n}\left(\mathcal{T}_{\mathcal{W}}^{(n)}
		(\mu_{XYZ}|x^{n},y^{n})\Big|x^{n}\right)&\leq&
		2^{-n(I(\mathbf{y};\mathbf{z}|\mathbf{x})-\epsilon)}.
	\end{eqnarray*}
	for all $(x^{n},y^{n},z^{n})\in X^{n}\times Y^{n}\times Z^{n}$ and $n\in\Z^{+}$, and 
	for any $\mu_{XYZ}$-typicality criterion $\mathcal{W}\leq\mathcal{W}_{0}$, there exists 
	a $\mu_{XY}$-typicality criterion $\mathcal{V}$ such that 
	\begin{displaymath}
		2^{-n(I(\mathbf{y};\mathbf{z}|\mathbf{x})+\epsilon)}\leq
		\kappa_{Z|X}^{n}\left(\mathcal{T}_{\mathcal{W}}^{(n)}
		(\mu_{XYZ}|x^{n},y^{n})\Big|x^{n}\right)
	\end{displaymath}
	whenever $n\geq n_{1}$ for some $n_{1}\in\Z^{+}$ and 
	$(x^{n},y^{n})\in\mathcal{T}_{\mathcal{V}}^{(n)}(\mu_{XY})$. Fix 
	$\mathcal{W}\leq\mathcal{W}_{0}$ and find such $\mathcal{V}$. Then by the conditional 
	typicality lemma, there exists a $\mu_{X}$-typicality criterion $\mathcal{U}$ such that 
	\begin{displaymath}
		\kappa_{Y|X}^{n}\left(\mathcal{T}_{\mathcal{V}}^{(n)}
		(\mu_{XY}|x^{n})\Big|x^{n}\right)\geq 1-\delta
	\end{displaymath}
	whenever $n\geq n_{2}$ for some $n_{2}\in\Z^{+}$ and 
	$x^{n}\in\mathcal{T}_{\mathcal{U}}^{(n)}(\mu_{X})$, for some given $\delta\in(0,1)$. 
	
	Fix $n\geq\max\set{n_{1},n_{2}}$ and take $\mathbf{x}^{n}$, $I_{n}$, 
	$\mathbf{y}^{n}(m_{1})$'s, $J_{n}$, $\mathbf{z}^{n}(m_{2})$'s as above. For each 
	$(m_{1},m_{2})\in I_{n}\times J_{n}$, define $\mathbf{e}(m_{1},m_{2})$ be the indicator 
	random variable of the event 
	\begin{displaymath}
		\set{(\mathbf{x}^{n},\mathbf{y}^{n}(m_{1}),\mathbf{z}^{n}(m_{2}))
		\in\mathcal{T}_{\mathcal{W}}^{(n)}(\mu_{XYZ})}
	\end{displaymath}
	and define $\mathbf{N}\defas\sum_{m_{1},m_{2}}\mathbf{e}(m_{1},m_{2})$. 
	Since $R_{1},R_{2}>0$, we have $|I_{n}|,|J_{n}|\geq2$. For each 
	$x^{n}\in X^{n}$, define 
	\begin{eqnarray*}
		p_{1}(x^{n})&\defas&\Pr\left((\mathbf{y}^{n}(1),\mathbf{z}^{n}(1))
		\in\mathcal{T}_{\mathcal{W}}^{(n)}(\mu_{XYZ}|x^{n})\bigg|
		\mathbf{x}^{n}=x^{n}\right),\\
		p_{2}(x^{n})&\defas&\Pr\left((\mathbf{y}^{n}(1),\mathbf{z}^{n}(1)),
		(\mathbf{y}^{n}(1),\mathbf{z}^{n}(2))\in\mathcal{T}_{\mathcal{W}}^{(n)}
		(\mu_{XYZ}|x^{n})\bigg|\mathbf{x}^{n}=x^{n}\right),\\
		p_{3}(x^{n})&\defas&\Pr\left((\mathbf{y}^{n}(1),\mathbf{z}^{n}(1)),
		(\mathbf{y}^{n}(2),\mathbf{z}^{n}(1))\in\mathcal{T}_{\mathcal{W}}^{(n)}
		(\mu_{XYZ}|x^{n})\bigg|\mathbf{x}^{n}=x^{n}\right),\\
		p_{4}(x^{n})&\defas&\Pr\left((\mathbf{y}^{n}(1),\mathbf{z}^{n}(1)),
		(\mathbf{y}^{n}(2),\mathbf{z}^{n}(2))\in\mathcal{T}_{\mathcal{W}}^{(n)}
		(\mu_{XYZ}|x^{n})\bigg|\mathbf{x}^{n}=x^{n}\right)=p_{1}(x^{n})^{2}.
	\end{eqnarray*}
	Note that 
	\begin{eqnarray*}
		p_{1}(\mathbf{x}^{n})&=&\mathrm{E}[\mathbf{e}(m_{1},m_{2})|\mathbf{x}^{n}]
		=\mathrm{E}[\mathbf{e}(m_{1},m_{2})^{2}|\mathbf{x}^{n}], \\
		p_{2}(\mathbf{x}^{n})&=&\mathrm{E}[\mathbf{e}(m_{1},m_{2})
		\mathbf{e}(m_{1},m_{2}')|\mathbf{x}^{n}], \\
		p_{3}(\mathbf{x}^{n})&=&\mathrm{E}[\mathbf{e}(m_{1},m_{2})
		\mathbf{e}(m_{1}',m_{2})|\mathbf{x}^{n}], \\
		p_{2}(\mathbf{x}^{n})&=&\mathrm{E}[\mathbf{e}(m_{1},m_{2})
		\mathbf{e}(m_{1}',m_{2}')|\mathbf{x}^{n}]
	\end{eqnarray*}
	almost surely, for $m_{1},m_{1}'\in I_{n}$ and $m_{2},m_{2}'\in J_{n}$ with 
	$m_{1}\neq m_{1}'$, $m_{2}\neq m_{2}'$. By Chevyshev's inequality, 
	\begin{eqnarray*}
		\Pr(\mathbf{N}=0|\mathbf{x}^{n})&\leq&\Pr\left(
		(\mathbf{N}-\mathrm{E}[\mathbf{N}|\mathbf{x}^{n}])^{2}\geq
		(\mathrm{E}[\mathbf{N}|\mathbf{x}^{n}])^{2}
		\Big|\mathbf{x}^{n}\right)\\&\leq&
		\frac{\mathrm{E}\left[(\mathbf{N}-\mathrm{E}[\mathbf{N}|\mathbf{x}^{n}])^{2}
		\Big|\mathbf{x}^{n}\right]}{(\mathrm{E}[\mathbf{N}|\mathbf{x}^{n}])^{2}}=
		\frac{\mathrm{E}[\mathbf{N}^{2}|\mathbf{x}^{n}]-
		(\mathrm{E}[\mathbf{N}|\mathbf{x}^{n}])^{2}}
		{(\mathrm{E}[\mathbf{N}|\mathbf{x}^{n}])^{2}}
	\end{eqnarray*}
	almost surely. We compute $\mathrm{E}[\mathbf{N}|\mathbf{x}^{n}]$ and 
	$\mathrm{E}[\mathbf{N}^{2}|\mathbf{x}^{n}]$ as follows: 
	\begin{eqnarray*}
		\mathrm{E}[\mathbf{N}|\mathbf{x}^{n}]&=&\sum_{m_{1},m_{2}}
		\mathrm{E}[\mathbf{e}(m_{1},m_{2})|\mathbf{x}^{n}]
		=|I_{n}||J_{n}|p_{1}(\mathbf{x}^{n}),\\
		\mathrm{E}[\mathbf{N}^{2}|\mathbf{x}^{n}]&=&\sum_{m_{1},m_{2}}
		\mathrm{E}[\mathbf{e}(m_{1},m_{2})^{2}|\mathbf{x}^{n}]\\&&
		+\sum_{m_{1},m_{2}}\sum_{m_{2}'\neq m_{2}}\mathrm{E}[\mathbf{e}(m_{1},m_{2})
		\mathbf{e}(m_{1},m_{2}')|\mathbf{x}^{n}]\\&&
		+\sum_{m_{1},m_{2}}\sum_{m_{1}'\neq m_{1}}\mathrm{E}[\mathbf{e}(m_{1},m_{2})
		\mathbf{e}(m_{1}',m_{2})|\mathbf{x}^{n}]\\&&
		+\sum_{m_{1},m_{2}}\sum_{m_{1}'\neq m_{1},m_{2}'\neq m_{2}}
		\mathrm{E}[\mathbf{e}(m_{1},m_{2})\mathbf{e}(m_{1}',m_{2}')|\mathbf{x}^{n}]
		\\&\leq&|I_{n}||J_{n}|p_{1}(\mathbf{x}^{n})+|I_{n}||J_{n}|^{2}
		p_{2}(\mathbf{x}^{n})+|I_{n}|^{2}|J_{n}|p_{3}(\mathbf{x}^{n})
		+|I_{n}|^{2}|J_{n}|^{2}p_{4}(\mathbf{x}^{n}),
	\end{eqnarray*}
	almost surely, thus 
	\begin{displaymath}
		\Pr(\mathbf{N}=0|\mathbf{x}^{n})\leq
		\frac{2^{-n(R_{1}+R_{2})}}{p_{1}(\mathbf{x}^{n})}
		+\frac{2^{-nR_{1}}p_{2}(\mathbf{x}^{n})}{p_{1}(\mathbf{x}^{n})^{2}}
		+\frac{2^{-nR_{2}}p_{3}(\mathbf{x}^{n})}{p_{1}(\mathbf{x}^{n})^{2}}
	\end{displaymath}
	almost surely. Note that 
	\begin{eqnarray*}
		p_{1}(x^{n})&=&(\kappa_{Y|X}^{n}\times\kappa_{Z|X}^{n})
		\left(\mathcal{T}_{\mathcal{W}}^{(n)}(\mu_{XYZ}|x^{n})\Big|x^{n}\right)\\
		&=&\int_{Y^{n}}\kappa_{Z|X}^{n}
		\left(\mathcal{T}_{\mathcal{W}}^{(n)}(\mu_{XYZ}|x^{n},y^{n})\Big|x^{n}\right)
		\,d\kappa_{Y|X}^{n}(y^{n}|x^{n})\\&\geq&
		\int_{\mathcal{T}_{\mathcal{V}}^{(n)}(\mu_{XY}|x^{n})}
		\kappa_{Z|X}^{n}\left(\mathcal{T}_{\mathcal{W}}^{(n)}
		(\mu_{XYZ}|x^{n},y^{n})\Big|x^{n}\right)
		d\kappa_{Y|X}^{n}(y^{n}|x^{n})\\
		&\geq&(1-\delta)2^{-n(I(\mathbf{y};\mathbf{z}|\mathbf{x})+\epsilon)}
	\end{eqnarray*}
	whenever $x^{n}\in\mathcal{T}_{\mathcal{U}}^{(n)}(\mu_{X})$, and similarly, 
	\begin{eqnarray*}
		p_{2}(x^{n})&=&\int_{Y^{n}}\left[\kappa_{Z|X}^{n}\left(
		\mathcal{T}_{\mathcal{W}}^{(n)}(\mu_{XYZ}|x^{n},y^{n})\Big|x^{n}\right)
		\right]^{2}\,d\kappa_{Y|X}^{n}(y^{n}|x^{n})\leq
		2^{-n(2I(\mathbf{y};\mathbf{z}|\mathbf{x})-2\epsilon)},\\
		p_{3}(x^{n})&=&\int_{Z^{n}}\left[\kappa_{Y|X}^{n}\left(
		\mathcal{T}_{\mathcal{W}}^{(n)}(\mu_{XYZ}|x^{n},z^{n})\Big|x^{n}\right)
		\right]^{2}\,d\kappa_{Z|X}^{n}(z^{n}|x^{n})\leq
		2^{-n(2I(\mathbf{y};\mathbf{z}|\mathbf{x})-2\epsilon)}
	\end{eqnarray*}
	whenever $x^{n}\in\mathcal{T}_{\mathcal{U}}^{(n)}(\mu_{X})$, so we have 
	\begin{eqnarray*}
		\Pr\left(\mathbf{N}=0,\ \mathbf{x}^{n}\in\mathcal{T}_{\mathcal{U}}^{(n)}(\mu_{X})
		\right)&=&\int_{\mathbf{x}^{n}\in\mathcal{T}_{\mathcal{U}}^{(n)}
		(\mu_{X})}\Pr\left(\mathbf{N}=0|\mathbf{x}^{n}\right)\,
		d\Pr\\&\leq&
		\frac{2^{-n(R_{1}+R_{2}-I(\mathbf{y};\mathbf{z}|\mathbf{x})-\epsilon)}}
		{1-\delta}+\frac{2^{-n(R_{1}-4\epsilon)}}{(1-\delta)^{2}}
		+\frac{2^{-n(R_{2}-4\epsilon)}}{(1-\delta)^{2}}.
	\end{eqnarray*}
	Since all exponents are positive, we get the result. 
\end{IEEEproof}

%%%%%%%%%%%%%%%%%%%%%%%%%%%%%%%%%%%%%%%%%%%%%%%%%%%%%%%%%%%%%%%%%%%%%%%%%%%%%%%%%%%%%%%%%%%%%%%%%%
\section{Applications to Coding Problems}
In this section, some applications of the new typicality to  
coding problems are given. Derivations of many outer bounds do not rely on 
the alphabet size, excluding the cardinality-bound-part of auxiliary variables. In 
\cite{Wyner:1978}, a general definition of conditional mutual information for 
arbitrary alphabets is given, and some basic properties such as the chain rule are 
derived. Hence, outer bounds often can be extended to the general alphabet case 
with only some minor obstacles. On the other hand, derivations of inner bounds are 
often based on strong typicality, so the new notion of typicality makes possible for those 
inner bounds to be extended to the general alphabet case also. These generalizations 
are quite straightforward; just replace usual strong typical sets 
appearing in the proofs with new typical sets. 

However, still there are some technical subtleties remaining. First, quantities 
easily become infinite or even undefined, when alphabets are not finite. 
For example, the differential entropy may not be defined for a general real-valued 
random variable. Even when quantities are well-defined but becomes infinity, problems can happen 
when differences of such quantities are involved. Thus, sometimes a transfer procedure of a proof 
for the finite alphabet case to the general case is not completely transparent. 
Second, even when we can prove the same result for a coding theorem as in the finite alphabet case, 
numerical evaluation of the obtained region is rarely possible, because in general it 
is an infinite-dimensional optimization problem. Only for some special cases (such as 
additive Gaussian noise channel) this evaluation is computationally possible. 

Despite of those subtleties, it is theoretically satisfactory that one does not need to 
pay much additional efforts on proving coding theorems for general alphabets. 
The point-to-point channel coding theorem and the point-to-point lossy source coding theorem 
are given as examples to explicitly show that \emph{there is essentially no additional thing 
required to prove coding theorems with infinite alphabets}. 
Proofs given here basically follow those given in \cite{ElGamalNetwork:2011}. 
We will first describe precise mathematical formulations of those problems before proving them. 
One can see that there is almost no complicated assumptions about regularity to make the proofs 
mathematically rigorous. 

\subsection{Point-to-point channel coding theorem}
Let $(X,\mathscr{A})$, $(Y,\mathscr{B})$ be standard Borel spaces. 

\begin{defn}[Memoryless channels]\ \\
	A \emph{memoryless channel with a cost function} is a quadruple 
	$(X,Y,\kappa,t)$, where $\kappa\in\mathcal{K}(X;Y)$ and $t:X\ra[0,\infty]$ 
	is a measurable function. Here, $X$ is called the \emph{input alphabet}, $Y$ is called 
	the \emph{output alphabet}, $\kappa$ is called the \emph{channel transition kernel}, 
	and $t$ is called the \emph{cost function}. For $n,M\in\Z^{+}$, an 
	\emph{$(n,M)$-channel code} for this memoryless channel consists 
	of two measurable mappings $f:[1:M]\ra X^{n}$ and $g:Y^{n}\ra[0:M]$, respectively called 
	an \emph{encoder} and a \emph{decoder}; here, $0\in[0:M]$ represents error declared 
	by the decoder. An $(n,M)$-channel code $(f,g)$ is said to 
	\emph{satisfy average cost constraint $B$} for some $B\in[0,\infty]$, if 
	\begin{displaymath}
		t^{(n)}(f(m))\leq B
	\end{displaymath}
	for all $m\in[1:M]$, where $t^{(n)}:X^{n}\ra[0,\infty]$ is defined as 
	$t^{(n)}:x^{n}\mapsto\frac{1}{n}\sum_{i=1}^{n}t(x_{i})$. The \emph{error probability} 
	of an $(n,M)$-channel code $(f,g)$ associated to the message $m\in[1:M]$ is defined as 
	\begin{displaymath}
		P_{e,m}(f,g)\defas\kappa^{n}\left(
		\setbc{y^{n}\in Y^{n}}{g(y^{n})\neq m}|f(m)\right).
	\end{displaymath}
	The average error probability of this channel code is defined as 
	\begin{displaymath}
		P_{e}(f,g)\defas\frac{1}{M}\sum_{m\in[1:M]}P_{e,m}(f,g).
	\end{displaymath}
	For $R\in[0,\infty)$ and $B\in[0,\infty]$, the pair $(R,B)$ is said to be 
	\emph{achievable}, if for any $\epsilon>0$, there exists an $(n,M)$-channel code 
	$(f,g)$ satisfying average cost constraint $B+\epsilon$ such that 
	\begin{displaymath}
		\frac{\log M}{n}\geq R\quad\textrm{and}\quad
		P_{e}(f,g)\leq\epsilon. 
	\end{displaymath}
	The \emph{operational capacity-cost function} 
	$\mathrm{C}_{o}:[0,\infty]\ra[0,\infty]$ is defined as 
	\begin{displaymath}
		\mathrm{C}_{o}:B\mapsto\sup\setbc{R\in[0,\infty)}
		{\textrm{$(R,B)$ is achievable}}.
	\end{displaymath}
	We define the supremum of $\emptyset$ to be $0$ as a convention. 
	On the other hand, the \emph{information capacity-cost function} 
	$\mathrm{C}_{i}:[0,\infty]\ra[0,\infty]$ is defined as 
	\begin{displaymath}
		\mathrm{C}_{i}:B\mapsto\sup_{\mu\in\Delta(X);\int t\,d\mu\leq B}
		I(\mu,\kappa)
	\end{displaymath}
	where $I(\mu,\kappa)$ is defined as 
	$D\left(\mu\kappa\|\mu\times\kappa_{*}\mu\right)$; see 
	\refremk{remk:marginals}. 
\end{defn}

\begin{remk}\ 
	\begin{enumerate}
		\item Define 
		\begin{displaymath}
			B_{\min}:=\inf_{x\in X}t(x),
		\end{displaymath}
		then it is clear that $\mathrm{C}_{i}(B)=0$ if $B<B_{\min}$. Also, for that 
		case $(R,B)$ is never achievable for any $R$. Hence, for $B<B_{\min}$, we 
		have $\mathrm{C}_{o}(B)=\mathrm{C}_{i}(B)=0$. 
		\item On the other hand, when $B\geq B_{\min}$, $(0,B)$ is always achievable; 
		choose $x_{0}\in X$ such that $B_{\min}\leq t(x_{0})<B+\epsilon$, and 
		consider 	the encoder $f:m\mapsto x_{0}$. 
	\end{enumerate}
\end{remk}

\begin{exam}\ \\
	In some literatures such as \cite{ElGamalNetwork:2011}, achievability of a 
	rate is defined in terms of codes satisfying cost constraint $B$, not $B+\epsilon$ 
	as in ours. This difference is just a minor issue, since 
	two different operational capacity-cost functions arising from different definitions of 
	achievability indeed coincide on $(B_{\min},\infty]$. However, at $B=B_{\min}$, 
	the capacity-cost function defined in terms of codes satisfying average cost constraint 
	$B$ (which will be denoted as $\mathrm{C}_{o}'$) may not be lower-semicontinuous 
	in general. For example, let $X=Y=\R$ and $\kappa(x)=\mathfrak{d}_{x}$, 
	$t(x)=x^{2}$ for each $x\in X$. This is the noiseless channel with real alphabet and 
	quadratic cost function. Then one can easily verify that $\mathrm{C}_{o}'(0)=0$ while 
	$\mathrm{C}_{o}'(B)=\infty$ for $B>0$. On the other hand, $\mathrm{C}_{o}(B)=\infty$ 
	for all $B\geq0$. 
\end{exam}

\begin{lemm}\ \\
	The information capacity-cost function $\mathrm{C}_{i}:[0,\infty]\ra[0,\infty]$ 
	defined above satisfies the followings: 
	\begin{enumerate}
		\item $\mathrm{C}_{i}$ is an increasing function. 
		\item $\mathrm{C}_{i}$ is concave on $[B_{\min},\infty]$ and 
		continuous on $(B_{\min},\infty)$. 
	\end{enumerate}
\end{lemm}

We call a function $f:I\ra[0,\infty]$ \emph{convex (concave, respectively)} where 
$I$ is a sub-interval of $[0,\infty]$, if $f(\lambda x+(1-\lambda)y)
\leq\lambda f(x)+(1-\lambda)f(y)$ ($f(\lambda x+(1-\lambda)y)
\geq\lambda f(x)+(1-\lambda)f(y)$, respectively) for all $x,y\in I$ and 
$\lambda\in[0,1]$. 

\begin{IEEEproof}
	As others are trivial consequences, we only prove concavity. It is also 
	sufficient to show that $\mathrm{C}_{i}$ is concave on $(B_{\min},\infty)$; 
	concavity at $B_{\min}$ and $\infty$ easily follows since $\mathrm{C}_{i}$ is 
	increasing. Let $B_{1},B_{2}\in(B_{\min},\infty)$ with $B_{1}<B_{2}$ and 
	$\lambda\in[0,1]$. We first assume that $\mathrm{C}_{i}(B_{1})$ and 
	$\mathrm{C}_{i}(B_{2})$ are both finite. Let $\epsilon>0$ be given and choose 
	$\mu_{1},\mu_{2}\in\Delta(X)$ with $\int t\,d\mu_{1}\leq B_{1}$, 
	$\int t\,d\mu_{2}\leq B_{2}$, $\mathrm{C}_{i}(B_{1})\leq I(\mu_{1},\kappa)
	+\epsilon$, and $\mathrm{C}_{i}(B_{2})\leq I(\mu_{2},\kappa)+\epsilon$. 
	Let $\mu=\lambda\mu_{1}+(1-\lambda)\mu_{2}$ and $B=\lambda B_{1}+(1-\lambda)B_{2}$, 
	then $\int t\,d\mu=\lambda\int t\,d\mu_{1}+(1-\lambda)\int t\,d\mu_{2}\leq B$, 
	and due to concavity of the function $I(\,\cdot\,,\kappa)$ 
	(see \reflemm{lemm:convexity mutual}), 
	\begin{displaymath}
		\mathrm{C}_{i}(B)\geq I(\mu,\kappa)\geq\lambda I(\mu_{1},\kappa)
		+(1-\lambda)I(\mu_{2},\kappa)
		\geq\lambda\mathrm{C}_{i}(B_{1})+(1-\lambda)\mathrm{C}_{i}(B_{2})-\epsilon.
	\end{displaymath}
	Since $\epsilon>0$ is arbitrary, it follows that 
	\begin{displaymath}
		\mathrm{C}_{i}(B)\geq\lambda\mathrm{C}_{i}(B_{1})
		+(1-\lambda)\mathrm{C}_{i}(B_{2}).
	\end{displaymath}
	Next, assume that $\mathrm{C}_{i}(B_{2})$ is infinite and $\lambda\neq1$. 
	Then for any $M\geq0$, one can choose $\mu_{2}$ so that 
	$I(\mu_{2},\kappa)\geq\frac{M}{1-\lambda}$. Then, 
	\begin{displaymath}
		\mathrm{C}_{i}(B)\geq I(\mu,\kappa)\geq \lambda I(\mu_{1},\kappa)
		+(1-\lambda)I(\mu_{2},\kappa)\geq\lambda I(\mu_{1},\kappa)+M\geq M, 
	\end{displaymath}
	so this shows that $\mathrm{C}_{i}(B)=\infty$. This concludes that either 
	but not both of the followings should be hold: 
	\begin{enumerate}
		\item $\mathrm{C}_{i}$ is identically $\infty$ on $(B_{\min},\infty)$, or 
		\item $\mathrm{C}_{i}$ is everywhere finite on $(B_{\min},\infty)$ and concave. 
	\end{enumerate}
	For both cases we have the result. 
\end{IEEEproof}

\begin{thrm}[Point-to-point channel coding theorem with average cost constraint]\ \\
	Let $(X,Y,\kappa,t)$ be a memoryless channel with a cost function. Then, 
	\begin{displaymath}
		\mathrm{C}_{o}(B)=\mathrm{C}_{i}^{+}(B)
		:=\lim_{\epsilon\ra0^{+}}\mathrm{C}_{i}(B+\epsilon)
	\end{displaymath}
	for all $B\in[0,\infty]$. 
\end{thrm}

\begin{remk}\ 
	\begin{enumerate}
		\item Since $\mathrm{C}_{i}$ is increasing, the limit on the right-hand 
		side always exists. 
		\item Note that $\mathrm{C}_{o}(B)=\mathrm{C}_{i}(B)=\mathrm{C}_{i}^{+}(B)=0$ for 
		$B<B_{\min}$. Hence, we may assume $B\geq B_{\min}$. Of course, we have 
		$\mathrm{C}_{i}^{+}(B)=\mathrm{C}_{i}(B)$ for $B>B_{\min}$ by continuity. 
	\end{enumerate}
\end{remk}

\begin{IEEEproof}[Proof of converse $(\mathrm{C}_{o}\leq\mathrm{C}_{i}^{+})$]
	Assume that $(R,B)$ is achievable for some $R\in[0,\infty)$. Then for given 
	$\epsilon\in(0,\frac{1}{2})$, there exists an $(n,M)$-channel code $(f,g)$ 
	satisfying the average cost constraint $B+\epsilon$ with 
	\begin{displaymath}
		\frac{\log M}{n}\geq R\quad\textrm{and}\quad
		P_{e}(f,g)\leq\epsilon. 
	\end{displaymath}
	Construct a uniformly distributed random variable 
	$\mathbf{m}$ taking values in $[1:M]$ and a random variable $\mathbf{y}^{n}$ taking 
	values in $Y^{n}$ which follows a conditional distribution $\kappa^{n}$ given 
	$\mathbf{x}^{n}$, where $\mathbf{x}^{n}\defas f(\mathbf{m})$. Let 
	$\hat{\mathbf{m}}\defas g(\mathbf{y}^{n})$, 
	then $P_{e}(f,g)=\Pr(\hat{\mathbf{m}}\neq\mathbf{m})$. 
	We proceed in a similar way to the case of discrete memoryless channel: 
	\begin{eqnarray*}
		\log M=H(\mathbf{m})
		=H(\mathbf{m}|\hat{\mathbf{m}})+I(\mathbf{m};\hat{\mathbf{m}})
		\leq H(P_{e}(f,g))+P_{e}(f,g)\log M+I(\mathbf{m};\mathbf{y}^{n})
	\end{eqnarray*}
	by Fano's inequality, so 
	\begin{eqnarray*}
		(1-P_{e}(f,g))\log M&\leq&H(P_{e}(f,g))+I(\mathbf{m};\mathbf{y}^{n})\\
		&=&H(P_{e}(f,g))+\sum_{i=1}^{n}I(\mathbf{m};\mathbf{y}_{i}|\mathbf{y}^{i-1})\\
		&\leq&H(P_{e}(f,g))+\sum_{i=1}^{n}I(\mathbf{m},\mathbf{y}^{i-1};\mathbf{y}_{i})\\
		&=&H(P_{e}(f,g))+\sum_{i=1}^{n}I(\mathbf{x}_{i};\mathbf{y}_{i})
		\leq H(P_{e}(f,g))+\sum_{i=1}^{n}\mathrm{C}_{i}
		\left(\mathrm{E}[t(\bold{x}_{i})]\right).
	\end{eqnarray*}
	Since $\mathrm{C}_{i}$ is increasing and concave on $[B_{\min},\infty]$, 
	\begin{eqnarray*}
		(1-P_{e}(f,g))\frac{\log M}{n}&\leq&
		\frac{H(P_{e}(f,g))}{n}+\frac{1}{n}\sum_{i=1}^{n}\mathrm{C}_{i}\left(
		\mathrm{E}[t(\bold{x}_{i})]\right)\\
		&\leq&\frac{H(P_{e}(f,g))}{n}+\mathrm{C}_{i}\left(
		\mathrm{E}[t^{(n)}(\bold{x}^{n})]\right)\leq
		\frac{H(P_{e}(f,g))}{n}+\mathrm{C}_{i}(B+\epsilon),
	\end{eqnarray*}
	so
	\begin{displaymath}
		R\leq\frac{\log M}{n}\leq\frac{1}{1-\epsilon}\left(H(\epsilon)
		+\mathrm{C}_{i}(B+\epsilon)\right).
	\end{displaymath}
	Since $\epsilon\in(0,\frac{1}{2})$ is arbitrary, we get $R\leq\mathrm{C}_{i}^{+}(B)$. 
	This shows that $\mathrm{C}_{o}(B)\leq\mathrm{C}_{i}^{+}(B)$. 
\end{IEEEproof}

\begin{IEEEproof}[Proof of achievability $(\mathrm{C}_{o}\geq\mathrm{C}_{i}^{+})$] 
	Let $R<\mathrm{C}_{i}^{+}(B)$ be given, so that $R<\mathrm{C}_{i}(B+\epsilon)$ 
	whenever $\epsilon>0$ is sufficiently small. Take any such $\epsilon>0$. 
	
	\textbf{(Codebook generation)} We find $\mu\in\Delta(X)$ and a $\mu$-typicality 
	criterion $\mathcal{U}$ as follows: 
	\begin{enumerate}
		\item If $B=\infty$, then take $\mu\in\Delta(X)$ with $R<I(\mu,\kappa)$. 
		Let $\mathcal{U}=(\emptyset;1;\emptyset)$. 
		\item If $B<\infty$, then take $\mu\in\Delta(X)$ with $R<I(\mu,\kappa)$ and 
		$\int t\,d\mu\leq B+\frac{\epsilon}{2}$. In this case, $t\in\mathscr{L}^{1}(\mu)$. 
		Let $\mathcal{U}=(t;\frac{\epsilon}{2};\emptyset)$. 
	\end{enumerate}
	Note that for both of the cases, $t^{(n)}(x^{n})\leq B+\epsilon$ whenever $x^{n}\in
	\mathcal{T}_{\mathcal{U}}^{(n)}(\mu)$. Apply the packing lemma with 
	$(X,\mathscr{A}_{X})\leftarrow(\set{*},\wp(\set{*}))$, $(Y,\mathscr{A}_{Y})\leftarrow
	(Y,\mathscr{B})$, $(Z,\mathscr{A}_{Z})\leftarrow(X,\mathscr{A})$, and 
	$\mu_{XZY}\leftarrow\mathfrak{d}_{*}\times\mu\kappa$ 
	to get a $\mu\kappa$-typicality criterion $\mathcal{V}$ and a positive constant $c>0$ 
	for given $R$. Fix sufficiently large $n\in\Z^{+}$, and randomly and independently generate 
	$\lceil2^{nR}\rceil$ i.i.d. sequences $\set{\bold{x}^{n}(m)}_{m\in[1:2^{nR}]}$ according 
	to $\mu^{n}$; exactly how $n$ should be large to be specified later. Now, for given 
	realization $\omega=\set{x^{n}(m)}_{m\in[1:2^{nR}]}$ of 
	$\set{\bold{x}^{n}(m)}_{m\in[1:2^{nR}]}$, we define an encoder 
	$f_{\omega}:[1:2^{nR}]\ra X^{n}$ and a decoder $g_{\omega}:Y^{n}\ra[0:2^{nR}]$ as follows: 
	
	\textbf{(Encoding)} If $x^{n}(m)\in\mathcal{T}_{\mathcal{U}}^{(n)}(\mu)$, then 
	$f_{\omega}(m)\defas x^{n}(m)$, while $f_{\omega}(m)\defas x_{0}^{n}$ otherwise, where 
	$x_{0}$ is a fixed element in $X$ such that $t(x_{0})\leq B+\epsilon$ 
	(such $x_{0}$ exists since $B\geq B_{\min}$). Then clearly this encoder satisfies 
	average cost constraint $B+\epsilon$. 
	
	\textbf{(Decoding)} If there uniquely exists $\hat{m}\in[1:2^{nR}]$ 
	such that $(x^{n}(\hat{m}),y^{n})\in\mathcal{T}_{\mathcal{V}}^{(n)}(\mu\kappa)$ for 
	given $y^{n}$, then $g_{\omega}(y^{n})\defas\hat{m}$, while $g_{\omega}(y^{n})\defas0$ 
	otherwise. Note that $g_{\omega}$ is measurable. 
	
	\textbf{(Analysis of the probability of error)} Let $\bold{f},\bold{g}$ be the 
	encoder and the decoder corresponding to $\bold{x}^{n}(m)$'s. Define 
	$P_{e}^{av}\defas\mathrm{E}[P_{e}(\bold{f},\bold{g})]$. Note that by symmetry 
	\begin{displaymath}
		\mathrm{E}[P_{e,m}(\bold{f},\bold{g})]=\mathrm{E}[P_{e,1}(\bold{f},\bold{g})]
	\end{displaymath}
	for each $m\in[1:2^{nR}]$, so we may assume that the message is chosen to be $1$; that is, 
	\begin{displaymath}
		P_{e}^{av}=\mathrm{E}[P_{e,1}(\bold{f},\bold{g})]
		=\Pr(\set{\bold{g}(\bold{y}^{n})\neq1})
	\end{displaymath}
	where $\bold{y}^{n}$ is the received sequence when the message is chosen to be $1$. 
	The error event $\set{\bold{g}(\bold{y}^{n})\neq1}$ is contained in the union of the 
	following events: 
	\begin{enumerate}
		\item $\Epsilon_{1}\defas\set{(\mathbf{x}^{n}(1),\mathbf{y}^{n})
		\notin\mathcal{T}_{\mathcal{V}}^{(n)}(\mu\kappa)}$, 
		\item $\Epsilon_{2}\defas\set{(\mathbf{x}^{n}(m),\mathbf{y}^{n})
		\in\mathcal{T}_{\mathcal{V}}^{(n)}(\mu\kappa)\ \textrm{for some $m\neq1$}}$, 
	\end{enumerate}
	so $P_{e}^{av}\leq\Pr(\Epsilon_{1})+\Pr(\Epsilon_{2})$. 
	Since $\mathbf{x}^{n}(m)$ and $\mathbf{y}^{n}$ are independent when $m\neq 1$, by the 
	assumption on $\mathcal{V}$, we know that $\Pr(\Epsilon_{2})\leq2^{-cn}$. 
	On the other hand, 
	\begin{eqnarray*}
		\Pr(\Epsilon_{1})&\leq&\Pr\left(\mathbf{x}^{n}(1)\notin
		\mathcal{T}_{\mathcal{U}}^{(n)}(\mu)\right)+\Pr\left(
		\mathbf{x}^{n}(1)\in\mathcal{T}_{\mathcal{U}}^{(n)}(\mu),\ 
		(\mathbf{x}^{n},\mathbf{y}^{n})\notin\mathcal{T}_{\mathcal{V}}^{(n)}(\mu\kappa)\right)
		\\&\leq&\left(1-\mu^{n}\left(\mathcal{T}_{\mathcal{U}}^{(n)}(\mu)\right)\right)
		+\left(1-(\mu\kappa)^{n}\left(
		\mathcal{T}_{\mathcal{V}}^{(n)}(\mu\kappa)\right)\right)
	\end{eqnarray*}
	also can be made sufficiently small when $n$ is large, by the asymptotic 
	equipartition property. Therefore, we can take $n$ sufficiently large to make 
	$P_{e}^{av}\leq\epsilon$. Hence, there exists $\omega$ such that 
	the $(n,\lceil 2^{nR}\rceil)$-channel code $(f_{\omega},g_{\omega})$ 
	(which is shown to satisfy average cost constraint $B+\epsilon$) having the property 
	\begin{displaymath}
		\frac{\log\lceil 2^{nR}\rceil}{n}\geq R\quad\textrm{and}\quad
		P_{e}(f_{\omega},g_{\omega})\leq\epsilon.
	\end{displaymath}
	Since $\epsilon>0$ can be taken to be arbitrarily small, $(R,B)$ is achievable. 
	Therefore, $\mathrm{C}_{o}(B)\geq\mathrm{C}_{i}^{+}(B)$. 
\end{IEEEproof}

\subsection{Point-to-point lossy source coding theorem}
Let $(X,\mathscr{A})$, $(Y,\mathscr{B})$ be standard Borel spaces. 

\begin{defn}[Memoryless sources]\ \\
	A \emph{memoryless source with a distortion function} is a quadruple 
	$(X,Y,\mu,t)$, where $\mu\in\Delta(X)$ and $t:X\times Y\ra[0,\infty]$ is a 
	measurable function. Here, $X$ is called the \emph{source alphabet}, $Y$ is called 
	the \emph{reconstruction alphabet}, $\mu$ is called the 
	\emph{source probability measure}, and $t$ is called the 
	\emph{distortion function}. We always assume that 
	$D_{\max}\defas\inf_{y\in Y}\int t(x,y)\,d\mu(x)<\infty$. 
	For $n,M\in\Z^{+}$, an \emph{$(n,M)$-source code} for this memoryless source consists 
	of two measurable mappings $f:X^{n}\ra[0:M)$ and $g:[0:M)\ra Y^{n}$, respectively 
	called an \emph{encoder} and a \emph{decoder}; here, $0\in[0:M)$ denotes the encoding 
	error. For $R\in[0,\infty)$ and $D\in[0,\infty]$, the pair $(R,D)$ is said to be 
	\emph{achievable}, if for any $\epsilon>0$, there exists an $(n,M)$-source code 
	$(f,g)$ such that 
	\begin{displaymath}
		\frac{\log M}{n}\leq R\quad\textrm{and}\quad
		\int t^{(n)}(x^{n},g\circ f(x^{n}))
		\,d\mu^{n}(x^{n})\leq D+\epsilon, 
	\end{displaymath}
	where $t^{(n)}:X^{n}\times Y^{n}\ra[0,\infty]$ is defined as 
	$t^{(n)}:(x^{n},y^{n})\mapsto\frac{1}{n}\sum_{i=1}^{n}t(x_{i},y_{i})$. 
	The \emph{operational rate-distortion function} 
	$\mathrm{R}_{o}:[0,\infty]\ra[0,\infty]$ is defined as 
	\begin{displaymath}
		\mathrm{R}_{o}:D\mapsto\inf\setbc{R\in[0,\infty)}{\textrm{$(R,D)$ is achievable}}.
	\end{displaymath}
	The infimum of $\emptyset$ is defined to be $\infty$. On the other hand, 
	the \emph{information rate-distortion function} 
	\mbox{$\mathrm{R}_{i}:[0,\infty]\ra[0,\infty]$} is defined as 
	\begin{displaymath}
		\mathrm{R}_{i}:D\mapsto\inf_{\kappa\in\mathcal{K}(X;Y);
		\int t\,d\mu\kappa\leq D}I(\mu,\kappa).
	\end{displaymath}
\end{defn}

\begin{remk}\ 
	\begin{enumerate}
		\item Define 
		\begin{displaymath}
			D_{\min}:=\inf_{\kappa\in\mathcal{K}(X;Y)}\int t\,d\mu\kappa,
		\end{displaymath}
		then it is clear that $\mathrm{R}_{i}(D)=\infty$ if $D<D_{\min}$. 
		Also, for that case $(R,D)$ is never achievable for any $R$: for any 
		encoder $f$ and decoder $g$, 
		\begin{displaymath}
			\int t^{(n)}(x^{n},g\circ f(x^{n}))\,d\mu^{n}(x^{n})
			=\frac{1}{n}\sum_{i=1}^{n}\int\left[
			\int t(x_{i},g_{i}\circ f(x^{n}))\,d\mu^{n-1}(x_{1}^{i-1},x_{i+1}^{n})\right]d\mu(x_{i})
		\end{displaymath}
		where $g=(g_{1},\ \cdots\ ,g_{n})$ by 
		Tonelli's theorem~\cite[Chapter 4]{PollardUserGuide:1981}. Define $\kappa_{i}(x_{i})$ as 
		\begin{displaymath}
			\kappa_{i}(B|x_{i}):=\mu^{n-1}\left(\setbc{(x_{1}^{i-1},x_{i+1}^{n})}
			{g_{i}\circ f(x^{n})\in B}\right),
		\end{displaymath}
		then 
		\begin{displaymath}
			\int t^{(n)}(x^{n},g\circ f(x^{n}))\,d\mu^{n}(x^{n})
			=\frac{1}{n}\sum_{i=1}^{n}\int\left[\int t(x_{i},y)\,d\kappa_{i}(y|x_{i})\right]
			d\mu(x_{i})=\frac{1}{n}\sum_{i=1}^{n}\int t\,d\mu\kappa_{i}\geq D_{\min}.
		\end{displaymath}
		Hence, for $D<D_{\min}$, we have $\mathrm{R}_{o}(D)=\mathrm{R}_{i}(D)=\infty$. 
		\item On the other hand, when $D>D_{\max}$, we have 
		$\mathrm{R}_{o}(D)=\mathrm{R}_{i}(D)=0$. To show that, take $y_{c}\in Y$ such 
		that $\int t(x,y_{c})\,d\mu(x)\leq D$. Then for $\mathrm{R}_{o}(D)=0$, consider 
		the decoder $g:m\mapsto y_{c}$, and for 
		$\mathrm{R}_{i}(D)=0$, consider the constant kernel 
		$\kappa:x\mapsto\mathfrak{d}_{y_{c}}$. 
	\end{enumerate}
\end{remk}

\begin{exam}\ \\
	Again, boundary behavior of rate-distortion functions can be pathological: 
	information rate-distortion function is in general not upper-semicontinuous at boundary, 
	while the operational rate-distortion function is upper-semicontinuous 
	due to effect of ``$+\epsilon$'' on the distortion criteria. Here is an example of 
	$\mathrm{R}_{o}\neq\mathrm{R}_{i}$: let $X=\set{1,2}$, $Y=\Z^{+}$, 
	$\mu(\set{1})=\mu(\set{2})=1/2$, and 
	\begin{displaymath}
		t:(x,y)\mapsto\begin{cases}0&\textrm{if $x=y$}\\\frac{1}{y}&\textrm{otherwise}\end{cases}.
	\end{displaymath}
	Then, one can easily see that $\mathrm{R}_{o}(0)=0$ by considering a constant decoder. 
	However, $\mathrm{R}_{i}(0)=1$, since any kernel $\kappa\in\mathcal{K}(X;Y)$ 
	satisfying $\int t\,d\mu\kappa\leq0$ should satisfy $\mu\kappa(\setbc{(x,y)}{x\neq y})=0$, 
	so $I(\mu,\kappa)=1$. Still, one can easily see that $\mathrm{R}_{i}(0)=0$. 
\end{exam}

\begin{lemm}\ \\
	The information rate-distortion function $\mathrm{R}_{i}:[0,\infty]\ra[0,\infty]$ 
	defined above satisfies the followings: 
	\begin{enumerate}
		\item $\mathrm{R}_{i}$ is a decreasing function. 
		\item $\mathrm{R}_{i}$ is convex on $[D_{\min},\infty]$ and 
		continuous on $(D_{\min},\infty]$. 
	\end{enumerate}
\end{lemm}

\begin{IEEEproof}
	It suffices to show convexity on $(D_{\min},\infty)$, because $\mathrm{R}_{i}$ is 
	clearly decreasing and $\mathrm{R}_{i}(D)=0$ for $D>D_{\max}$. 
	Let $D_{1},D_{2}\in(D_{\min},\infty)$ with $D_{1}<D_{2}$ and $\lambda\in[0,1]$. 
	We may assume that $\mathrm{R}_{i}(D_{1})$ 
	and $\mathrm{R}_{i}(D_{2})$ are both finite, since the result is trivial when 
	one of them is infinite. Let $\epsilon>0$ be given and choose 
	$\kappa_{1},\kappa_{2}\in\mathcal{K}(X;Y)$ with 
	$\int t\,d\mu\kappa_{1}\leq D_{1}$, $\int t\,d\mu\kappa_{2}\leq D_{2}$, 
	$\mathrm{R}_{i}(D_{1})\geq I(\mu,\kappa_{1})-\epsilon$, and 
	$\mathrm{R}_{i}(D_{2})\geq I(\mu,\kappa_{2})-\epsilon$. 
	Let $\kappa=\lambda\kappa_{1}+(1-\lambda)\kappa_{2}$ and 
	$D=\lambda D_{1}+(1-\lambda)D_{2}$, then $\int t\,d\mu\kappa
	=\lambda\int t\,d\mu\kappa_{1}+(1-\lambda)\int t\,
	d\mu\kappa_{2}\leq D$, and due to convexity of the function $I(\mu,\,\cdot\,)$ 
	(see \reflemm{lemm:convexity mutual}), 
	\begin{displaymath}
		\mathrm{R}_{i}(D)\leq I(\mu,\kappa)\leq\lambda I(\mu,\kappa_{1})
		+(1-\lambda)I(\mu,\kappa_{2})\leq\lambda\mathrm{R}_{i}(D_{1})
		+(1-\lambda)\mathrm{R}_{i}(D_{2})+\epsilon.
	\end{displaymath}
	Since $\epsilon>0$ is arbitrary, it follows that 
	\begin{displaymath}
		\mathrm{R}_{i}(D)\leq\lambda\mathrm{R}_{i}(D_{1})
		+(1-\lambda)\mathrm{R}_{i}(D_{2}).
	\end{displaymath}
	Therefore, we get convexity. 
\end{IEEEproof}

\begin{thrm}[Point-to-point lossy source coding theorem]\ \\
	Let $(X,Y,\mu,t)$ be a memoryless source with a distortion measure. Then, 
	\begin{displaymath}
		\mathrm{R}_{o}(D)=\mathrm{R}_{i}^{+}(D):=\lim_{\epsilon\ra0^{+}}
		\mathrm{R}_{i}(D+\epsilon).
	\end{displaymath}
	for all $D\in[0,\infty]$. 
\end{thrm}

\begin{remk}
	\item Since $\mathrm{R}_{i}$ is decreasing, the limit on the right-hand 
	side always exists. 
	\item Note that $\mathrm{R}_{o}(D)=\mathrm{R}_{i}(D)=\mathrm{R}_{i}^{+}(D)=\infty$ 
	for $D<D_{\min}$ and $\mathrm{R}_{o}(D)=\mathrm{R}_{i}(D)
	=\mathrm{R}_{i}^{+}(D)=0$ for $D>D_{\max}$. Hence, we may assume $D\geq D_{\min}$. 
	Of course, we have $\mathrm{R}_{i}^{+}(D)=\mathrm{R}_{i}(D)$ for 
	$D>D_{\min}$ by continuity. 
\end{remk}

\begin{IEEEproof}[Proof of converse $(\mathrm{R}_{o}\geq\mathrm{R}_{i}^{+})$]
	Assume that $(R,D)$ is achievable for some $R\in[0,\infty)$. 
	Then for given $\epsilon\in(0,1)$, there exists an $(n,M)$-source code 
	$(f,g)$ such that 
	\begin{displaymath}
		\frac{\log M}{n}\leq R\quad\textrm{and}\quad
		\int t^{(n)}(x^{n},g\circ f(x^{n}))\,d\mu^{n}(x^{n})\leq D+\epsilon. 
	\end{displaymath}
	Construct a random variable $\mathbf{x}^{n}$ following the 
	distribution $\mu^{n}$ and define $\mathbf{m}\defas f(\mathbf{x}^{n})$,  
	$\mathbf{y}^{n}\defas g(\mathbf{m})$. We proceed just as the case of 
	discrete memoryless source: 
	\begin{eqnarray*}
		\log M&\geq& I(\mathbf{m};\mathbf{m})\\&\geq&
		I(\mathbf{x}^{n};\mathbf{y}^{n})\\
		&=&\sum_{i=1}^{n}I(\mathbf{x}_{i};\mathbf{y}^{n}|\mathbf{x}^{i-1})\\
		&=&\sum_{i=1}^{n}I(\mathbf{x}_{i};\mathbf{y}^{n},\mathbf{x}^{i-1})\\
		&\geq&\sum_{i=1}^{n}I(\mathbf{x}_{i};\mathbf{y}_{i})
		\geq\sum_{i=1}^{n}\mathrm{R}_{i}\left(
		\mathrm{E}[t(\mathbf{x}_{i},\mathbf{y}_{i})]\right).
	\end{eqnarray*}
	Since $\mathrm{R}_{i}$ is decreasing and convex on $[D_{\min},\infty]$, 
	\begin{eqnarray*}
		\frac{\log M}{n}&\geq&
		\frac{1}{n}\sum_{i=1}^{n}\mathrm{R}_{i}\left(
		\mathrm{E}[t(\mathbf{x}_{i},\mathbf{y}_{i})]\right)\\
		&\geq&\mathrm{R}_{i}\left(\mathrm{E}[t^{(n)}(\bold{x}^{n},\bold{y}^{n})]\right)
		=\mathrm{R}_{i}\left(\int t^{(n)}(x^{n},g\circ f(x^{n}))
		\,d\mu^{n}(x^{n})\right)\geq\mathrm{R}_{i}(D+\epsilon),
	\end{eqnarray*}
	so 
	\begin{displaymath}
		R\geq\frac{\log M}{n}\geq\mathrm{R}_{i}(D+\epsilon).
	\end{displaymath}
	Then since $\epsilon\in(0,1)$ is arbitrary, we get $R\geq\mathrm{R}_{i}^{+}(D)$. 
	Hence, $\mathrm{R}_{o}(D)\geq\mathrm{R}_{i}^{+}(D)$. 
\end{IEEEproof}

\begin{IEEEproof}[Proof of achievability $(\mathrm{R}_{o}\leq\mathrm{R}_{i}^{+})$] 
	We may assume that $\mathrm{R}_{i}^{+}(D)<\infty$ since if 
	$\mathrm{R}_{i}^{+}(D)=\infty$ then we clearly have 
	$\mathrm{R}_{o}(D)\leq\mathrm{R}_{i}^{+}(D)$. Let $R>\mathrm{R}_{i}^{+}(D)$ be given, 
	so that $R>\mathrm{R}_{i}(D+\frac{\epsilon}{4})$ whenever $\epsilon>0$ is 
	sufficiently small. Take any such $\epsilon>0$. 
	
	\textbf{(Codebook generation)} We find $\kappa\in\mathcal{K}(X;Y)$ 
	with $R>I(\mu;\kappa)$ and $\int t\,d\mu\kappa\leq D+\frac{\epsilon}{4}$; 
	note that $t\in\mathscr{L}^{1}(\mu\kappa)$. Let 
	$\mathcal{V}_{1}=\left(t;\frac{\epsilon}{4};\emptyset\right)$, then 
	$t^{(n)}(x^{n},y^{n})\leq D+\frac{\epsilon}{2}$ whenever 
	$(x^{n},y^{n})\in\mathcal{T}_{\mathcal{V}_{1}}^{(n)}(\mu\kappa)$. 
	Apply the covering lemma with $(X,\mathscr{A}_{X})\leftarrow
	(\set{*},\wp(\set{*}))$, $(Y,\mathscr{A}_{Y})\leftarrow
	(X,\mathscr{A})$, $(Z,\mathscr{A}_{Z})\leftarrow(Y,\mathscr{B})$, 
	$\mu_{XYZ}\leftarrow\mathfrak{d}_{*}\times\mu\kappa$ to get a 
	$\mu\kappa$-typicality criterion $\mathcal{V}\leq\mathcal{V}_{1}$, 
	a $\mu$-typicality criterion $\mathcal{U}$, and $c>0$, for given $R$. 	
	Let $\nu:=\kappa_{*}\mu$. Fix sufficiently large $n\in\Z^{+}$, and 
	randomly and independently generate $\lfloor2^{nR}\rfloor$ i.i.d. sequences 
	$\set{y^{n}(m)}_{m\in[1:2^{nR})}$ according to $\nu^{n}$; 
	exactly how $n$ should be large to be specified later. 
	Now, for given realization $\omega:=\set{y^{n}(m)}_{m\in[1:2^{nR})}$ of 
	$\set{\bold{y}^{n}(m)}_{m\in[1:2^{nR})}$, we define an encoder 
	$f_{\omega}:X^{n}\ra[0:2^{nR})$ and a decoder $g_{\omega}:[0:2^{nR})\ra X^{n}$ as follows: 
	
	\textbf{(Encoding)} If $(x^{n},y^{n}(m))\in\mathcal{T}_{\mathcal{V}}^{(n)}(\mu\kappa)$ 
	for some $m\in[1:2^{nR})$, then define $f_{\omega}(x^{n})$ to be the minimum of such $m$, 
	while $f_{\omega}(x^{n})\defas 0$ otherwise. Then $f_{\omega}$ is measurable. 
	
	\textbf{(Decoding)} Define $g_{\omega}(m)\defas y^{n}(m)$ for each $m\in[1:2^{nR})$ and 
	$g_{\omega}(0)\defas y_{c}^{n}$, where $y_{c}\in Y$ is a fixed element in $Y$ such that 
	$\int t(x,y_{c})\,d\mu(x)\leq D_{\max}+1$. 
	
	\textbf{(Analysis of expected distortion)} Let $\bold{x}^{n}$ be the random 
	variable representing the input to the encoder which is independent to 
	$\bold{y}^{n}(m)$'s, and let $\bold{f}$, $\bold{g}$ be the encoder and the decoder 
	corresponding to $\bold{y}^{n}(m)$'s. Let $\bold{y}^{n}$ be the reconstructed codeword 
	$\bold{g}(\bold{f}(\bold{x}^{n}))$. Define the following 
	error events: 
	\begin{enumerate}
		\item $\Epsilon_{1}\defas\set{\mathbf{x}^{n}
		\notin\mathcal{T}_{\mathcal{U}}^{(n)}(\mu)}$, 
		\item $\Epsilon_{2}\defas\set{\mathbf{x}^{n}\in\mathcal{T}_{\mathcal{U}}^{(n)}(\mu)
		\ \textrm{and}\ (\mathbf{x}^{n},\mathbf{y}^{n}(m))\notin
		\mathcal{T}_{\mathcal{V}}^{(n)}(\mu\kappa)\ \textrm{for all $m\in[1:2^{nR}]$}}$. 
	\end{enumerate}
	Since $\mathbf{x}^{n}$ and $\mathbf{y}^{n}(m)$'s are independent, by the assumption on 
	$\mathcal{U}$ and $\mathcal{V}$, we know that $\Pr(\Epsilon_{2})\leq2^{-cn}$ provided 
	that $n$ is sufficiently large. Together with the asymptotic equipartition property 
	applied to $\mu$, it follows that 
	\begin{displaymath}
		\mathrm{Pr}\left(\Epsilon_{1}\cup\Epsilon_{2}\right)
		\leq\frac{\epsilon}{2(D_{\max}+1)}
	\end{displaymath}
	provided that $n$ is sufficiently large. Therefore, 
	\begin{eqnarray*}
		\mathrm{E}\left[t^{(n)}(\bold{x}^{n},\bold{g}(\bold{f}(\bold{x}^{n})))
		\right]&\leq&\mathrm{\Pr}(\Epsilon_{1}\cup\Epsilon_{2})(D_{\max}+1)+
		D+\frac{\epsilon}{2}\leq D+\epsilon.
	\end{eqnarray*}
	Hence, there exists $\omega$ such that the $(n,\lfloor2^{nR}\rfloor)$-code 
	$(f_{\omega},g_{\omega})$ satisfies 
	\begin{displaymath}
		\frac{\log\lfloor 2^{nR}\rfloor}{n}\leq R\quad\textrm{and}\quad
		\int t^{(n)}(x^{n},g_{\omega}\circ f_{\omega}(x^{n}))
		\,d\mu^{n}(x^{n})\leq D+\epsilon. 
	\end{displaymath}
	Since $\epsilon>0$ can be taken to be arbitrarily small, $(R,D)$ is achievable. 
	Therefore, $\mathrm{R}_{o}(D)\leq\mathrm{R}_{i}^{+}(D)$. 
\end{IEEEproof}

%%%%%%%%%%%%%%%%%%%%%%%%%%%%%%%%%%%%%%%%%%%%%%%%%%%%%%%%%%%%%%%%%%%%%%%%%%%%%%%%%%%%%%%%%%%%%%%%%%
\section{Markov Lemma}
Along with conditional typicality lemma, joint typicality lemma, and packing and covering 
lemmas, there is another fundamental lemma used for derivations of inner bounds, called 
Markov lemma. It says that whenever we have a Markov chain 
$\mathbf{y}-\mathbf{x}-\mathbf{z}$, joint typicality of $(\mathbf{x}^{n},\mathbf{y}^{n})$ 
together with joint typicality of $(\mathbf{x}^{n},\mathbf{z}^{n})$ implies joint typicality of 
$(\mathbf{x}^{n},\mathbf{y}^{n},\mathbf{z}^{n})$ with high probability. 
The important difference from conditional typicality lemma is that 
$\bold{z}^{n}$ does not need to be conditionally i.i.d. given 
$(\bold{x}^{n},\bold{y}^{n})$. Markov lemma does not seem to be obtained in 
our setting with its full generality yet; however, it becomes a simple corollary 
of the bounded conditional typicality lemma when involved test functions are bounded. 
This includes the finite alphabet case as a special case since any integrable 
function on a finite measurable space should be bounded almost everywhere. 

Let $(X,\mathscr{A})$, $(Y,\mathscr{B})$, and $(Z,\mathscr{C})$ be measurable spaces. 
The following is a generalization of Lemma 12.1 of ~\cite[p.296]{ElGamalNetwork:2011}: 

\begin{thrm}[Bounded Markov lemma]\label{thrm:bdd Markov}\ \\
	Let $\mu\in\Delta(X)$, $\kappa\in\mathcal{K}(X;Y)$, and 
	$\lambda\in\mathcal{K}(X;Z)$. For each $n\in\Z^{+}$ and a $\mu\lambda$-typicality 
	criterion $\mathcal{S}$, let $\lambda_{\mathcal{S}}^{(n)}\in\mathcal{K}(X^{n};Z^{n})$. 
	Assume that, for any $\epsilon>0$, there exist a $\mu\lambda$-typicality criterion 
	$\mathcal{S}_{0}$ so that for any $\mu\lambda$-typicality criterion 
	$\mathcal{S}\leq\mathcal{S}_{0}$, one can find a $\mu$-typicality criterion 
	$\mathcal{U}$, satisfying 
	\begin{displaymath}
		\lambda_{\mathcal{S}}^{(n)}(E|x^{n})
		\leq2^{\epsilon n}\lambda^{n}(E|x^{n})
	\end{displaymath}
	for all $x^{n}\in\mathcal{T}_{\mathcal{U}}^{(n)}(\mu)$ and a measurable subset 
	$E$ of $\mathcal{T}_{\mathcal{S}}^{(n)}(\mu\lambda|x^{n})$, whenever 
	$n$ is sufficiently large. Then for any $\mu(\kappa\times\lambda)$-bounded 
	typicality criterion $\mathcal{W}$, there exists a $\mu\lambda$-typicality 
	criterion $\mathcal{S}_{0}$ and a positive 
	number $c>0$ such that, for any $\mu\lambda$-typicality criterion 
	$\mathcal{S}\leq\mathcal{S}_{0}$, there exists a $\mu\kappa$-typicality 
	criterion $\mathcal{V}$ so that 
	\begin{displaymath}
		\sup_{(x^{n},y^{n})\in\mathcal{T}_{\mathcal{V}}^{(n)}(\mu\kappa)}
		\lambda_{\mathcal{S}}^{(n)}\left(\mathcal{T}_{\mathcal{S}}^{(n)}
		(\mu\lambda|x^{n})\setminus\mathcal{T}_{\mathcal{W}}^{(n)}
		(\mu(\kappa\times\lambda)|x^{n},y^{n})\Big|x^{n}\right)\leq2^{-cn}.
	\end{displaymath}
	for all sufficiently large $n$. 
\end{thrm}

To avoid potential confusion, the dependency relation in the condition is 
written here formally: 
\begin{displaymath}
	\forall\epsilon;\exists\mathcal{S}_{0};\forall\mathcal{S}\leq\mathcal{S}_{0};
	\exists\mathcal{U};\exists n_{0};\forall n\geq n_{0};
	\forall x^{n}\in\mathcal{T}_{\mathcal{U}}^{(n)}(\mu);\forall E
	\subseteq\mathcal{T}_{\mathcal{S}}^{(n)}(\mu\lambda|x^{n})
	\left(\lambda_{\mathcal{S}}^{(n)}(E|x^{n})
	\leq2^{\epsilon n}\lambda^{n}(E|x^{n})\right)
\end{displaymath}
where some restrictions on the domains of variables are understood implicitly. 

\begin{IEEEproof}
	Let $\mathcal{W}$ be a $\mu(\kappa\times\lambda)$-bounded typicality criterion. 
	Then we can find a $\mu\kappa$-typicality criterion $\mathcal{V}_{1}$ and a 
	positive number $c_{1}>0$ such that 
	\begin{displaymath}
		\sup_{(x^{n},y^{n})\in\mathcal{T}_{\mathcal{V}_{1}}^{(n)}(\mu\kappa)}
		\lambda^{n}\left(Z^{n}\setminus\mathcal{T}_{\mathcal{W}}^{(n)}
		(\mu(\kappa\times\lambda)|x^{n},y^{n})\Big|x^{n}\right)\leq2^{-c_{1}n}
	\end{displaymath}
	for all $n\in\Z^{+}$ larger than some $n_{1}\in\Z^{+}$ by applying the 
	bounded conditional typicality lemma. Next, pick any $\epsilon\in(0,c_{1})$ 
	and apply the assumption on $\lambda_{\mathcal{S}}^{(n)}$'s to get a 
	$\mu\lambda$-typicality criterion $\mathcal{S}_{0}$. Let $c:=c_{1}-\epsilon$. 
	Now, choose any $\mu\lambda$-typicality criterion $\mathcal{S}\leq\mathcal{S}_{0}$. 
	Then there exists a $\mu$-typicality criterion $\mathcal{U}$ so that 
	\begin{displaymath}
		\lambda_{\mathcal{S}}^{(n)}(E|x^{n})
		\leq2^{\epsilon n}\lambda^{n}(E|x^{n})
	\end{displaymath}
	for all $x^{n}\in\mathcal{T}_{\mathcal{U}}^{(n)}(\mu)$ and a measurable subset 
	$E$ of $\mathcal{T}_{\mathcal{S}}^{(n)}(\mu\lambda|x^{n})$, whenever 
	$n$ is sufficiently large, say, larger than some $n_{0}\in\Z^{+}$. 
	Choose a $\mu\kappa$-typicality criterion $\mathcal{V}\leq\mathcal{V}_{1}$ 
	such that $(x^{n},y^{n})\in\mathcal{T}_{\mathcal{V}}^{(n)}(\mu\kappa)$ implies 
	$x^{n}\in\mathcal{T}_{\mathcal{U}}^{(n)}(\mu)$ for all $n\in\Z^{+}$ 
	(using \refprop{prop:typ pullback} with the projection onto $X$). 
	Fix $n\geq\max\set{n_{0},n_{1}}$ and $(x^{n},y^{n})\in
	\mathcal{T}_{\mathcal{V}}^{(n)}(\mu\kappa)$, then it follows that 
	\begin{eqnarray*}
		&&\lambda_{\mathcal{S}}^{(n)}\left(\mathcal{T}_{\mathcal{S}}^{(n)}
		(\mu\lambda|x^{n})\setminus\mathcal{T}_{\mathcal{W}}^{(n)}
		(\mu(\kappa\times\lambda)|x^{n},y^{n})\Big|x^{n}\right)\\&&\quad\quad\leq
		2^{n\epsilon}\lambda^{n}\left(\mathcal{T}_{\mathcal{S}}^{(n)}
		(\mu\lambda|x^{n})\setminus\mathcal{T}_{\mathcal{W}}^{(n)}
		(\mu(\kappa\times\lambda)|x^{n},y^{n})\Big|x^{n}\right)
		\leq2^{\epsilon n}2^{-c_{1}n}=2^{-cn}. 
	\end{eqnarray*}
\end{IEEEproof}

Note that, in the proof above, rather than Markovity (which is, the fact that 
$\lambda$ and $\lambda_{\mathcal{S}}^{(n)}$ are only functions of $x^{n}$, not $y^{n}$), 
exponential decay of the probability of error was the crucial concern, which is a result of 
boundedness of test functions. Unfortunately, we do not have this property for general 
integrable test functions, so validity of the theorem for that case is still not clear. 
As noted in Chapter 1, boundedness is quite a strong condition. The theorem cannot be applied 
directly to even the simplest case with Gaussian measures and quadratic functions, because 
quadratic functions are not bounded. Therefore, it is highly desired to extend 
the theorem to more general situations. 

\begin{defn}[Log-exponential typicality criteria]\ \\
	Let $\mu\in\Delta(X)$ and $\kappa\in\mathcal{K}(X;Y)$. A measurable function 
	$g:X\times Y\ra\R$ is said to be an \emph{log-exponential test function} with respect to 
	$(\mu,\kappa)$, if there exists a positive real number $\delta>0$ such that 
	\begin{displaymath}
		\int\log\left[\int2^{\delta\abs{g(x,y)}}\,d\kappa(y|x)\right]d\mu(x)<\infty.
	\end{displaymath}
	A $\mu\kappa$-typicality criterion $\mathcal{V}:=(\mathscr{G};\epsilon;K)$ is 
	said to be \emph{log-exponential} with respect to $(\mu,\kappa)$, if each 
	$g\in\mathscr{G}$ is log-exponential with respect to $(\mu,\kappa)$. 
\end{defn}

The definition above seems quite artificial, but it appears naturally when one tries to 
make an exponential decay of error probability in conditional typicality lemma. Also, note 
that when pointwise values of $\kappa$ are Gaussian where variances are 
uniformly bounded, a function of at most quadratic order will become a 
log-exponential test function. 

One can easily see by using Jensen's inequality that an exponentially integrable function is 
always log-exponential with respect to any decomposition of the measure into a marginal and a 
corresponding conditional distribution (a function $f$ is \emph{exponentially integrable}, 
if $2^{\delta\abs{f}}$ is integrable for some $\delta>0$). 
A sum of a log-exponential function and a bounded measurable function is again 
log-exponential. Now we extend the bounded conditional typicality lemma in terms of 
log-exponential typical sets: 

\begin{thrm}[Log-exponential conditional typicality lemma]\label{thrm:log-exp cond typ}\ \\
	Let $\mu\in\Delta(X)$ and $\kappa\in\mathcal{K}(X;Y)$. Then for any $\mu\kappa$-typicality 
	criterion $\mathcal{V}$ log-exponential with respect to $(\mu,\kappa)$, there exists a 
	$\mu$-typicality criterion $\mathcal{U}$ and a positive number $c>0$ such that 
	\begin{displaymath}
		\sup_{x^{n}\in\mathcal{T}_{\mathcal{U}}^{(n)}(\mu)}
		\kappa^{n}\left(Y^{n}\setminus\mathcal{T}_{\mathcal{V}}^{(n)}(\mu\kappa|x^{n})
		\Big|x^{n}\right)\leq2^{-cn}
	\end{displaymath}
	for all sufficiently large $n\in\Z^{+}$. 
\end{thrm}

\begin{IEEEproof}
	We may assume that $\mathcal{V}=(g;\epsilon;K)$ as usual. Find a positive real 
	number $\delta>0$ such that 
	\begin{displaymath}
		\int\log\left[\int2^{\delta\abs{g(x,y)}}\,d\kappa(y|x)\right]d\mu(x)<\infty.
	\end{displaymath}
	For each $k\in\Z^{+}$, define the truncation $g_{k}:X\times Y\ra\R$ as 
	\begin{displaymath}
		g_{k}:(x,y)\mapsto\begin{cases}g(x,y)&\textrm{if $\abs{g(x,y)}\leq k$}\\
		0&\textrm{otherwise}\end{cases},
	\end{displaymath}
	and also define a measurable function $h_{k}:X\ra\R$ as 
	\begin{displaymath}
		h_{k}:x\mapsto\log\left(
		\int2^{\delta\abs{g(x,y)-g_{k}(x,y)}}\,d\kappa(y|x)\right),
	\end{displaymath}
	then the Lebesgue dominated convergence theorem guarantees that we can choose 
	$k\in\Z^{+}$ such that 
	\begin{displaymath}
		\abs{\int g\,d\mu\kappa-\int g_{k}\,d\mu\kappa}\leq\frac{\epsilon}{3}\quad
		\textrm{and}\quad\int h_{k}\,d\mu\leq\frac{\epsilon\delta}{12}.
	\end{displaymath}
	Let $\mathcal{V}_{k}:=\left(g_{k};\frac{\epsilon}{3};K\right)$, then there 
	exists a bounded $\mu$-typicality criterion $\mathcal{U}_{k}$ and a positive 
	real number $c_{1}>0$ such that 
	\begin{displaymath}
		\sup_{x^{n}\in\mathcal{T}_{\mathcal{U}_{k}}^{(n)}(\mu)}
		\kappa^{n}\left(Y^{n}\setminus\mathcal{T}_{\mathcal{V}_{k}}^{(n)}
		(\mu\kappa|x^{n})\Big|x^{n}\right)\leq2^{-c_{1}n}
	\end{displaymath}
	for sufficiently large $n\in\Z^{+}$ by applying the bounded conditional 
	typicality lemma. Define $\mathcal{U}:=\mathcal{U}_{k}\wedge
	\left(h_{k};\frac{\epsilon\delta}{12};\emptyset\right)$. Now, fix $n\in\Z^{+}$ 
	large enough so that the above inequality holds, and let 
	$x^{n}\in\mathcal{T}_{\mathcal{U}}^{(n)}(\mu)$. Consider the set 
	\begin{displaymath}
		Z:=\setbc{y^{n}\in Y^{n}}{\abs{\frac{1}{n}\sum_{i=1}^{n}g(x_{i},y_{i})
		-\frac{1}{n}\sum_{i=1}^{n}g_{k}(x_{i},y_{i})}\geq\frac{\epsilon}{3}}, 
	\end{displaymath}
	then clearly $Z$ is contained in 
	\begin{displaymath}
		\setbc{y^{n}\in Y^{n}}{\prod_{i=1}^{n}
		2^{\delta\abs{g(x_{i},y_{i})-g_{k}(x_{i},y_{i})}}
		\geq2^{\frac{\epsilon\delta}{3}n}},
	\end{displaymath}
	so by Chevychev's inequality, 
	\begin{displaymath}
		\kappa^{n}(Z|x^{n})\leq2^{-\frac{\epsilon\delta}{3}n}\prod_{i=1}^{n}
		\int2^{\delta\abs{g(x_{i},y_{i})-g_{k}(x_{i},y_{i})}}\,d\kappa(y_{i}|x_{i})
		=2^{-\frac{\epsilon\delta}{3}n}2^{\sum_{i=1}^{n}h_{k}(x_{i})}. 
	\end{displaymath}
	Since $x^{n}\in\mathcal{T}_{\mathcal{U}}^{(n)}(\mu)$, we know that 
	\begin{displaymath}
		\frac{1}{n}\sum_{i=1}^{n}h_{k}(x_{i})\leq
		\int h_{k}\,d\mu+\frac{\epsilon\delta}{12}\leq\frac{\epsilon\delta}{6},
	\end{displaymath}
	so it follows that $\kappa^{n}(Z|x^{n})\leq2^{-\frac{\epsilon\delta}{6}n}$. 
	Note that if $y^{n}\in\mathcal{T}_{\mathcal{V}_{k}}^{(n)}(\mu\kappa)\setminus Z$, 
	then $(x_{i},y_{i})\notin K$ for $i=1,\ \cdots\ ,n$ and 
	\begin{eqnarray*}
		&&\abs{\frac{1}{n}\sum_{i=1}^{n}g(x_{i},y_{i})-\int g\,d\mu\kappa}\leq
		\abs{\frac{1}{n}\sum_{i=1}^{n}g(x_{i},y_{i})-
		\frac{1}{n}\sum_{i=1}^{n}g_{k}(x_{i},y_{i})}\\
		&&\quad\quad\quad\quad
		+\abs{\frac{1}{n}\sum_{i=1}^{n}g_{k}(x_{i},y_{i})-
		\int g_{k}\,d\mu\kappa}+\abs{\int g_{k}\,d\mu\kappa-\int g\,d\mu\kappa}\\
		&&\quad\quad\quad\leq
		\frac{\epsilon}{3}+\frac{\epsilon}{3}+\frac{\epsilon}{3}=\epsilon. 
	\end{eqnarray*}
	Therefore, we get 
	\begin{eqnarray*}
		\kappa^{n}\left(Y^{n}\setminus\mathcal{T}_{\mathcal{V}}^{(n)}(\mu\kappa|x^{n})
		\Big|x^{n}\right)&\leq&
		\kappa^{n}\left(Y^{n}\setminus\mathcal{T}_{\mathcal{V}_{k}}^{(n)}
		(\mu\kappa|x^{n})\Big|x^{n}\right)+\kappa^{n}(Z|x^{n})\\
		&\leq&2^{-c_{1}n}+2^{-\frac{\epsilon\delta}{6}n}
	\end{eqnarray*}
	for all $x^{n}\in\mathcal{T}_{\mathcal{U}}^{(n)}(\mu)$ provided that 
	$n$ is sufficiently large. 
\end{IEEEproof}

By using the log-exponential conditional typicality lemma instead of the 
bounded conditional typicality lemma in the proof of \refthrm{thrm:bdd Markov}, 
we get the following: 

\begin{coro}[Log-exponential Markov lemma]\ \\
	\refthrm{thrm:bdd Markov} is still true when $\mathcal{W}$ is log-exponential 
	with respect to $(\mu\kappa,\lambda)$. 
\end{coro}

From Jensen's inequality, one can easily see that a test function is log-exponential 
with respect to $(\mu\kappa,\lambda)$ if it is log-exponential with respect to 
$(\mu,\kappa\times\lambda)$. 

%%%%%%%%%%%%%%%%%%%%%%%%%%%%%%%%%%%%%%%%%%%%%%%%%%%%%%%%%%%%%%%%%%%%%%%%%%%%%%%%%%%%%%%%%%%%%%%%%%
\section{The Gaussian case}
Most of results about typical sets given in this paper were just described in terms of 
\emph{existence} of typical sets satisfying some properties. It was not necessary to be 
careful about the actual contents in the typicality criteria. Unfortunately, to 
apply the log-exponential Markov lemma, we should keep track the list of test functions inside 
the given typicality criteria, because we have to know whether or not those 
functions are log-exponential. The aim of this section is to establish a claim 
saying that \emph{we still do not need to care about those things when everything is 
Gaussian}. As discussed in the previous section, a function of at most quadratic order is 
log-exponential with respect to Gaussian measures. What we will show here is that indeed 
functions of at most quadratic order are sufficient to build the whole theory 
when every measure is Gaussian. 

The main motivation of this is the fact that the 
test function which appears in the proof of the joint typicality lemma was the 
logarithm of a Radon-Nikodym derivative; when measures are Gaussian, this function 
may become a quadratic function. To be precise, let us formalize our discussions. 
We will consider (possibly singular) Gaussian measures on Euclidean spaces. 
Let us denote the Gaussian measure on $\R^{d}$ of mean $m$ and (possibly singular) 
covariance matrix $\Sigma$ as $\mathrm{N}^{d}(m,\Sigma)$. 
We review some simple facts about Gaussian measures: 
\begin{enumerate}
	\item Let $\mu\in\Delta(\R^{d_{1}})$ and $\kappa\in\mathcal{K}(\R^{d_{1}};\R^{d_{2}})$. 
	Then $\mu\kappa$ is a Gaussian measure on $\R^{d_{1}+d_{2}}$, if and only if, $\mu$ is 
	a Gaussian measure on $\R^{d_{1}}$ and $\kappa:x\mapsto\mathrm{N}^{d_{2}}(Ax+b,\Lambda)$ 
	for some $d_{2}\times d_{1}$ matrix $A$ and a vector $b\in\R^{d_{2}}$, and a 
	$d_{2}\times d_{2}$ positive-semidefinite matrix $\Lambda$~\cite{Ramon:1997}. 
	\item Any Gaussian measure on $\R^{d_{1}}$ is an affine transformation of the standard 
	Gaussian measure $\mathrm{N}^{d_{2}}(0,I)$ (here, $I$ is the $d_{2}\times d_{2}$ 
	identity matrix) where $d_{2}$ is the rank of the covariance matrix. To see why, let 
	$\mu=\mathrm{N}^{d_{1}}(m,\Sigma)$ and let $d_{2}$ be the rank of $\Sigma$. Since 
	$\Sigma$ is symmetric, 
	it is orthogonally diagonalizable~\cite[p.247]{RomanAlgebra:2007}; hence, we can 
	write $\Sigma=P^{T}DP$ for some orthogonal matrix $P$ and a diagonal matrix $D$. 
	We may assume that $D=\begin{bmatrix}N&0\\0&0\end{bmatrix}$ where $N$ is a 
	$d_{2}\times d_{2}$ diagonal matrix of positive entries. Then the affine transform 
	$y\mapsto P^{T}\begin{bmatrix}N^{1/2}\\0\end{bmatrix}y+m$ ($y$ is a column vector 
	of length $d_{2}$) sends the standard Gaussian 
	measure $\mathrm{N}^{d_{2}}(0,I)$ into $\mu$. 
	\item If $\mu,\nu$ are Gaussian measures on $\R^{d_{1}}$ and $\mu\ll\nu$, then 
	$\mu$ and $\nu$ should have the same support. To see why, first find an 
	affine map $T:\R^{d_{2}}\ra\R^{d_{1}}$ sending the standard Gaussian meaure 
	to $\nu$, then from this it is clear that the support of $\nu$ is the 
	affine subspace $T[\R^{d_{2}}]$ of $\R^{d_{1}}$. Then since 
	$\nu\ll T_{*}\mathrm{m}$ where $\mathrm{m}$ is the Lebesgue measure on $\R^{d_{2}}$, 
	it follows that $\mu\ll T_{*}\mathrm{m}$, so the support of $\mu$ is contained in 
	$T[\R^{d_{2}}]$. Since $T^{-1}:T[\R^{d_{2}}]\ra\R^{d_{2}}$ is an affine isomorphism, 
	we can think of the pushforward $(T^{-1})_{*}\mu$. This is a Gaussian measure 
	on $\R^{d_{2}}$ which is absolutely continuous with respect to the Lebesgue 
	measure; hence, it should be non-singular. Thus, in fact, $\nu\ll\mu$ as well. 
	We can also compute the Radon-Nikodym derivative $\frac{d\mu}{d\nu}$: let 
	$\mu':=(T^{-1})_{*}\mu$ and $\nu':=(T^{-1})_{*}\nu$, then 
	$\frac{d\mu}{d\nu}\circ T=\frac{d\mu'}{d\nu'}$ since 
	\begin{displaymath}
		\int_{B}\left(\frac{d\mu}{d\nu}\circ T\right)d\nu'
		=\int_{T[B]}\frac{d\mu}{d\nu}\,dT_{*}\nu'
		=\int_{T[B]}\frac{d\mu}{d\nu}d\nu=\mu(T[B])=\mu'(B)
	\end{displaymath}
	for any Borel subset $B$ of $\R^{d_{2}}$. Now, let 
	$\mu'=\mathrm{N}^{d_{2}}(m,\Sigma)$ then 
	\begin{displaymath}
		\frac{d\mu}{d\nu}\circ T(z)=\frac{1}{\abs{\Sigma}^{1/2}}
		\exp\left(\frac{1}{2}\left(
		\norm{z}^{2}-(z-m)^{T}\Sigma^{-1}(z-m)\right)\right).
	\end{displaymath}
	\item The product of Gaussian measures is Gaussian: 
	$\mathrm{N}^{d_{1}}(a,\Sigma)\times\mathrm{N}^{d_{2}}(b,\Lambda)
	=\mathrm{N}^{d_{1}+d_{2}}((a,b),\Sigma\oplus\Lambda)$, where 
	$\Sigma\oplus\Lambda=\begin{bmatrix}\Sigma&0\\0&\Lambda\end{bmatrix}$. 
\end{enumerate}

We first formally define functions of quadratic order 
as functions which grow not faster than sum of a constant with the norm-square function. 

\begin{defn}[Quadratic typical sets]\ \\
	Let $\mu\in\Delta(\R^{d})$. A $\mu$-integrable function 
	$f:\R^{d}\ra\R$ is said to be \emph{$\mu$-quadratic}, if there exists a $\mu$-null set 
	$N$ and a constant $M\geq0$ such that $\abs{f(x)}\leq M(1+\norm{x}^{2})$ whenever 
	$x\in\R^{d}\setminus N$. A $\mu$-typicality criterion $\mathcal{U}$ is said to be a 
	\emph{$\mu$-quadratic typicality criterion}, if each test function in $\mathcal{U}$ 
	is $\mu$-quadratic. A $\mu$-typical set with respect to a $\mu$-quadratic typicality 
	criterion is called a \emph{$\mu$-quadratic typical set}. 
\end{defn}

The pullback of a quadratic typicality criteria under an affine map is again quadratic. 

\begin{prop}\ \\
	Let $\mu\in\Delta(\R^{d_{1}})$ and $T:\R^{d_{1}}\ra\R^{d_{2}}$ be an affine map. 
	If $g:\R^{d_{2}}\ra\R$ is $T_{*}\mu$-quadratic, then $g\circ T$ is 
	$\mu$-quadratic. Hence for a $T_{*}\mu$-quadratic typicality criterion $\mathcal{V}$, 
	the pullback $T^{*}\mathcal{V}$ is a $\mu$-quadratic typicality criterion. 
\end{prop}

\begin{IEEEproof}
	Take a $T_{*}\mu$-null set $K$ and a constant $M_{1}\geq0$ such that 
	$\abs{g(y)}\leq M_{1}(1+\norm{y}^{2})$. 
	Clearly, $T^{-1}[K]$ is a $\mu$-null set and there exists $M_{2}\geq0$ such 
	that $\norm{T(x)}^{2}\leq M_{2}(1+\norm{x}^{2})$ for all $x\in\R^{d_{1}}$ 
	since $T$ is affine. Then for $x\in\R^{d_{1}}\setminus T^{-1}[K]$, 
	\begin{displaymath}
		\abs{g\circ T(x)}\leq M_{1}(1+\norm{T(x)}^{2})
		\leq M_{1}(1+M_{2})+M_{1}M_{2}\norm{x}^{2},
	\end{displaymath}
	thus $g\circ T$ is $\mu$-quadratic. 
\end{IEEEproof}

We prove that any quadratic test function is indeed log-exponential with respect to 
a Gaussian measure. In fact, it is even exponentially integrable. 

\begin{lemm}\ \\
	Let $\mu$ be a Gaussian measure on $\R^{d}$ and $f:\R^{d}\ra\R$ a 
	$\mu$-quadratic function. Then, $f$ is exponentially integrable 
	with respect to $\mu$. 
\end{lemm}

\begin{IEEEproof}
	Let $T:\R^{d'}\ra\R^{d}$ be an affine map sending the standard Gaussian 
	measure $\lambda:=\mathrm{N}^{d'}(0,I)$ to $\mu$. Since $f$ is $\mu$-quadratic, 
	$f\circ T$ is $\lambda$-quadratic, so there exists a $\lambda$-null set 
	$N$ and a constant $M\geq0$ such that 
	$\abs{f\circ T(z)}\leq M(1+\norm{z}^{2})$ for all $z\in\R^{d'}\setminus N$. Then, 
	\begin{eqnarray*}
		\int_{\R^{d}}2^{\delta\abs{f(x)}}\,d\mu(x)
		&=&\int_{\R^{d}}2^{\delta\abs{f(x)}}\,dT_{*}\lambda(x)
		=\int_{\R^{d'}\setminus N}2^{\delta\abs{f\circ T(z)}}\,d\lambda(z)\\
		&\leq&\int_{\R^{d'}}\exp\left(\delta M+\delta M\norm{z}^{2}\right)
		\frac{1}{(2\pi)^{d'/2}}\exp\left(-\frac{\norm{z}^{2}}{2}\right)\,dz\\
		&=&\frac{e^{\delta M}}{(2\pi)^{d'/2}}\int_{\R^{d'}}\exp\left(
		-\frac{1}{2}(1-2\delta M)\norm{z}^{2}\right)\,dz<\infty
	\end{eqnarray*}
	when $\delta\in\left(0,\frac{1}{2M}\right)$. 
\end{IEEEproof}

From now on, we prove that all the results derived in Section III, Section IV, and 
Section VI can be written in terms of quadratic typical sets when all the involved 
measures are jointly Gaussian.  

\begin{thrm}[Gaussian conditional typicality lemma]\ \\
	Let $\mu\in\Delta(\R^{d_{1}})$ and $\kappa\in\mathcal{K}(\R^{d_{1}};\R^{d_{2}})$ 
	be jointly Gaussian. Then for any $\mu\kappa$-quadratic typicality 
	criterion $\mathcal{V}$, there exists a $\mu$-quadratic typicality 
	criterion $\mathcal{U}$ and a positive number $c>0$ such that 
	\begin{displaymath}
		\sup_{x^{n}\in\mathcal{T}_{\mathcal{U}}^{(n)}(\mu)}
		\kappa^{n}\left(Y^{n}\setminus\mathcal{T}_{\mathcal{V}}^{(n)}(\mu\kappa|x^{n})
		\Big|x^{n}\right)\leq2^{-cn}
	\end{displaymath}
	for sufficiently large $n\in\Z^{+}$. 
\end{thrm}

\begin{IEEEproof}
	Since bounded functions are clearly $\mu$-quadratic, we only need to check that 
	the function 
	\begin{displaymath}
		h_{k}:x\mapsto\log\left(\int2^{\delta\abs{g(x,y)-g_{k}(x,y)}}\,d\kappa(y|x)\right)
	\end{displaymath}
	appearing in the proof of \refthrm{thrm:log-exp cond typ} is $\mu$-quadratic 
	when we have chosen $\delta>0$ sufficiently small. First, find a constant $M\geq0$ 
	such that $\abs{g(x,y)-g_{k}(x,y)}\leq M(1+\norm{x}^{2}+\norm{y}^{2})$ whenever 
	$(x,y)\in(\R^{d_{1}}\times\R^{d_{2}})\setminus K$ for some $\mu\kappa$-null set 
	$K$. For each $x\in\R^{d_{1}}$, define $K_{x}:=\setbc{y\in\R^{d_{2}}}{(x,y)\in K}$, 
	then there exists a $\mu$-null set $N$ so that $\kappa(K_{x}|x)=0$ 
	whenever $x\in\R^{d_{1}}\setminus N$. Then for such $x$, 
	\begin{eqnarray*}
		h_{k}(x)&=&\log\left(\int_{\R^{d_{2}}\setminus K_{x}}2^{\delta\abs{
		g(x,y)-g_{k}(x,y)}}\,d\kappa(y|x)\right)\\
		&\leq&\delta M(1+\norm{x}^{2})+\log\left(\int_{\R^{d_{2}}}
		2^{\delta M\norm{y}^{2}}\,d\kappa(y|x)\right).
	\end{eqnarray*}
	Since $\mu$ and $\kappa$ are jointly Gaussian, we can write 
	$\kappa(x)=\mathrm{N}^{d_{2}}(Ax+b,\Lambda)$ for each $x\in\R^{d_{1}}$. 
	Therefore, there is a linear map $B:\R^{d}\ra\R^{d_{2}}$ such that 
	the affine map $T_{x}:z\mapsto Bz+Ax+b$ maps the standard Gaussian measure 
	$\lambda:=\mathrm{N}^{d}(0,I)$ to $\kappa(x)$, where $d$ is the rank of 
	$\Lambda$. Note that $d,A,B,b$ does not depend on $x$. Hence, we can write 
	\begin{eqnarray*}
		\int_{\R^{d_{2}}}2^{\delta M\norm{y}^{2}}\,d\kappa(y|x)&=&
		\int_{\R^{d_{2}}}2^{\delta M\norm{y}^{2}}\,dT_{x*}\lambda(y)\\
		&=&\int_{\R^{d}}2^{\delta M\norm{Bz+Ax+b}^{2}}\,d\lambda(z)\\
		&\leq&2^{2\delta M\norm{Ax+b}^{2}}
		\int_{\R^{d}}2^{2\delta M\norm{Bz}^{2}}\,d\lambda(z). 
	\end{eqnarray*}
	Since $\lambda$ is the standard Gaussian measure, one can show by direct 
	computation that whenever $\delta>0$ is sufficiently small, we have 
	\begin{displaymath}
		M_{1}:=\int_{\R^{d}}2^{2\delta M\norm{Bz}^{2}}\,d\lambda(z)<\infty. 
	\end{displaymath}
	A precise upper bound on $\delta$ only depends on $B$ and $M$, 
	so it follows that 
	\begin{displaymath}
		h_{k}(x)\leq\delta M(1+\norm{x}^{2})
		+2\delta M\norm{Ax+b}^{2}+\log M_{1}
		\leq C(1+\norm{x}^{2})
	\end{displaymath}
	for some constant $C\geq0$, whenever $x\in\R^{d}\setminus N$. 
	Therefore, $h_{k}$ is $\mu$-quadratic. 
\end{IEEEproof}

\begin{thrm}[Gaussian conditional divergence lemma]\ \\
	In the statement of the conditional divergence lemma, let $X=\R^{d_{1}}$, 
	$Y=\R^{d_{2}}$, and both $\mu\kappa$ and $\mu\lambda$ be Gaussian. 
	Then $D(\mu\kappa\|\mu\lambda)$ always exists and nonnegative, and 
	$D(\mu\kappa\|\mu\lambda)=\infty$ if and only if $\mu\kappa\not\ll\mu\lambda$. 
	Also, $\mathcal{V}_{0}$ in the statement can be found as a 
	$\mu\kappa$-quadratic typicality criterion for all cases. Furthermore, for the 
	case when $D(\mu\kappa\|\mu\lambda)$ is finite, $\mathcal{U}$, which depends on 
	$\mathcal{V}$, can be also found to be $\mu$-quadratic whenever $\mathcal{V}$ 
	is a $\mu\kappa$-quadratic typicality criterion. 
\end{thrm}

\begin{IEEEproof}
	It is trivial that $D(\mu\kappa\|\mu\lambda)$ always exists and nonnegative, 
	since both $\mu\kappa$ and $\mu\lambda$ are probability measures. 
	\begin{enumerate}
		\item (For $D(\mu\kappa\|\mu\lambda)<\infty$) The only test 
		function involved in $\mathcal{V}_{0}$ in the proof of the conditional 
		divergence lemma is $\log\frac{d\mu\kappa}{d\mu\lambda}$. We show that 
		this is a $\mu\kappa$-quadratic function. As remarked before, 
		there is an affine map $T:\R^{d}\ra\R^{d_{1}+d_{2}}$ sending a 
		Gaussian measure $\mathrm{N}^{d}(m,\Sigma)$ to $\mu\kappa$ 
		and sending the standard Gaussian measure $\mathrm{N}^{d}(0,I)$ to 
		$\mu\lambda$, and  
		\begin{displaymath}
			\log\frac{d\mu\kappa}{d\mu\lambda}\circ T(z)
			=\frac{1}{2}\left(\norm{z}^{2}-(z-m)^{T}\Sigma^{-1}(z-m)\right)
			-\frac{1}{2}\log\abs{\Sigma}.
		\end{displaymath}
		Let $K=\R^{d_{1}+d_{2}}\setminus T[\R^{d}]$, then $K$ is a 
		$\mu\kappa$-null set, and write $T^{-1}:T[\R^{d}]\ra\R^{d}$ 
		as $T^{-1}(x,y)=A(x,y)+b$ for a $d\times(d_{1}+d_{2})$ matrix $A$ and a 
		column vector $b\in\R^{d}$. Then for $(x,y)\in\R^{d_{1}+d_{2}}\setminus K$, 
		\begin{displaymath}
			\log\frac{d\mu\kappa}{d\mu\lambda}(x,y)
			=\frac{1}{2}\left(\norm{A(x,y)+b}^{2}-(A(x,y)+b-m)^{T}\Sigma^{-1}
			(A(x,y)+b-m)\right)-\frac{1}{2}\log\abs{\Sigma}, 
		\end{displaymath}
		so $\log\frac{d\mu\kappa}{d\mu\lambda}$ is clearly $\mu\kappa$-quadratic. 
		Therefore, $\mathcal{V}_{0}$ can be chosen to be $\mu\kappa$-quadratic. 
		To show the claim about $\mathcal{U}$, note that in the proof of 
		the conditional divergence lemma, $\mathcal{U}$ can be taken to be 
		$\mu$-quadratic by applying the Gaussian conditional typicality lemma 
		instead of the usual conditional typicality lemma, whenever $\mathcal{V}$ 
		is given to be $\mu\kappa$-quadratic. 
		
		\item (For $\mu\kappa\not\ll\mu\lambda$) The test function chosen in 
		the proof of the conditional divergence lemma is a bounded function, 
		so the conclusion is trivial. 
		
		\item (For $\mu\kappa\ll\mu\lambda$ but $D(\mu\kappa\|\mu\lambda)=\infty$) 
		This case cannot happen, since any $\mu\kappa$-quadratic 
		function is $\mu\kappa$-integrable, and $\log\frac{d\mu\kappa}{d\mu\lambda}$ 
		is $\mu\kappa$-quadratic as proved in the case $1$. 
	\end{enumerate}
\end{IEEEproof}

Since joint typicality lemma is just a specialization of conditional divergence lemma, 
it can be also stated in terms of quadratic typical sets. Packing and covering lemmas 
(as well as their ``mutual versions'') are consequences of conditional typicality 
lemma and joint typicality lemma, so they also can be stated in terms of 
quadratic typical sets. Now Markov lemma is the only remaining: 

\begin{thrm}[Gaussian Markov lemma]\ \\
	Let $\mu\in\Delta(\R^{d_{1}})$, $\kappa\in\mathcal{K}(\R^{d_{1}};\R^{d_{2}})$, and 
	$\lambda\in\mathcal{K}(\R^{d_{1}};\R^{d_{3}})$ so that both $\mu\kappa$ and 
	$\mu\lambda$ are Gaussian. For each $n\in\Z^{+}$ and a $\mu\lambda$-quadratic 
	typicality criterion $\mathcal{S}$, let $\lambda_{\mathcal{S}}^{(n)}
	\in\mathcal{K}(\R^{d_{1}n};\R^{d_{3}n})$ (which is not necessarily Gaussian). 
	Assume that, for any $\epsilon>0$, there exists a $\mu\lambda$-quadratic typicality 
	criterion $\mathcal{S}_{0}$ so that for any $\mu\lambda$-quadratic typicality 
	criterion $\mathcal{S}\leq\mathcal{S}_{0}$, one can find a $\mu$-quadratic 
	typicality criterion $\mathcal{U}$, satisfying 
	\begin{displaymath}
		\lambda_{\mathcal{S}}^{(n)}(E|x^{n})
		\leq2^{\epsilon n}\lambda^{n}(E|x^{n})
	\end{displaymath}
	for all $x^{n}\in\mathcal{T}_{\mathcal{U}}^{(n)}(\mu)$ and a measurable subset 
	$E$ of $\mathcal{T}_{\mathcal{S}}^{(n)}(\mu\lambda|x^{n})$, whenever 
	$n$ is sufficiently large. Then for any $\mu(\kappa\times\lambda)$-quadratic 
	typicality criterion $\mathcal{W}$, there exists a $\mu\lambda$-quadratic 
	typicality criterion $\mathcal{S}_{0}$ and a positive 
	number $c>0$ such that, for any $\mu\lambda$-quadratic typicality criterion 
	$\mathcal{S}\leq\mathcal{S}_{0}$, there exists a $\mu\kappa$-quadratic 
	typicality criterion $\mathcal{V}$ so that 
	\begin{displaymath}
		\sup_{(x^{n},y^{n})\in\mathcal{T}_{\mathcal{V}}^{(n)}(\mu\kappa)}
		\lambda_{\mathcal{S}}^{(n)}\left(\mathcal{T}_{\mathcal{S}}^{(n)}
		(\mu\lambda|x^{n})\setminus\mathcal{T}_{\mathcal{W}}^{(n)}
		(\mu(\kappa\times\lambda)|x^{n},y^{n})\Big|x^{n}\right)\leq2^{-cn}.
	\end{displaymath}
	for sufficiently large $n$. 
\end{thrm}

\begin{IEEEproof}
	Use the Gaussian conditional typicality lemma instead of the 
	bounded conditional typicality lemma in the proof of the bounded Markov lemma. 
\end{IEEEproof}

It is now clear that there should be no problem to directly apply the same 
derivation of an inner bound of a given discrete memoryless coding problem 
relying on those fundamental lemmas to the corresponding Gaussian memoryless coding problem. 
However, this does not mean that we have the same formula for an achievable region. For example, 
consider the quadratic Gaussian distributed source coding problem~\cite{Oohama:1997}: we have a 
jointly Gaussian random sources $\bold{x}_{1},\bold{x}_{2}$, which are 
encoded separately at rates $R_{1}$, $R_{2}$, respectively, and then decoded jointly. 
The distortion criteria is given as 
\begin{displaymath}
	\mathrm{E}\left[\frac{1}{n}\sum_{i=1}^{n}
	(\bold{x}_{1i}-\hat{\bold{x}}_{1i})^{2}\right]\leq D_{1},\quad\quad
	\mathrm{E}\left[\frac{1}{n}\sum_{i=1}^{n}
	(\bold{x}_{2i}-\hat{\bold{x}}_{2i})^{2}\right]\leq D_{2}
\end{displaymath}
while $\hat{\bold{x}}_{1}^{n}$ and $\hat{\bold{x}}_{2}^{n}$ are reconstructions of 
$\bold{x}_{1}^{n}$ and $\bold{x}_{2}^{n}$ at the decoder, respectively. 
Here, the theory of quadratic typical sets does not immediately give the 
following Berger-Tung inner bound~\cite{Berger:1978}\cite{Tung:1978}:
\begin{eqnarray*}
	R_{1}&>&I(\bold{x}_{1};\bold{u}_{1}|\bold{u}_{2}),\\
	R_{2}&>&I(\bold{x}_{2};\bold{u}_{2}|\bold{u}_{1}),\\
	R_{1}+R_{2}&>&I(\bold{x}_{1},\bold{x}_{2};\bold{u}_{1},\bold{u}_{2})
\end{eqnarray*}
for some auxiliary random variables $\bold{u}_{1},\bold{u}_{2}$ satisfying 
the Markov chain $\bold{u}_{1}-\bold{x}_{1}-\bold{x}_{2}-\bold{u}_{2}$ and 
measurable functions $\hat{x}_{1},\hat{x}_{2}$ such that 
$\mathrm{E}\left[\norm{\bold{x}_{1}-\hat{x}_{1}(\bold{u}_{1},\bold{u}_{2})}^{2}\right]
\leq D_{1}$ and $\mathrm{E}\left[\norm{\bold{x}_{2}-\hat{x}_{2}
(\bold{u}_{1},\bold{u}_{2})}^{2}\right]\leq D_{2}$. What we can say immediately using the 
theory of quadratic typical sets is that, the above inner bound holds when the joint 
distribution of $(\bold{x}_{1},\bold{x}_{2},\bold{u}_{1},\bold{u}_{2},
\hat{x}_{1}(\bold{u}_{1},\bold{u}_{2}),\hat{x}_{2}(\bold{u}_{1},\bold{u}_{2}))$ 
is Gaussian. That is, all variables including not only the variables 
stated in the problem but also auxiliary variables, should have a jointly Gaussian distribution. 
For the case of quadratic Gaussian distributed source coding problem, the optimal choice 
of auxiliary variables are indeed Gaussian~\cite{Wagner:2008}, but one cannot be sure 
that this will always be the case for other problems. Yet, when Markov lemma was not 
necessary, we can apply the theory of general typical sets rather than quadratic typical 
sets so such restriction need not to be concerned. 

%%%%%%%%%%%%%%%%%%%%%%%%%%%%%%%%%%%%%%%%%%%%%%%%%%%%%%%%%%%%%%%%%%%%%%%%%%%%%%%%%%%%%%%%%%%%%%%%%%
\section{Some Remarks on Sources with Memory}
We have discussed a generalization of strong typicality which can be applied to a 
wide range of sources without memory. Perhaps, it is possible to extend the concept 
of typical sets to sources with memory. Such an extension will enable generalization of 
many results about memoryless problems into problems containing sources or channels with memory. 
It is not certain whether such generalizations are useful in practice or not, because the 
obtained results will be multi-letter characterizations; however, finding the ``ultimate'' 
definition of typical sets which can be applied to a very large range of sources is theoretically 
appealing. The idea of the extension will be the same: consider a finite collection 
of test functions. However, it is not obvious to say \emph{what} are test functions. 
The Shannon-McMillan-Breiman theorem~\cite{Breiman:1957} and its extension to 
random sequences of continuous variables~\cite{Barron:1985} suggests that it is natural to define 
weak typical sets of a stationary ergodic stochastic process 
$\bold{x}:=\seq{\bold{x}_{k}}_{k\in\Z}$ with well-defined joint densities as 
\begin{displaymath}
	\mathcal{A}_{\epsilon}^{(n)}\left(\bold{x}\right)
	:=\setbc{(x_{0},\ \cdots\ ,x_{n-1})\in\R^{n}}{\abs{-\frac{1}{n}
	\log p_{n}(x_{0},\ \cdots\ ,x_{n-1})-h(\bold{x})}\leq\epsilon}
\end{displaymath}
where $p_{n}$ is the joint pdf of $\seq{\bold{x}_{i}}_{i=0}^{n-1}$ and 
\begin{displaymath}
	h(\bold{x}):=\lim_{n\ra\infty}\frac{1}{n}h(\bold{x}_{0},\ \cdots\ ,\bold{x}_{n-1})
\end{displaymath}
is the differential entropy rate. Taking this as a motivating example, we can conclude that, 
rather than to consider a single test function, we should consider a 
\emph{sequence of test functions} for sources with memory. 

For a memoryless source, we have defined typical sets with respect to only the marginal 
probability distribution. For a source with memory (that is, a random sequence), we should 
deal with the whole probability distribution on the space of \emph{sequences of symbols}. 
This space can be viewed as a single probability space endowed with a measurable self-map 
called the \emph{shift map}, representing the flow of time. One may argue that this dynamical 
system is ``the essence'' of the random sequence, so it seems natural that we should 
think of the definition of typical sets that can be given for general dynamical systems. 

Let us restrict ourselves to consider only invertible ergodic measure-preserving dynamical 
systems~\cite{PetersenErgodic:1989} (for example, bidirectional stationary ergodic random 
sequences). In the motivating example, we can write 
\begin{displaymath}
	\frac{1}{n}\log p_{n}(x_{0},\ \cdots\ ,x_{n-1})
	=\frac{1}{n}\sum_{i=0}^{n-1}\log p_{i}(x_{i}|x_{0},\ \cdots\ ,x_{i-1})
\end{displaymath}
where $p_{i}(x_{i}|x_{0},\ \cdots\ ,x_{i-1})$ is the conditional pdf of $\bold{x}_{i}$ 
given $(\bold{x}_{0},\ \cdots\ ,\bold{x}_{i-1})$. If we define 
\begin{displaymath}
	f_{i}\left(\seq{x_{k}}_{k\in\Z}\right):=
	\log p_{i}(x_{0}|x_{-1},x_{-2},\ \cdots\ ,x_{-i+1})
\end{displaymath}
for each $i$, then 
\begin{displaymath}
	\frac{1}{n}\log p_{n}(x_{0},\ \cdots\ ,x_{n-1})
	=\frac{1}{n}\sum_{i=0}^{n-1}f_{i}\left(T^{i}\seq{x_{k}}_{k\in\Z}\right)
\end{displaymath}
where 
\begin{displaymath}
	T:\seq{x_{k}}_{k\in\Z}\mapsto\seq{x_{k+1}}_{k\in\Z}
\end{displaymath}
is the shift map. Note also that 
\begin{displaymath}
	\mathrm{E}\left[f_{i}\left(\seq{\bold{x}_{k}}_{k\in\Z}\right)\right]
	=-h(\bold{x}_{0}|\bold{x}_{-1},\ \cdots\ ,\bold{x}_{-i+1})
	=-h(\bold{x}_{i}|\bold{x}_{1},\ \cdots\ ,\bold{x}_{i-1}), 
\end{displaymath}
so 
\begin{displaymath}
	\lim_{i\ra\infty}\mathrm{E}\left[f_{i}
	\left(\seq{\bold{x}_{k}}_{k\in\Z}\right)\right]
	=-\lim_{i\ra\infty}h(\bold{x}_{i}|\bold{x}_{1},\ \cdots\ ,\bold{x}_{i-1})
	=-h(\bold{x}).
\end{displaymath}
Therefore, the weak typical set is the projection onto $\R^{n}$ of the following set: 
\begin{displaymath}
	\setbc{\seq{x_{k}}_{k\in\Z}\in\R^{\Z}}{\abs{\frac{1}{n}\sum_{i=0}^{n-1}
	f_{i}\left(T^{i}\seq{x_{k}}_{k\in\Z}\right)-\lim_{i\ra\infty}\mathrm{E}\left[
	f_{i}\left(\seq{\bold{x}_{k}}_{k\in\Z}\right)\right]}\leq\epsilon}. 
\end{displaymath}
Thus, a typical set for an invertible ergodic measure-preserving 
dynamical system $(X,\mathscr{A},\mu,T)$ may look like 
\begin{displaymath}
	\mathcal{T}_{\mathcal{U}}^{(n)}(\mu,T):=
	\setbc{x\in X\setminus N}{\abs{\frac{1}{n}\sum_{i=0}^{n}
	f_{i}(T^{i}x)-\lim_{i\ra\infty}\int f_{i}\,d\mu}\leq\epsilon\quad
	\textrm{for all $\seq{f_{i}}_{i=0}^{\infty}\in\mathscr{F}$}}
\end{displaymath}
where $N$ is a $\mu$-null set, $\mathscr{F}$ is a finite collection of 
``test sequences'' $\seq{f_{i}}_{i=0}^{\infty}$ of measurable functions on $X$, and 
$\mathcal{U}=(\mathscr{F};\epsilon;N)$. A test sequence may not be an arbitrary 
sequence of measurable functions, and there should be some conditions to be satisfied. 
The following generalization of the Birkhoff's ergodic theorem given 
in~\cite{Breiman:1957} suggests a possible class of test sequences: 

\begin{thrm}[Breiman, 1957]\label{thrm:Breiman}\ \\
	Let $(X,\mathscr{A},\mu,T)$ be an ergodic measure-preserving dynamical system. 
	Let $\seq{f_{i}}_{i=0}^{\infty}$ be a sequence of measurable functions on 
	$X$ such that $\int\sup_{i}\abs{f_{i}}\,d\mu<\infty$ that is convergent 
	$\mu$-almost everywhere to some function $f$. Then, 
	\begin{displaymath}
		\lim_{n\ra\infty}\frac{1}{n}\sum_{i=0}^{n-1}f_{i}(T^{i}x)=\int f\,d\mu
	\end{displaymath}
	for $\mu$-almost every $x\in X$. 
\end{thrm}

This theorem gives a sort of asymptotic equipartition property. 
According to~\cite{Barron:1985}, some results discussed in this paper 
(for example, the divergence lemma) are expected to be generalized to the case of 
stationary ergodic sources (in fact, as depicted in~\cite{Barron:1985}, 
\refthrm{thrm:Breiman} can be stated for possibly non-ergodic stationary sources in terms of 
conditional expectations, so it is possible to think of an even more general case 
of such sources). However, the situation is more 
complicated than the memoryless case, because the Hoeffding's inequality does not hold in 
general for dependent random variables. There are some generalizations of the Hoeffding's 
inequality, such as the Azuma's inequality~\cite{Azuma:1967}, but it is still not clear that what 
restrictions on the class of test sequences lead us to the most natural definition of 
typical sets for sources with memory. 

%%%%%%%%%%%%%%%%%%%%%%%%%%%%%%%%%%%%%%%%%%%%%%%%%%%%%%%%%%%%%%%%%%%%%%%%%%%%%%%%%%%%%%%%%%%%%%%%%%
\section{Conclusion}
A new notion of typical sets for a general class of memoryless sources was defined, which properly 
generalizes the conventional notion of strong typical sets. It turns out that the weak typicality 
is also a special case of the proposed notion. The definition is based on an observation that 
typical average lemma is the one validating most of useful properties of strong 
typical sets. Some similar approaches already exist, including \cite{Mitran:2010} and 
\cite{Raginsky:2013}, but the new notion will be more appropriate for network 
information theory in the sense that, many technical lemmas, including 
conditional typicality lemma, joint typicality lemma, and packing and covering lemmas, 
can be easily generalized in a completely rigorous manner. Together with Markov lemma 
introduced in \cite{Berger:1978} and \cite{Tung:1978}, these lemmas have been the main tools 
for deriving inner bounds of many multi-terminal coding problems. It was explicitly shown 
that some classical coding theorems can be generalized in a straightforward way 
only with very little technical assumptions. On the other hand, Markov lemma also has 
been generalized in restrictive ways, but this limitation causes no problem especially 
when the joint probability distribution is Gaussian and every involved test function 
is at most of quadratic order. However, still more improvements are desired to get a 
better theory. Also, there may be a notion of typicality generalizing the introduced 
notion further to include sources with memory, but this task is not seem to be simple.

\section*{Acknowledgment}

This work was supported by MSIP as GFP/(CISS-2012M3A6A6054195). 
The author would like to thank Prof. Sae-Young Chung for his guidance and useful 
discussions with him. The author also would like to thank Seung uk Jang for his 
careful verification of statements and proofs. Suggestion of the terminology 
``test functions'' of an anonymous reviewer is appreciated as well. 

\appendix
Here, several folklore lemmas are collected. 

\begin{lemm}\label{lemm:joint RND}\ \\
	Let $(X,\mathscr{A})$ be a measurable space and $(Y,\mathscr{B})$ be a 
	countably-generated measurable space. 
	Let $\mu\in\Delta(X)$ and $\kappa:X\ra\Delta(Y)$ be a 
	probability kernel. Let $\lambda:X\ra\mathcal{P}(Y)$ be a 
	\mbox{$\sigma$-finite} positive measure kernel with 
	$\mu\kappa\ll\mu\lambda$. Fix a Radon-Nikodym derivative 
	$g=\frac{d\mu\kappa}{d\mu\lambda}$, then there exists a 
	$\mu$-null set $N$ so that $\kappa(x)\ll\lambda(x)$ and $g(x,\cdot)$ is a 
	Radon-Nikodym derivative of $\kappa(x)$ with respect to $\lambda(x)$ 
	for all $x\in X\setminus N$. 
\end{lemm}

\begin{IEEEproof}
	Let $\mathscr{B}_{0}$ be the algebra generated by a countable generator 
	of $\mathscr{B}$. Then $\mathscr{B}_{0}$ is countable. 
	Fix $B\in\mathscr{B}_{0}$, then for any $A\in\mathscr{A}$ we have 
	\begin{eqnarray*}
		\int_{A}\left[\int_{B}g(x,y)\,d\lambda(y|x)\right]d\mu(x)
		&=&\int_{A\times B}\frac{d\mu\kappa}
		{d\mu\lambda}\,d\mu\lambda\\&=&\mu\kappa(A\times B)=
		\int_{A}\kappa(B|x)\,d\mu(x),
	\end{eqnarray*}
	so there exists a $\mu$-null set $N_{B}$ such that 
	\begin{displaymath}
		\int_{B}g(x,y)\,d\lambda(y|x)=\kappa(B|x)
	\end{displaymath}
	for all $x\in X\setminus N_{B}$. Let $N\defas\bigcup_{B\in\mathscr{B}_{0}}N_{B}$ 
	and fix $x\in X\setminus N$. Define 
	\begin{displaymath}
		\mathscr{C}\defas\setbc{B\in\mathscr{B}}
		{\int_{B}g(x,y)\,d\lambda(y|x)=\kappa(B|x)}
	\end{displaymath}
	then we have proved that $\mathscr{B}_{0}\subseteq\mathscr{C}$. We claim that 
	$\mathscr{C}=\mathscr{B}$. Since $\mathscr{B}_{0}$ is an algebra, it suffices 
	to show that $\mathscr{C}$ is a monotone class, by the monotone 
	class theorem~\cite[p.18]{RobertProb:2000}. Let $\seq{B_{k}}_{k\in\Z^{+}}$ be an 
	increasing sequence in $\mathscr{C}$ and $B\defas\bigcup_{k\in\Z^{+}}B_{k}$, then it 
	follows by monotone convergence theorem and countable-additivity of $\kappa(x)$ that 
	\begin{displaymath}
		\int_{B}g(x,y)\,d\lambda(y|x)=\lim_{k\ra\infty}
		\int_{B_{k}}g(x,y)\,d\lambda(y|x)
		=\lim_{k\ra\infty}\kappa(B_{k}|x)=\kappa(B|x),
	\end{displaymath}
	so $B\in\mathscr{C}$. Similarly, let $\seq{B_{k}}_{k\in\Z^{+}}$ be a 
	decreasing sequence in $\mathscr{C}$ and $B\defas\bigcap_{k\in\Z^{+}}B_{k}$, then 
	it follows by the Lebesgue dominated convergence theorem 
	and the countable-additivity of $\kappa(x)$ that 
	\begin{displaymath}
		\int_{B}g(x,y)\,d\lambda(y|x)=\lim_{k\ra\infty}
		\int_{B_{k}}g(x,y)\,d\lambda(y|x)
		=\lim_{k\ra\infty}\kappa(B_{k}|x)=\kappa(B|x),
	\end{displaymath}
	so $B\in\mathscr{C}$. This proves the claim, so we have 
	\begin{displaymath}
		\int_{B}g(x,y)\,d\lambda(y|x)=\kappa(B|x)
	\end{displaymath}
	for all $B\in\mathscr{B}$. Therefore, it follows that 
	$\kappa(x)\ll\lambda(x)$ and $g(x,\cdot)$ is a 
	Radon-Nikodym derivative of $\kappa(x)$ with respect to $\lambda(x)$. Since 
	$\mathscr{B}_{0}$ is countable, $N$ is a $\mu$-null set. Hence, we get the conclusion. 
\end{IEEEproof}

\begin{lemm}\label{lemm:convexity mutual}\ \\
	Let $(X,\mathscr{A})$ and $(Y,\mathscr{B})$ be measurable spaces. 
	Then, the function $I:\Delta(X)\times\mathcal{K}(X;Y)\ra[0,\infty]$ defined as 
	\begin{displaymath}
		I:(\mu,\kappa)\mapsto 
		D(\mu\kappa\|\mu\times\kappa_{*}\mu)
	\end{displaymath}
	is concave in the first variable and convex in the second variable. 
\end{lemm}

\begin{IEEEproof}
	Let $\Pi(X)$ be the set of all canonical projections from $X$ onto 
	finite measurable partitions of $X$. Let $\Pi(Y)$ be similarly defined. 
	Then we can write~\cite{GrayEntropy:2011} 
	\begin{displaymath}
		I:(\mu,\kappa)\mapsto\sup_{\mathcal{P}\in\Pi(X),
		\mathcal{Q}\in\Pi(Y)}I(\mathcal{P}_{*}\mu,\mathcal{Q}_{*}\kappa)
	\end{displaymath}
	where we define $\mathcal{Q}_{*}\kappa:x\mapsto\mathcal{Q}_{*}\kappa(x)$. 
	To prove concavity in the first variable, let $\kappa\in\mathcal{K}(X;Y)$, 
	$\mu_{1},\mu_{2}\in\Delta(X)$, $\lambda\in[0,1]$, and 
	$\mu\defas\lambda\mu_{1}+(1-\lambda)\mu_{2}$. We may assume that 
	$I(\mu,\kappa)<\infty$, then for given $\epsilon>0$, there exists $\mathcal{P}\in\Pi(X)$ 
	and $\mathcal{Q}\in\Pi(Y)$ such that 
	\begin{displaymath}
		I(\mu,\kappa)\leq I(\mathcal{P}_{*}\mu,\mathcal{Q}_{*}\kappa)+\epsilon
		=I(\lambda\mathcal{P}_{*}\mu_{1}+(1-\lambda)\mathcal{P}_{*}\mu_{2},
		\mathcal{Q}_{*}\kappa)+\epsilon.
	\end{displaymath}
	Since $I(\,\cdot\,,\,\cdot\,)$ is concave in the first variable 
	when the alphabets are finite~\cite[p.33]{CoverElements:2006}, 
	\begin{eqnarray*}
		I(\mu,\kappa)&\leq&I(\lambda\mathcal{P}_{*}\mu_{1}
		+(1-\lambda)\mathcal{P}_{*}\mu_{2},\mathcal{Q}_{*}\kappa)+\epsilon\\
		&\leq&\lambda I(\mathcal{P}_{*}\mu_{1},\mathcal{Q}_{*}\kappa)+
		(1-\lambda)I(\mathcal{P}_{*}\mu_{2},\mathcal{Q}_{*}\kappa)+\epsilon\\
		&\leq&\lambda I(\mu_{1},\kappa)+(1-\lambda)I(\mu_{2},\kappa)+\epsilon.
	\end{eqnarray*}
	Since $\epsilon>0$ is arbitrary, concavity of $I$ in the first variable 
	is proved. To prove convexity in the second variable, let $\mu\in\Delta(X)$, 
	$\kappa_{1},\kappa_{2}\in\mathcal{K}(X;Y)$, $\lambda\in[0,1]$, and 
	$\kappa\defas\lambda\kappa_{1}+(1-\lambda)\kappa_{2}$. Then, 
	\begin{eqnarray*}
		I(\mu,\kappa)&=&\sup_{\mathcal{P}\in\Pi(X),\mathcal{Q}\in\Pi(Y)}
		I(\mathcal{P}_{*}\mu,\mathcal{Q}_{*}\kappa)\\&=&
		\sup_{\mathcal{P}\in\Pi(X),\mathcal{Q}\in\Pi(Y)}
		I(\mathcal{P}_{*}\mu,\lambda\mathcal{Q}_{*}\kappa_{1}
		+(1-\lambda)\mathcal{Q}_{*}\kappa_{2})\\&\leq&
		\sup_{\mathcal{P}\in\Pi(X),\mathcal{Q}\in\Pi(Y)}\left(
		\lambda I(\mathcal{P}_{*}\mu,\mathcal{Q}_{*}\kappa_{1})+(1-\lambda)
		I(\mathcal{P}_{*}\mu,\mathcal{Q}_{*}\kappa_{2})\right)\\&\leq&
		\lambda\sup_{\mathcal{P}\in\Pi(X),\mathcal{Q}\in\Pi(Y)}
		I(\mathcal{P}_{*}\mu,\mathcal{Q}_{*}\kappa_{1})+(1-\lambda)
		\sup_{\mathcal{P}\in\Pi(X),\mathcal{Q}\in\Pi(Y)}
		I(\mathcal{P}_{*}\mu,\mathcal{Q}_{*}\kappa_{2})\\
		&=&\lambda I(\mu,\kappa_{1})+(1-\lambda)I(\mu,\kappa_{2}),
	\end{eqnarray*}
	thus convexity of $I$ in the second variable is also proved. 
\end{IEEEproof}



%\bibliographystyle{IEEEtran}
%\begin{thebibliography}{1}
%\end{thebibliography}

\bibliographystyle{IEEEtran}
\bibliography{IEEEabrv,References}

\begin{thebibliography}{10}
\providecommand{\url}[1]{#1}
\csname url@samestyle\endcsname
\providecommand{\newblock}{\relax}
\providecommand{\bibinfo}[2]{#2}
\providecommand{\BIBentrySTDinterwordspacing}{\spaceskip=0pt\relax}
\providecommand{\BIBentryALTinterwordstretchfactor}{4}
\providecommand{\BIBentryALTinterwordspacing}{\spaceskip=\fontdimen2\font plus
\BIBentryALTinterwordstretchfactor\fontdimen3\font minus
  \fontdimen4\font\relax}
\providecommand{\BIBforeignlanguage}[2]{{%
\expandafter\ifx\csname l@#1\endcsname\relax
\typeout{** WARNING: IEEEtran.bst: No hyphenation pattern has been}%
\typeout{** loaded for the language `#1'. Using the pattern for}%
\typeout{** the default language instead.}%
\else
\language=\csname l@#1\endcsname
\fi
#2}}
\providecommand{\BIBdecl}{\relax}
\BIBdecl

\bibitem{CoverElements:2006}
T.~M. Cover and J.~A. Thomas, \emph{Elements of Information Theory},
  2nd~ed.\hskip 1em plus 0.5em minus 0.4em\relax Wiley, 2006.

\bibitem{OrlitskyRoche:2001}
A.~Orlitsky and J.~R. Roche, ``Coding for computing,'' \emph{{IEEE} Trans. Inf.
  Theory}, vol.~47, no.~3, pp. 903--917, Mar. 2001.

\bibitem{ElGamalNetwork:2011}
A.~{El Gamal} and Y.-H. Kim, \emph{Network Information Theory}.\hskip 1em plus
  0.5em minus 0.4em\relax Cambridge University Press, 2011.

\bibitem{Mitran:2010}
\BIBentryALTinterwordspacing
P.~Mitran, ``Typical sequences for {P}olish alphabets,'' \emph{arXiv.org
  preprint}, May. 2010. [Online]. Available:
  \url{http://arxiv.org/abs/1005.2321}
\BIBentrySTDinterwordspacing

\bibitem{Raginsky:2013}
M.~Raginsky, ``Empirical processes, typical sequences and coordinated actions
  in standard {B}orel spaces,'' \emph{{IEEE} Trans. Inf. Theory}, vol.~59,
  no.~3, pp. 1288--1301, Mar. 2013.

\bibitem{CohnMeasure:1980}
D.~L. Cohn, \emph{Measure Theory}.\hskip 1em plus 0.5em minus 0.4em\relax
  Birkh{\"a}user, 1980.

\bibitem{Siu-Wai:2010}
S.-W. Ho and R.~W. Yeung, ``On information divergence measures and a unified
  typicality,'' \emph{{IEEE} Trans. Inf. Theory}, vol.~56, no.~12, pp.
  5893--5905, Dec. 2010.

\bibitem{RoydenReal:2010}
H.~L. Royden and P.~M. Fitzpatrick, \emph{Real Analysis}, 4th~ed.\hskip 1em
  plus 0.5em minus 0.4em\relax Pearson, 2010.

\bibitem{LangReal:1993}
S.~Lang, \emph{Real and Functional Analysis}, 3rd~ed., ser. Graduate Texts in
  Mathematics.\hskip 1em plus 0.5em minus 0.4em\relax Springer-Verlag, 1993.

\bibitem{RobertProb:2000}
R.~B. Ash and C.~A. Dol{\'e}ans-Dade, \emph{Probability \& Measure Theory},
  2nd~ed.\hskip 1em plus 0.5em minus 0.4em\relax Academic Press, 2000.

\bibitem{GrayEntropy:2011}
R.~M. Gray, \emph{Entropy and Information Theory}, 2nd~ed.\hskip 1em plus 0.5em
  minus 0.4em\relax Springer, 2011.

\bibitem{PollardUserGuide:1981}
D.~Pollard, \emph{A User's Guide to Measure Theoretic Probability}.\hskip 1em
  plus 0.5em minus 0.4em\relax Cambridge University Press, 2002.

\bibitem{Hoeffding:1963}
W.~Hoeffding, ``Probability inequalities for sums of bounded random
  variables,'' \emph{J. Amer. Statist. Assoc.}, vol.~58, no. 301, pp. 13--30,
  Mar. 1963.

\bibitem{ParthasarathyMeas:1967}
K.~R. Parthasarathy, \emph{Probability Measures on Metric Spaces}.\hskip 1em
  plus 0.5em minus 0.4em\relax Academic Press Inc., 1967.

\bibitem{Wyner:1978}
A.~D. Wyner, ``A definition of conditional mutual information for arbitrary
  ensembles,'' \emph{Inf. Control}, vol.~38, no.~1, pp. 51--59, Jul. 1978.

\bibitem{Ramon:1997}
R.~A. Albajar and J.~F.~L. Fidalgo, ``Characterizing the general multivariate
  normal distribution through the conditional distributions,'' \emph{Extracta
  Math.}, vol.~12, no.~1, pp. 15--18, 1997.

\bibitem{RomanAlgebra:2007}
S.~Roman, \emph{Advanced Linear Algebra}, 3rd~ed., ser. Graduate Texts in
  Mathematics.\hskip 1em plus 0.5em minus 0.4em\relax Springer, 2007.

\bibitem{Oohama:1997}
Y.~Oohama, ``Gaussian multiterminal source coding,'' \emph{{IEEE} Trans. Inf.
  Theory}, vol.~43, no.~6, pp. 1912--1922.

\bibitem{Berger:1978}
T.~Berger, ``Multiterminal source coding,'' in \emph{The Information Theory
  Approach to Communications}, G.~Longo, Ed.\hskip 1em plus 0.5em minus
  0.4em\relax Springer-Verlag, New York, 1978, pp. 171--231.

\bibitem{Tung:1978}
S.-Y. Tung, ``Multiterminal source coding,'' Ph.D. dissertation, Cornell
  University, May, 1978.

\bibitem{Wagner:2008}
A.~B. Wagner, S.~Tavildar, and P.~Viswanath, ``Rate region of the quadratic
  gaussian two-encoder source-coding problem,'' \emph{{IEEE} Trans. Inf.
  Theory}, vol.~54, no.~5, pp. 1938--1961.

\bibitem{Breiman:1957}
L.~Breiman, ``The individual ergodic theorem of information theory,''
  \emph{Ann. Math. Stat.}, vol.~28, no.~3, pp. 809--811, Sep. 1957.

\bibitem{Barron:1985}
A.~R. Barron, ``The strong ergodic theorem for densities: Generalized
  shannon-mcmillan-breiman theorem,'' \emph{Ann. Prob.}, vol.~13, no.~4, pp.
  1292--1303, Nov. 1985.

\bibitem{PetersenErgodic:1989}
K.~Petersen, \emph{Ergodic Theory}, reprint~ed.\hskip 1em plus 0.5em minus
  0.4em\relax Cambridge University Press, 1989.

\bibitem{Azuma:1967}
K.~Azuma, ``Weighted sums of certain dependent random variables,'' \emph{Tohoku
  Math. J.}, vol.~19, no.~3, pp. 357--367, 1967.

\end{thebibliography}

\end{document}